\documentclass{article}

\usepackage{hyperref}
\pdfoutput=1

\usepackage[utf8]{inputenc}
\usepackage{amsmath}
\usepackage{amssymb}
\usepackage{booktabs}
\usepackage{graphicx}
\usepackage{grffile}
\usepackage{subfigure}
\usepackage{marvosym} 
\usepackage{color}
\usepackage{textcomp} 
\usepackage{array}
\usepackage{wasysym} 
\usepackage{tikz}
\usepackage{natbib}
\usepackage[nottoc]{tocbibind}
\usepackage{longtable}
\usepackage{multicol}
\usepackage{multirow}
\usepackage{geometry}

\newcommand{\wPlot}{0.45\textwidth}

\newcommand{\wOnePointFive}{0.8\textwidth}
\newcommand{\wTwo}{0.45\textwidth}

\newcommand{\wFour}{0.24\textwidth}

\def\mathnote#1{%
  \tag*{\rlap{\hspace\marginparsep\smash{\parbox[t]{\marginparwidth}{#1}}}}
}

\definecolor{colorAffiliation}{rgb}{1, 0.4, 0.4}
\definecolor{colorAnimal}{rgb}{0.24, 0.84, 0.32}
\definecolor{colorAuthorship}{rgb}{0.8, 0, 1}
\definecolor{colorCitation}{rgb}{0.10, 0.16, 0.50}
\definecolor{colorCoauthorship}{rgb}{0.81, 0.90, 0.12}
\definecolor{colorCocitation}{rgb}{0.79, 0.58, 0.15}
\definecolor{colorCommunication}{rgb}{0, 0, 1}
\definecolor{colorComputer}{rgb}{0.56, 0.32, 0.69}
\definecolor{colorFeature}{rgb}{0.55, 0.55, 1}
\definecolor{colorHumanContact}{rgb}{0.86, 0.81, 0.59}
\definecolor{colorHumanSocial}{rgb}{0.42, 0.42, 0.37}
\definecolor{colorHyperlink}{rgb}{1, 0.62, 0}
\definecolor{colorInfrastructure}{rgb}{1.0, 0.2, 0.2}
\definecolor{colorInteraction}{rgb}{0.25, 0.75, 0.25}
\definecolor{colorLexical}{rgb}{0, 1, 0}
\definecolor{colorMetabolic}{rgb}{0.58, 0.70, 0.67}
\definecolor{colorMisc}{rgb}{0.85, 0.95, 0.59}
\definecolor{colorNeural}{rgb}{0.11, 0.56, 0.33}
\definecolor{colorOnlineContact}{rgb}{0.27, 0.54, 0.95}
\definecolor{colorRating}{rgb}{0.70, 0.70, 0}
\definecolor{colorSocial}{rgb}{0.10, 0.10, 0.38}
\definecolor{colorSoftware}{rgb}{0.10, 0.37, 0.10}
\definecolor{colorText}{rgb}{0.43, 0.52, 0.82}
\definecolor{colorTrophic}{rgb}{0.40, 0.42, 0.61}

\definecolor{InternalLinkColor}{rgb}{0.0, 0.35, 0.0}
\definecolor{ExternalLinkColor}{rgb}{0.0, 0.0, 0.69}
\hypersetup{ 
    colorlinks=true, 
    citecolor=InternalLinkColor, 
    urlcolor=ExternalLinkColor, 
    linkcolor=InternalLinkColor
}

\begin{document}

\title{
  \includegraphics[width=2.8cm]{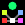} \\
  \vspace{0.9cm}
         {\Huge Handbook of Network Analysis\footnote{This handbook is
             continuously updated; an up-to-date version can always be found at
         \href{https://github.com/kunegis/konect-handbook/raw/master/konect-handbook.pdf}{https://github.com/kunegis/konect-handbook/raw/master/konect-handbook.pdf}.}} \\
         The KONECT Project \\
         \texttt{\href{http://KONECT.cc/}{KONECT.cc}}
}

\author{
  Jérôme Kunegis 
}

\newgeometry{top=4cm}
\maketitle

\section*{Abstract}
This is the handbook for the KONECT project, a scientific project to
archive network datasets, compute systematic network theoretic
statistics about them, visualize their properties, and provide
corresponding data and Free Software tools to programmers, researchers and teachers in 
fields related to network analysis and graph mining, by
Jérôme Kunegis.  The name \emph{KONECT} stands for
\emph{\underline{Ko}blenz \underline{Ne}twork \underline{C}ollec\underline{t}ion}, as parts of the KONECT project
were initiated for the PhD thesis of Jérôme Kunegis at
the University of Koblenz--Landau in Germany \citeyearpar{kunegis:phd}.  
This handbook
documents all methods, definitions and conventions used in the project,
and serves as a general handbook of network mining, with an emphasis on
spectral graph theoretical methods, i.e., such methods that are based on
the use of specific characteristic matrices of graphs. 
KONECT datasets and code is used in academia as the basis for research
on real-world datasets, in education as the basis for teaching, and
in particular it serves as the basis for research projects with the goal
to study large numbers of network datasets.  KONECT borrows 
datasets from many sources in and outside academia, and lends itself
datasets to other network dataset collection projects. 
\thispagestyle{empty}
\restoregeometry
\newpage

\section{Introduction}
Everything is a network~-- whenever we look at the interactions between
things, a network is formed implicitly.  In the areas of data mining,
machine learning, information retrieval, etc., networks are modeled
as \emph{graphs}.  Many, if not most problem types can be applied to
graphs: clustering, classification, prediction, pattern recognition, and
others.  Networks arise in almost all areas of research, commerce and
daily life in the form of social networks, road networks, communication
networks, trust networks, hyperlink networks, chemical interaction
networks, neural networks, collaboration networks and lexical networks.
The content of text documents is routinely modeled as document--word
networks, taste as person--item networks and trust as person--person
networks.  In recent years, whole database systems have appeared
specializing in storing networks.  In fact, a majority of research
projects in the areas of web mining, web science and related areas uses
datasets that can be understood as networks.  Unfortunately, results
from the literature can often not be compared easily because
they use different datasets. What is more, different network datasets
have slightly different properties, such as allowing multiple or only
single edges between two nodes.  In order to provide a unified view on
such network datasets, and to allow the application of network analysis
methods across disciplines, the KONECT project defines a comprehensive
network taxonomy and provides a consistent access to network datasets.
To validate this approach on real-world data from the Web, KONECT
also provides a large number (1,000+) of network datasets of different
types and different application areas. 

\begin{figure}
  \centering
  \includegraphics[width=0.8\textwidth]{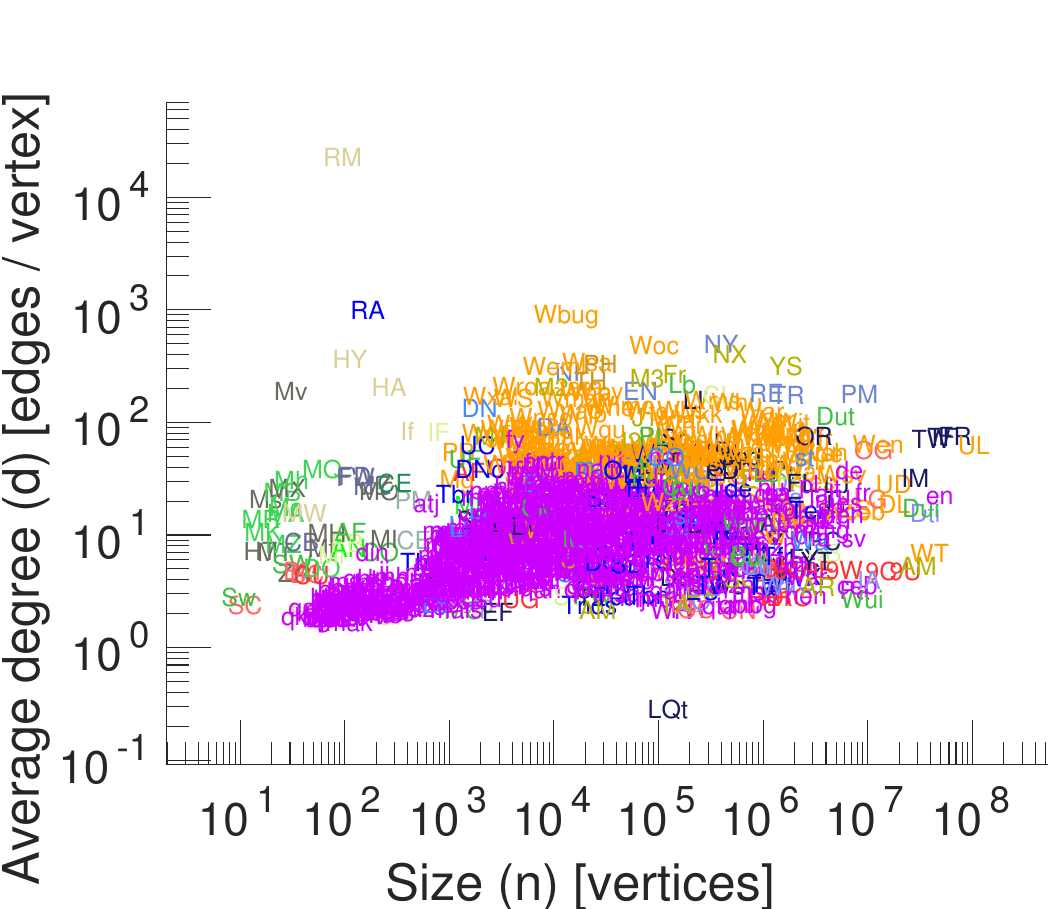}
  \caption[*]{
    All networks in KONECT
    arranged by the size (the number of nodes) and the
    average number of neighbors of all nodes.  Each network is
    represented by a two- or three-character code. The color of each
    code corresponds to the network 
    category as given in Table~\ref{tab:categories}.  
  }
  \label{fig:scatter.size.avgdegree}
\end{figure}

KONECT, the \underline{Ko}blenz \underline{Ne}twork \underline
Collec\underline tion, contains over 1,000 network datasets as of 2017.
In addition to these datasets, KONECT consists of Matlab code to
generate statistics and plots about them, which are shown on the
KONECT
website\footnote{\href{http://konect.cc/}{http://konect.cc/}}.
KONECT contains networks of all sizes, from small classical datasets
from the social sciences such as Kenneth Read's Highland Tribes network
with 16 vertices and 58 edges
(\href{http://konect.cc/networks/ucidata-gama/}{\textsf{HT}}),
to the Twitter social network with 52 million nodes and 1.9 billion
edges
(\href{http://konect.cc/networks/twitter_mpi/}{\textsf{TF}}).
Figure~\ref{fig:scatter.size.avgdegree} shows a scatter plot of all
networks by the number of nodes and the average degree in the network.
Each network in KONECT is represented by a unique two- or
three-character code which we write in a \textsf{sans-serif font}, and
is indicated in parentheses as used previously in this paragraph. The
full list of codes is given
online.\footnote{\href{http://konect.cc/networks/}{http://konect.cc/networks/}}

\subsection{Software and Software Packages}
\begin{figure}
  \centering
  \includegraphics[width=\wOnePointFive]{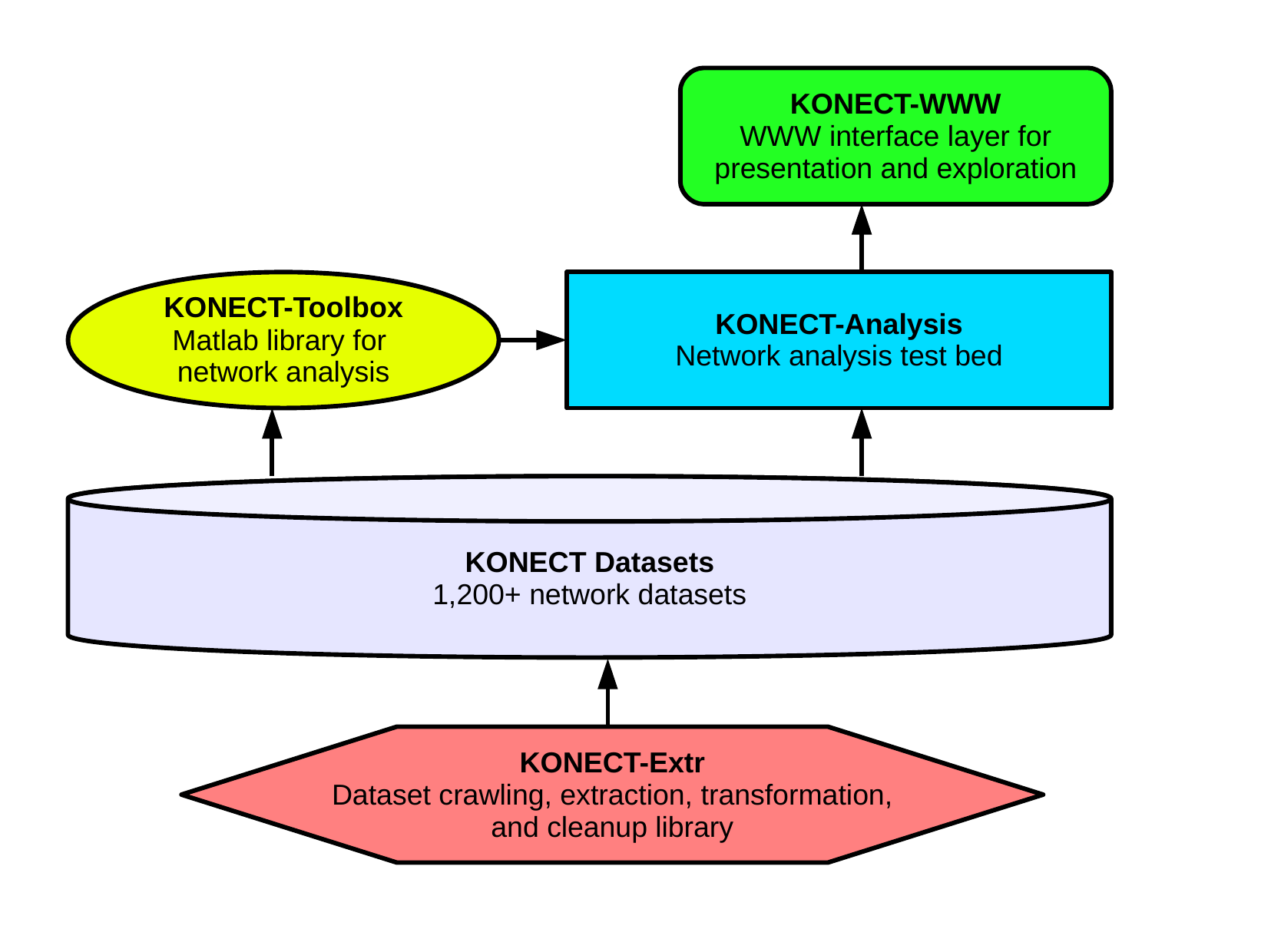}
  \caption{
    \label{fig:organization}
    Overview of KONECT's components. 
  }
\end{figure}

The KONECT project consists of several components, whose interactions is
summarized in Figure~\ref{fig:organization}.  Various parts of the
KONECT project are available at Github, including this Handbook.\footnote{\href{https://github.com/kunegis/konect-analysis}{github.com/kunegis/konect-analysis}}\footnote{\href{https://github.com/kunegis/konect-toolbox}{github.com/kunegis/konect-toolbox}}\footnote{\href{https://github.com/kunegis/konect-handbook}{github.com/kunegis/konect-handbook}}\footnote{\href{https://github.com/kunegis/konect-extr}{github.com/kunegis/konect-extr}}\footnote{\href{https://github.com/kunegis/konect-www}{github.com/kunegis/konect-www}}

\subsection{History of KONECT}
The roots of KONECT lie in the research of Jérôme Kunegis at the
Technical University of Berlin, within the DAI Laboratory.  The first
networks were bipartite rating graphs, collected to support Jérôme
Kunegis' work on collaborative filtering
\citep{kunegis:negative-resistance,kunegis:adapting-ratings}.  The
earliest networks include MovieLens, Jester, BookCrossing, and the Netflix Prize
dataset. 

What later became known as the KONECT project's collection of networks
properly started out in December 2008,
as evaluation for Jérôme Kunegis' ICML 2009
paper \emph{Learning Spectral Graph Transformations for Link Prediction}
\citep{kunegis:spectral-transformation}, codenamed \emph{Spectral
  Transformation}.  It then consisted of a collection of network
datasets and spectral link prediction methods, i.e., code based on the
decomposition of various characteristic graph matrices.  
The first unipartite networks appeared at this time, one of the earliest
being the trust network of Advogato. 
The first dataset crawled specifically for KONECT was the Slashdot Zoo,
for which crawling began in 2008, with the corresponding paper published
a year later \citep{kunegis:slashdot-zoo}. 

Later, more datasets
were added and the codebase was called the \emph{Graph Store}.  This was
at a time when the word \emph{store} was in fashion, in part due to the
emergence of \emph{app stores}. 
In that phase, the project was used for the experiments of several papers in the area of
collaborative filtering and recommender systems.  

When Jérôme moved from
TU Berlin to the University of Koblenz--Landau in Koblenz (Germany) the
project was renamed \emph{Web Store}, in line with Koblenz' Institute
for Web Science and Technologies (WeST).  The name \emph{KONECT --
  \underline{Ko}blenz \underline{Ne}twork
  \underline{C}ollec\underline{t}ion} was adopted sometime in 2011.  
Jérôme wrote his PhD thesis at the University of Koblenz--Landau in the
same year. 
In that phase, most of the mathematical notation of KONECT was
settled.  In particular, Jérôme Kunegis' PhD thesis contains a
mathematical glossary that contains many of the symbol in the present
document.  That PhD thesis was also the first time that a list of all
networks was generated, before the website went online. 
The first KONECT website was created in 2011 at \texttt{konect.uni-koblenz.de}.
In this phase of the project, extraction of datasets and maintenance of the
KONECT website was performed in collaboration with Daniel Dünker, Holger
Heinz, and Martina Sekulla. 

Code for dataset extraction and the Matlab Toolbox were first published
on the KONECT website as downloadable tarballs.  All published code was
licensed under the GNU General Public License (GPL) version~3 from the beginning. 
A short overview paper of the KONECT system was
published in 2013 at the International World Wide Web Conference (WWW),
as part of the Web Observatory Workshop \citep{kunegis:konect}\footnote{
  The Google Scholar citation page for that publication gives a list of
  papers using KONECT, see
  \href{https://scholar.google.com/scholar?cites=7174338004474749050}{https://scholar.google.com/scholar?cites=7174338004474749050}},
and remains the primary publication to cite when using KONECT. 
In 2015 and 2016, various parts of the KONECT project were placed on
GitHub, again under the GNU
General Public License version~3, including this handbook.  

From 2017 on, the KONECT project continued to be developed at the University of
Namur (Belgium), with web hosting provided by the Institute for Web
Science and Technologies (WeST) at the University of
Koblenz--Landau in Koblenz (Germany).  In that phase, the number of
networks, statistics, and plots was yet again increased, and a more
powerful computation server was acquired.  The domain name
\href{http://konect.cc/}{\texttt{konect.cc}} was adopted in October
2017, with hosting now switched to the University of Namur. 
A tutorial on the use of large network dataset collections, largely based
on KONECT, was given in the same year at the Conference on Information and Knowledge Management (CIKM~2017) under the name \emph{Network Analysis
  in the Age of Large Network Dataset Collections -- Challenges,
  Solutions and Applications}.\footnote{\href{http://xn.unamur.be/network-collection-tutorial-cikm2017/}{http://xn.unamur.be/network-collection-tutorial-cikm2017/}}

The Stu build
tool\footnote{\href{https://github.com/kunegis/stu}{https://github.com/kunegis/stu}}
was developed in parallel with KONECT.  KONECT used \texttt{make(1)} in
the first years of its operation.  The Stu build system was initiated in
2014, and KONECT was gradually switched to it, serving as its main use
case.  Stu was essentially completed in August 2017 with publication of version
2.5, but continues to see occasional improvements.  

As Jérôme Kunegis left academia in 2023 to join the industrial world in 2023,
the KONECT website continued to be available.  A few years later, the site stopped
operating.  This handbook has not been continued.

\subsection{Status of this Handbook}
This handbook is constantly updated.  The most up to date version can
always be found on
GitHub.\footnote{\href{https://github.com/kunegis/konect-handbook/raw/master/konect-handbook.pdf}{https://github.com/kunegis/konect-handbook/raw/master/konect-handbook.pdf}}  
Approximately every year, a version of the handbook is published on
arXiv, as a new version of the existing arXiv
entry.\footnote{\href{http://arxiv.org/abs/1402.5500}{http://arxiv.org/abs/1402.5500}} 
The handbook is explicitly \emph{not} peer reviewed, as we update it as
needed.  Since the handbook serves as the central place in which we add
new definitions, the reader will notice that the prose may vary greatly
from section to section.  Also, some notation used may be very establish
and will not change, but other choices in notation are not set in stone,
and we reserve the possibility to change our notation to suit our
needs.  However, any change in notation will be reflected in KONECT
source code and on the KONECT world wide web site.  While it is
perfectly acceptable to cite this handbook, we urge readers not to refer
to any equation or section number as these \emph{will} change
constantly.  

What will \emph{not} change, by design, are all internal
names.  For instance, the clustering coefficient will always be denoted
as \texttt{clusco}.  Thus, when citing statistics, decompositions or
other items defined in this handbook, please use internal names; these
are always given in a non-proportional font. 
Throughout the handbook, we use margin notes to give the internal 
\marginpar{\textlangle{}\texttt{name}\textrangle}
names of various parameters, as shown here on the side of this text.  
In KONECT, almost everything has an \emph{internal name}, i.e., a
systematic name that is used throughout code, and which is stable.
Internal names are never changed.  Therefore, some internal names may be
slightly inaccurate or inconsistent.  For purposes of backward
compatibility, this is preferred over changing names.  For instance,
KONECT's Enron email network is called \texttt{enron}, the edge weights
of networks with negative edges is called \texttt{signed}, the set
of all unipartite networks is called \texttt{SQUARE}, and the category
of online social networks is called \texttt{Social}.  Internal names are
case-sensitive. 

Certain parts of the handbook are based on previous work, and certain
parts have been published as papers by the authors.  These works are
always cited in the relevant sections, and should be cited in preference
when using the corresponding methods. 

\subsection{Structure of this Handbook}
This handbook serves both as a compendium of mathematical definitions
used in the KONECT project, as well as documentation of the software
used in KONECT. 
\begin{itemize}
\item Section \ref{sec:taxonomy} -- \textbf{Networks} gives the basic
  definition of a network as used in KONECT -- as networks exists in
  many different variants such as directed, bipartite, signed, etc., the
  exact definitions are often overlooked, but can be crucial for
  understanding individual relations. The taxonomy of networks given in
  this section serves for the rest of the project.  This section also
  attempts to illustrate the breadth of the concept of \emph{network},
  bridging many different disciplines. 
\item Section \ref{sec:definitions} -- \textbf{Graph Theory} gives the
  graph-theoretical definitions underlying all topics of network analysis.  Again,
  the devil is in the details and the graph theory literature is often
  split between multiple incompatible definitions for commonly used
  terms.  This sections presents the terminology and definitions as used
  in KONECT, which represents a compromise that has been proven useful
  in practice. 
\item Section \ref{sec:statistics} -- \textbf{Statistics} is devoted to
  network statistics, i.e., numerical measures that characterize a
  network as a whole.  These are central to network analysis, and most
  common network statistics are covered.  The section is structured by
  the type of analysis underlying the different statistics, roughly
  order from simple to complex.  Each subsection serves as an
  introduction to the underlying topic, although not all subsections
  (yet) contain enough exposition to server as a general introduction to
  these topics. 
\item Section \ref{sec:matrix} -- \textbf{Matrices and Decompositions} reviews
  characteristic matrices used to analyse graphs, with a focus on their
  decompositions.  These are crucial in various types of analyses,
  include pairwise node measures such as distances and similarities, as
  well as node-based measures such as centralities.  Due to the focus of
  the KONECT project on such decompositions, this section quite detailed
  and complete, although the emphasis is mainly on the eigenvalue and
  singular value decompositions, and matrix to which they can be
  applied. 
\item Section \ref{sec:plots} -- \textbf{Plots} reviews common ways to
  visualize properties of a network.  Some of those, such as the degree
  distribution, are ubiquitous in the literature.  Since KONECT plots
  are extensively used by the authors in published research, the plots
  cover quite distinct areas of network analysis. 
\item Section \ref{sec:other} -- \textbf{Other Definitions} covers other
  definitions used in KONECT, including node features,
  i.e., numerical measures that characterize individual nodes in a
  network, and error measures.  These are often used as measures of centrality or
  importance, etc. 
\item Section \ref{sec:toolbox} -- \textbf{The KONECT Toolbox} describes the GNU
  Octave and Matlab toolbox that is part of the KONECT project. 
\item Section \ref{sec:format} -- \textbf{File Formats} finally documents the
  different file formats used by KONECT. 
\end{itemize}

\section{Networks}
\label{sec:taxonomy}
Datasets in KONECT represent networks, i.e., a set of nodes connected by
links. Networks can be classified by their format
(directed/undirected/bipartite), by their edge weight types and
multiplicities, by the 
presence of metadata such as timestamps and node labels, and by
the types of objects represented by nodes and links. 
The full list of networks is given
online.\footnote{\href{http://konect.cc/networks/}{http://konect.cc/networks/}}  

\subsection{Format}
The format of a network is always one of the following.  The network
formats are summarized in Table~\ref{tab:format}. 
\begin{itemize}
\item 
  In \textbf{undirected networks} (U), 
  \marginpar{\texttt{sym}}
  edges are undirected.  That is,
  there is no difference between the edge from $u$ to $v$ and the edge
  from $v$ to $u$; both are the edge $\{u,v\}$. 
  An example of an undirected network is the social network of
  Facebook
  (\href{http://konect.cc/networks/facebook-wosn-wall/}{\textsf{Ow}}),
  in which there is no difference between the statements ``A 
  is a friend of B'' and ``B is a friend of A.''
\item In a \textbf{directed network} (D), 
  \marginpar{\texttt{asym}}
  the links are directed. That is, there is a
  difference between the edge $(u,v)$ and the edge $(u,v)$. 
  Directed networks are sometimes also called \emph{digraphs} (for \emph{directed
    graphs}), and their edges \emph{arcs}. 
  An example of a directed social network is the follower network of
  Twitter
  (\href{http://konect.cc/networks/twitter_mpi/}{\textsf{TF}}),
  in which the fact that user A follows user B does not imply 
  that user B follows user A. 
\item \textbf{Bipartite networks} (B) 
  \marginpar{\texttt{bip}}
  include two types of nodes, and all edges
  connect one node type with the other. An example of a bipartite
  network is a rating graph, consisting of the node types \emph{user}
  and \emph{movie}, and each rating connects a user and a movie
  (\href{http://konect.cc/networks/movielens-10m_rating/}{\textsf{M3}}).  
  Bipartite networks are always undirected in KONECT.  Datasets that can
  be represented as hypergraphs can also be represented without loss as
  bipartite networks, explaining why KONECT does not have a
  \emph{hypergraph} format, and also why bipartite networks are common. 
\end{itemize}

\begin{table}
  \caption{
    The network formats allowed in KONECT.
    Each network dataset is exactly of one type.  
    \label{tab:format}
  }
  \centering
\makebox[\textwidth]{
  \begin{tabular}{ c l llll }
    \toprule
    \textbf{\#} & \textbf{Icon} & \textbf{Type} & \textbf{Edge partition} & \textbf{Edge types} & \textbf{Internal name} \\
    \midrule
    1 & U & Undirected	& Unipartite	& Undirected	& \texttt{sym}  \\
    2 & D & Directed	& Unipartite	& Directed	& \texttt{asym} \\
    3 & B & Bipartite	& Bipartite	& Undirected	& \texttt{bip}  \\
    \bottomrule
  \end{tabular}
}
\end{table}

\subsection{Weights}
The edge weight and multiplicity types of networks are represented by
one of the following types. 
The types of edge weights and multiplicities are summarized in
Table~\ref{tab:weights}. 
\begin{itemize}
\item An \textbf{unweighted network} ($-$) 
  \marginpar{\texttt{unweighted}}
  has edges that are
  unweighted, and only a 
  single edge is allowed between any two nodes.  
\item In a \textbf{network with multiple edges} ($=$), 
  \marginpar{\texttt{positive}}
  two nodes can be
  connected by any number of edges, and all edges are unweighted. This
  type of network is also called a multigraph.  
\item In a \textbf{positive network} ($+$), 
  \marginpar{\texttt{posweighted}}
  edges are annotated
  with positive weights, and only a single edge is allowed between
  any node pair.  The weight of zero is identified with the lack of an edge
  and thus, we require that each edge has a weight strictly larger than
  zero, except when the special tag \texttt{\#zeroweight} is defined.
\item In a \textbf{signed network} ($\pm$), 
  \marginpar{\texttt{signed}}
  both positive and negative edges are 
  allowed. Positive and negative edges are represented by positive and
  negative edge weights. Many networks of this type have only the
  weights $\pm 1$, but in the general case we allow any nonzero weight;
  this distinction is not made. 
\item \textbf{Networks with multiple signed edges} ($\stackrel{+}{=}$) 
  \marginpar{\texttt{multisigned}}
  allow multiple edges between two nodes, which may have the same values
  as edges in a signed network.  
\item \textbf{Rating networks} ($*$) 
  \marginpar{\texttt{weighted}}
  have arbitrary real edge weights.  They
  differ from positive and signed networks in that the edge weights are
  interpreted as an interval scale, and thus the value zero has no
  special meaning.  Adding a constant to all edge weights does not
  change the semantics of a rating network. 
  Ratings can be discrete, such as the one-to-five star ratings, or
  continuous, such as a rating given in percent. 
  This type of network allows only a single edge between two nodes. 
\item \textbf{Networks with multiple ratings} ($_*{}^*$) 
  \marginpar{\texttt{multiweighted}}
  have edges annotated
  with rating values, and allow multiple edges between two nodes.
\item \textbf{Dynamic networks} ($\rightleftharpoons$) are networks in
  \marginpar{\texttt{dynamic}}
  which edges can appear and disappear.  They are always
  temporal. Individual edges are not weighted. 
\end{itemize}

\begin{table}
  \caption{
    The edge weight and multiplicity types allowed in KONECT. Each
    network dataset is exactly of one type. 
    Note that due to historical reasons, networks with multiple
    unweighted edges have the internal name \texttt{positive}, while
    positively weighted networks have the internal
    \texttt{posweighted}. 
    For signed networks and positive edge weights, weights of zero are
    only allowed when the tag \texttt{\#zeroweight} is set. 
    \label{tab:weights}
  }
  \centering
\makebox[\textwidth]{
  \begin{tabular}{cllllll}
    \toprule
    \textbf{\#} & \textbf{Icon} &\textbf{Type} & \textbf{Multiple} & \textbf{Edge weight} &
    \textbf{Edge weight} &
    \textbf{Internal name} \\ 
    & & & \textbf{edges} & \textbf{range} & \textbf{scale}  \\
    \midrule
    1 & $-$   & Unweighted          & No  & $\{1\}$        & --    & \texttt{unweighted} \\
    2 & $=$   & Multiple unweighted & Yes & $\{1\}$         &  --  & \texttt{positive}   \\
    3 & $+$   & Positive weights    & No  & $(0, \infty)$      & Ratio scale     & \texttt{posweighted}\\
    4 & $\pm$ & Signed              & No  & $(-\infty,+\infty)$ & Ratio scale    & \texttt{signed}     \\
    5 & $\stackrel{+}{=}$ & Multiple signed  & Yes & $(-\infty,+\infty)$ & Ratio scale    & \texttt{multisigned} \\
    6 & $*$   & Rating              & No  & $(-\infty,+\infty)$ &Interval scale  & \texttt{weighted}   \\
    7 & $_*{}^*$ & Multiple ratings & Yes & $(-\infty,+\infty)$ & Interval scale  & \texttt{multiweighted}\\
    8 & $\rightleftharpoons$ & Dynamic & Yes & $\{1\}$            & --	     & \texttt{dynamic}      \\
    9 &  & Multiple positive weights & Yes & $(0, \infty)$      & Ratio scale & \texttt{multiposweighted} \\
    \bottomrule
  \end{tabular}
  }
\end{table}

\subsection{Temporal Networks}
Furthermore, networks can have one more property: 
\begin{itemize}
\item \textbf{Temporal networks} (\Clocklogo) include a timestamp for each
  edge, and thus the network can be reconstructed for any moment in the
  past.  By default, the timestamp is in Unix time, and gives one
  timestamp to each edge, denoting when that edge was added, or when the
  event represented by the edge took place.  
  It is unspecified whether timestamps include or not leap seconds.  In
  practice, no dataset has precise enough information for that to matter.
  When the tag \texttt{\#unspecifiedtime} is set, the timestamps are not in Unix
  time, but rather in an unspecified monotonous time scale. 
  The fact that a network is temporal is not saved in any special
  location, but is determined by the fact that timestamps are present in
  the TSV file, as will be described later. 
\end{itemize}

\subsection{Categories}
Finally, the network categories classify networks by the type of data they
represent.  
An overview of the categories is given in Table~\ref{tab:categories}. 

\begin{table*}
  \caption{
    The network categories in KONECT.  
    Each category is assigned a color, which is used in plots, for
    instance in Figure~\ref{fig:scatter.size.avgdegree}. The property
    icons are defined in Table \ref{tab:weights}.  U: Undirected
    network, D: Directed network, B: Bipartite network. 
    \label{tab:categories}
  }
  \centering
  \makebox[\textwidth]{
    \begin{tabular}{lllllr}
\toprule
& \textbf{Internal name} & \textbf{Vertices} & \textbf{Edges} & \textbf{Properties} & \textbf{Count} \\
\midrule
\textcolor{colorAffiliation}{$\newmoon$} &\texttt{Affiliation} & Actors, groups & Membership & \phantom{U} \phantom{D} B $-$ $=$ \phantom{$+$} \phantom{$\pm$} \phantom{$\stackrel{+}{=}$} \phantom{$*$} \phantom{$_*{}^*$} \phantom{$\rightleftharpoons$} \phantom{$++$}  &  17\\
\textcolor{colorAnimal}{$\newmoon$} &\texttt{Animal} & Animals & Tie & U D \phantom{B} $-$ \phantom{$=$} $+$ \phantom{$\pm$} \phantom{$\stackrel{+}{=}$} \phantom{$*$} \phantom{$_*{}^*$} \phantom{$\rightleftharpoons$} \phantom{$++$}  &  9\\
\textcolor{colorAuthorship}{$\newmoon$} &\texttt{Authorship} & Authors, works & Authorship & \phantom{U} \phantom{D} B $-$ $=$ \phantom{$+$} \phantom{$\pm$} \phantom{$\stackrel{+}{=}$} \phantom{$*$} \phantom{$_*{}^*$} \phantom{$\rightleftharpoons$} \phantom{$++$}  &  809\\
\textcolor{colorCitation}{$\newmoon$} &\texttt{Citation} & Documents & Citation & \phantom{U} D \phantom{B} $-$ \phantom{$=$} \phantom{$+$} \phantom{$\pm$} \phantom{$\stackrel{+}{=}$} \phantom{$*$} \phantom{$_*{}^*$} \phantom{$\rightleftharpoons$} \phantom{$++$}  &  7\\
\textcolor{colorCoauthorship}{$\newmoon$} &\texttt{Coauthorship} & Authors & Coauthorship & U \phantom{D} \phantom{B} $-$ $=$ \phantom{$+$} \phantom{$\pm$} \phantom{$\stackrel{+}{=}$} \phantom{$*$} \phantom{$_*{}^*$} \phantom{$\rightleftharpoons$} \phantom{$++$}  &  7\\
\textcolor{colorCocitation}{$\newmoon$} &\texttt{Cocitation} & Authors & Cocitation & U \phantom{D} \phantom{B} \phantom{$-$} $=$ \phantom{$+$} \phantom{$\pm$} \phantom{$\stackrel{+}{=}$} \phantom{$*$} \phantom{$_*{}^*$} \phantom{$\rightleftharpoons$} \phantom{$++$}  &  2\\
\textcolor{colorCommunication}{$\newmoon$} &\texttt{Communication} & Persons & Message & U D \phantom{B} $-$ $=$ \phantom{$+$} \phantom{$\pm$} \phantom{$\stackrel{+}{=}$} \phantom{$*$} \phantom{$_*{}^*$} \phantom{$\rightleftharpoons$} \phantom{$++$}  &  42\\
\textcolor{colorComputer}{$\newmoon$} &\texttt{Computer} & Computers & Connection & U D \phantom{B} $-$ $=$ \phantom{$+$} \phantom{$\pm$} \phantom{$\stackrel{+}{=}$} \phantom{$*$} \phantom{$_*{}^*$} \phantom{$\rightleftharpoons$} \phantom{$++$}  &  14\\
\textcolor{colorFeature}{$\newmoon$} &\texttt{Feature} & Items, features & Property & \phantom{U} \phantom{D} B $-$ $=$ $+$ \phantom{$\pm$} \phantom{$\stackrel{+}{=}$} \phantom{$*$} \phantom{$_*{}^*$} \phantom{$\rightleftharpoons$} \phantom{$++$}  &  17\\
\textcolor{colorHumanContact}{$\newmoon$} &\texttt{HumanContact} & Persons & Real-life contact & U \phantom{D} \phantom{B} $-$ $=$ $+$ \phantom{$\pm$} \phantom{$\stackrel{+}{=}$} \phantom{$*$} \phantom{$_*{}^*$} \phantom{$\rightleftharpoons$} \phantom{$++$}  &  7\\
\textcolor{colorHumanSocial}{$\newmoon$} &\texttt{HumanSocial} & Persons & Real-life tie & U D \phantom{B} $-$ \phantom{$=$} $+$ $\pm$ $\stackrel{+}{=}$ \phantom{$*$} \phantom{$_*{}^*$} \phantom{$\rightleftharpoons$} \phantom{$++$}  &  12\\
\textcolor{colorHyperlink}{$\newmoon$} &\texttt{Hyperlink} & Web page & Hyperlink & U D B $-$ $=$ \phantom{$+$} \phantom{$\pm$} \phantom{$\stackrel{+}{=}$} \phantom{$*$} \phantom{$_*{}^*$} $\rightleftharpoons$ \phantom{$++$}  &  197\\
\textcolor{colorInfrastructure}{$\newmoon$} &\texttt{Infrastructure} & Location & Connection & U D \phantom{B} $-$ $=$ $+$ \phantom{$\pm$} \phantom{$\stackrel{+}{=}$} \phantom{$*$} \phantom{$_*{}^*$} \phantom{$\rightleftharpoons$} \phantom{$++$}  &  23\\
\textcolor{colorInteraction}{$\newmoon$} &\texttt{Interaction} & Persons, items & Interaction & U D B $-$ $=$ \phantom{$+$} \phantom{$\pm$} $\stackrel{+}{=}$ \phantom{$*$} \phantom{$_*{}^*$} \phantom{$\rightleftharpoons$} \phantom{$++$}  &  26\\
\textcolor{colorLexical}{$\newmoon$} &\texttt{Lexical} & Words & Lexical relationship & U D \phantom{B} $-$ $=$ \phantom{$+$} \phantom{$\pm$} \phantom{$\stackrel{+}{=}$} \phantom{$*$} \phantom{$_*{}^*$} \phantom{$\rightleftharpoons$} \phantom{$++$}  &  5\\
\textcolor{colorMetabolic}{$\newmoon$} &\texttt{Metabolic} & Metabolites & Interaction & U D \phantom{B} $-$ $=$ \phantom{$+$} \phantom{$\pm$} \phantom{$\stackrel{+}{=}$} \phantom{$*$} \phantom{$_*{}^*$} \phantom{$\rightleftharpoons$} \phantom{$++$}  &  8\\
\textcolor{colorMisc}{$\newmoon$} &\texttt{Misc} & Various & Various & U D B $-$ $=$ $+$ \phantom{$\pm$} \phantom{$\stackrel{+}{=}$} \phantom{$*$} \phantom{$_*{}^*$} \phantom{$\rightleftharpoons$} \phantom{$++$}  &  15\\
\textcolor{colorNeural}{$\newmoon$} &\texttt{Neural} & Neurons & Connection & \phantom{U} D \phantom{B} \phantom{$-$} \phantom{$=$} $+$ \phantom{$\pm$} \phantom{$\stackrel{+}{=}$} \phantom{$*$} \phantom{$_*{}^*$} \phantom{$\rightleftharpoons$} \phantom{$++$}  &  1\\
\textcolor{colorOnlineContact}{$\newmoon$} &\texttt{OnlineContact} & Users & Online interaction & U D \phantom{B} $-$ $=$ \phantom{$+$} $\pm$ $\stackrel{+}{=}$ \phantom{$*$} \phantom{$_*{}^*$} $\rightleftharpoons$ \phantom{$++$}  &  15\\
\textcolor{colorRating}{$\newmoon$} &\texttt{Rating} & Users, items & Rating & \phantom{U} \phantom{D} B $-$ $=$ \phantom{$+$} \phantom{$\pm$} \phantom{$\stackrel{+}{=}$} $*$ $_*{}^*$ \phantom{$\rightleftharpoons$} \phantom{$++$}  &  21\\
\textcolor{colorSocial}{$\newmoon$} &\texttt{Social} & Persons & Online tie & U D \phantom{B} $-$ \phantom{$=$} $+$ $\pm$ \phantom{$\stackrel{+}{=}$} $*$ \phantom{$_*{}^*$} $\rightleftharpoons$ \phantom{$++$}  &  50\\
\textcolor{colorSoftware}{$\newmoon$} &\texttt{Software} & Software Component & Dependency & \phantom{U} D \phantom{B} $-$ $=$ \phantom{$+$} \phantom{$\pm$} \phantom{$\stackrel{+}{=}$} \phantom{$*$} \phantom{$_*{}^*$} \phantom{$\rightleftharpoons$} \phantom{$++$}  &  3\\
\textcolor{colorText}{$\newmoon$} &\texttt{Text} & Documents, words & Occurrence & \phantom{U} \phantom{D} B \phantom{$-$} $=$ \phantom{$+$} \phantom{$\pm$} \phantom{$\stackrel{+}{=}$} \phantom{$*$} \phantom{$_*{}^*$} \phantom{$\rightleftharpoons$} \phantom{$++$}  &  10\\
\textcolor{colorTrophic}{$\newmoon$} &\texttt{Trophic} & Species & Carbon exchange & U D \phantom{B} $-$ \phantom{$=$} $+$ \phantom{$\pm$} \phantom{$\stackrel{+}{=}$} \phantom{$*$} \phantom{$_*{}^*$} \phantom{$\rightleftharpoons$} \phantom{$++$}  &  4\\
\midrule
& \textbf{Total} &&&& \textbf{1321}\\
\bottomrule
\end{tabular}

  }
\end{table*}

%
%
\begin{description}
\item[Affiliation networks] are bipartite networks denoting the
  \marginpar{\texttt{Affiliation}}
  membership of actors in groups.  Groups can be defined as narrowly as
  individual online communities in which users have been active
  (\href{http://konect.cc/networks/flickr-groupmemberships/}{\textsf{FG}})
  or as broadly as countries
  (\href{http://konect.cc/networks/dbpedia-country/}{\textsf{CN}}). The
  actors are mainly persons, but can also be other actors such as musical
  groups. Note that in all affiliation networks we consider, each actor
  can be in more than one group, as otherwise the network cannot be
  connected.

\item[Animal networks] are networks of contacts between animals.  
  \marginpar{\texttt{Animal}}
  They are the animal equivalent to human social networks.  Note that
  datasets of websites such as Dogster
  (\href{http://konect.cc/networks/petster-friendships-dog/}{\textsf{Sd}})
  are \emph{not} included here but in the \texttt{Social} (online social
  network) category, since the networks are generated by humans. 

\item[Authorship networks] are unweighted bipartite networks consisting
  \marginpar{\texttt{Authorship}}
  of links between authors and their works.  In some authorship networks
  such as that of scientific literature
  (\href{http://konect.cc/networks/dblp-author/}{\textsf{Pa}}),
  works have typically only few authors, whereas works in other
  authorship networks may have many authors, as in Wikipedia articles
    (\href{http://konect.cc/networks/edit-enwiki/}{\textsf{en}}).

\item[Citation networks] consist of documents that reference each
  \marginpar{\texttt{Citation}}
  other.  The primary example are scientific publications, but the
  category also allow patents and other types of documents that
  reference each other.  The category does not include hyperlink
  networks, i.e., web pages that reference each other.  Those are in the
  category \texttt{Hyperlink}. 

\item[Co-authorship networks] are unipartite networks connecting authors
  \marginpar{\texttt{Coauthorship}}
  who have written works together, for instance academic literature, but
  also other types of works such as music or movies.  These are also
  often called collaboration networks.  

\item[Co-citation networks] are unipartite networks of documents,
  connected by an edge if they are both cited by a same other document.
  As a general rule, co-citation networks are the self-join of citation
  networks. 
  
\item[Communication networks] contain edges that represent
  \marginpar{\texttt{Communication}}
  individual messages between persons.  Communication networks are directed
  and allow multiple edges.  
  Examples of communication networks are those of
  emails (\href{http://konect.cc/networks/enron/}{\textsf{EN}})
  and those of
  Facebook messages
  (\href{http://konect.cc/networks/facebook-wosn-wall/}{\textsf{Ow}}). Note
  that in some instances, edge directions are not 
  known and KONECT can only provide an undirected network. 

\item[Computer networks] are networks of connected computers. 
  \marginpar{\texttt{Computer}}
  Nodes in them are computers, and edges are connections. 
  When speaking about \emph{networks} in a computer science context, one
  often means only computer networks.  An example is the internet
  topology network (\href{http://konect.cc/networks/topology/}{\textsf{TO}}).

\item[Feature networks] are bipartite, and denote any kind of feature
  \marginpar{\texttt{Feature}}
  assigned to entities. Feature networks are unweighted and have
  edges that are not annotated with edge creation times.  Examples are
  songs and their genres
  (\href{http://konect.cc/networks/dbpedia-genre/}{\textsf{GE}}).   

\item[Human contact networks] are unipartite networks of actual contact
  \marginpar{\texttt{HumanContact}}
  between persons, i.e., talking with each other, spending time
  together, or at least being physically close.  Usually, these datasets
  are collected by giving out RFID tags to people with chips that record
  which other people are in the vicinity.  Determining when an actual
  contact has happened (as opposed to for instance to persons standing
  back to back) is a nontrivial research problem. 
  An example is the Reality Mining dataset
  (\href{http://konect.cc/networks/mit/}{\textsf{RM}}). 

\item[Human social networks] are real-world social networks between
  humans.  \marginpar{\texttt{HumanSocial}} The ties are offline,
  and not from an online social network.  Also, the ties represent a
  state, as opposed to human contact networks, in which each edge
  represents an event.

\item[Hyperlink networks] are the
  networks \marginpar{\texttt{Hyperlink}} of pages on the World Wide Web
  or on another system of linked knowledge,
  connected by hyperlinks or similar types of links.  These are in general directed.  Since any
  type of information can be represented on the World Wide Web,
  hyperlink networks are often simultaneously in another category.  For
  instance, user pages linked to each other represent a hyperlink
  network and a social network at the same time.  In such cases, only
  the more specific (non-hyperlink) category is used in KONECT.

\item[Infrastructure networks] are networks of physical infrastructure.  
  \marginpar{\texttt{Infrastructure}}
  Examples are road networks
  (\href{http://konect.cc/networks/roadNet-CA/}{\textsf{RO}}), airline
  connection networks
  (\href{http://konect.cc/networks/opsahl-openflights/}{\textsf{OF}}), 
  and power grids
  (\href{http://konect.cc/networks/opsahl-powergrid/}{\textsf{UG}}).  
  
\item[Interaction networks] are bipartite networks consisting of people
  \marginpar{\texttt{Interaction}}
  and items or between people and other people, where each edge represents an interaction. 
  In interaction networks, we always allow multiple edges between the
  same person--item pair, and an interaction is always to be understood as an event. 
  Interaction networks can be online or offline.
  Examples are
  people writing in forums
  (\href{http://konect.cc/networks/opsahl-ucforum/}{\textsf{UF}}),
  commenting on movies
  (\href{http://konect.cc/networks/filmtipset_comment/}{\textsf{Fc}}),
  listening to songs
  (\href{http://konect.cc/networks/lastfm_song/}{\textsf{Ls}})
  and sports results. 

\item[Lexical networks] consist of words from natural 
  \marginpar{\texttt{Lexical}} languages and the relationships between
  them.  Relationships can be semantic (i.e, related to the meaning of
  words) such as the synonym relationship
  (\href{http://konect.cc/networks/wordnet-words/}{\textsf{WO}}),
  associative such as when two words are associated with each other by
  people in experiments
  (\href{http://konect.cc/networks/eat/}{\textsf{EA}}), or denote
  co-occurrence, i.e., the fact that two words co-occur in text
  (\href{http://konect.cc/networks/lasagne-spanishbook/}{\textsf{SB}}).

\item[Metabolic networks] model metabolic pathways,
  \marginpar{\texttt{Metabolic}} in which nodes a chemical substances
  and edges are often directed and represents chemical interactions.  A
  subset are protein--protein interaction networks (PPI), in which nodes
  are proteins, and which are usually undirected.

\item[Miscellaneous networks] are any networks that do not fit into one
  \marginpar{\texttt{Misc}}
  of the other categories. 

\item[Neural networks] are networks reprensentating the structure of the
  brain.  Nodes are neurons are higher-level groupings of the brain,
  while edges are connections between them.  The field concerned with
  the network analysis of such structures is called \emph{network neuroscience}. 
  
\item[Online contact networks] consist of people and interactions between
  \marginpar{\texttt{OnlineContact}} them.  Contact networks are
  unipartite and allow multiple edges, i.e., there can always be
  multiple interactions between the same two persons.  They can be both
  directed or undirected.

\item[Physical networks] represent physically existing network
  \marginpar{\texttt{Physical}} structures in the broadest sense.  This
  category covers such diverse data as physical computer networks
  (\href{http://konect.cc/networks/topology/}{\textsf{TO}}), transport
  networks
  (\href{http://konect.cc/networks/opsahl-openflights/}{\textsf{OF}}) and
  biological food networks
  (\href{http://konect.cc/networks/foodweb-baydry/}{\textsf{FD}}).

\item[Rating networks] consist of assessments given to items by users,
  \marginpar{\texttt{Rating}} weighted by a rating value.  Rating
  networks are bipartite.  Networks in which users can rate other users
  are not included here, but in the Social category instead.  If only a
  single type of rating is possible, for instance the ``favorite''
  relationship, then rating networks are unweighted.  Examples of items
  that are rated are movies
  (\href{http://konect.cc/networks/movielens-10m_rating/}{\textsf{M3}}),
  songs (\href{http://konect.cc/networks/yahoo-song/}{\textsf{YS}}),
  jokes (\href{http://konect.cc/networks/jester1/}{\textsf{J1}}), and even
  sexual escorts
  (\href{http://konect.cc/networks/escorts/}{\textsf{SX}}).

\item[Online social networks] represent ties between
  \marginpar{\texttt{Social}} persons in online social networking
  platforms.  Certain social networks allow negative edges, which denote
  enmity, distrust or dislike.  Examples are Facebook friendships
  (\href{http://konect.cc/networks/facebook-wosn-links/}{\textsf{Ol}}),
  the
  Twitter follower relationship
  (\href{http://konect.cc/networks/twitter_mpi/}{\textsf{TF}}), and
  friends and foes on Slashdot
  (\href{http://konect.cc/networks/slashdot-zoo/}{\textsf{SZ}}).  Note
  that some social networks can be argued to be rating networks, for
  instance the user--user rating network of a dating site
  (\href{http://konect.cc/networks/libimseti/}{\textsf{LI}}).  These
  networks are all included in the \textsf{Social} category.
  Online social networks may be undirected (such as on Facebook),
  or directed (such as on Twitter).
  For historical reasons, the internal name of this category is
  \textsf{Social}, even though it includes only online social networks. 

\item[Software networks] are networks of interacting software
  \marginpar{\texttt{Software}} component.  Node can be software
  packages connected by their dependencies, source files connected by
  includes, and classes connected by imports.

\item[Text networks] consist of text documents containing words.  They
  \marginpar{\texttt{Text}} are bipartite and their nodes are documents
  and words.  Each edge represents the occurrence of a word in a
  document. Document types are for instance newspaper articles
  (\href{http://konect.cc/networks/gottron-trec/}{\textsf{TR}}) and
  Wikipedia articles
  (\href{http://konect.cc/networks/gottron-excellent/}{\textsf{EX}}).

\item[Trophic networks] consist of biological 
  \marginpar{\texttt{Trophic}} species connected by edges denoting which
  pairs of species are subject to exchange of substances such as carbon
  or nitrogen.  In simple cases, these networks can be described as
  ``who eats whom'', but the category also includes the exchanges of
  more specific chemical species. 
  The term \emph{food chain} may describe such relationships,
  but note that in the general case, a trophic network is not a chain,
  i.e., it is not linear.  Trophic networks are directed.  Nodes may be
  individual species, may may also be broader or narrower classes of
  organisms. 

\end{description}

Note that the category system of KONECT is in flux.  As networks are
added to the collection, large categories are split into smaller ones. 

\subsection{What Is and Is Not a Network}
While the KONECT motto asserts that \emph{everything is a network}, this
does not imply that everything is a complex network.  Thus, certain networks
have a structure that is trivial in a way that render many network
analysis methods moot.  In KONECT, 
we do not include such networtks.
This includes networks without a giant connected component,
in which most nodes are not reachable from each other, and trees, in
which there is only a single path between any two nodes.  Note that
bipartite relationships extracted from n-to-1 relationships are
therefore excluded, as they lead to a disjoint network. For instance, a
bipartite person--city network containing \emph{was-born-in} edges would
not be included, as each city would form its own component disconnected
from the rest of the network.  On the other hand, a band--country
network where edges denote the country of origin of individual band
members is included, as members of a single band can have different
countries of origin. In fact the Countries network
(\href{http://konect.cc/networks/dbpedia-country/}{\textsf{CN}})
is of this form.  Another example is a bipartite song--genre network,
which would only be included in KONECT when songs can have multiple
genres.  As an example of the lack of complex structure when only a
single genre is allowed, the degree distribution in such a song--genre
network is skewed because all song nodes have degree one, the diameter
cannot be computed since the network is disconnected, and each connected
component trivially has a diameter of two or less.

Certain other types of structures are equivalent to networks.  For
instance, hypergraphs, in which each hyperedge may include any number of
nodes, can be presented equivalently as bipartite graphs, up to certain
small differences described in Section~\ref{sec:definitions}.  The same holds e.g.\ for partially-ordered sets and
directed graphs.  In KONECT, we choose the ``network'' aspect to model those. 

\subsection{Groups}
\label{sec:groups}
Types of networks are separated into groups, as given in
Table~\ref{tab:groups}.
Each groups represents a combination of metadata nased on network
format, weights and other attributes.  The importance of groups is that
for most network analysis methods, we can specify to which groups of
networks they apply. 

\begin{table}
  \centering
  \caption{
    \label{tab:groups}
    The list of groups in KONECT.  Groups represent sets of networks
    that have similar metadata, based on their format, weights and other
    attributes. 
  }
  \makebox[\textwidth]{
  \begin{tabular}{ l l }
    \toprule
    \textbf{Internal name} & \textbf{Definition} \\
    \midrule
    \texttt{ALL}		& All networks \\
    \texttt{SYM}		& Undirected, unipartite \\
    \texttt{ASYM}		& Directed, unipartite \\
    \texttt{BIP}		& Bipartite \\
    \texttt{SQUARE}		& Unipartite \\
    \texttt{NEGATIVE}		& Networks where edges can be interpreted to have negative edges \\
    \texttt{NONUNWEIGHTED}	& Networks where edges can be interpreted to have a weight or multiplicity \\
    \texttt{ASYMNEGATIVE}	& Directed networks where edges that can be negative \\
    \texttt{SQUARENEGATIVE}	& Unipartite networks where edges can be negative \\
    \texttt{TIME}		& Temporal networks \\
    \texttt{TIME\_NEGATIVE}	& Temporal networks where edges can be negative \\
    \texttt{MULTI}		& Networks with multiple edges \\
    \bottomrule
  \end{tabular}
  }
\end{table}

\subsection{Tags}
\label{sec:tags}
The following tags can be given to networks.  They are declared in the
\texttt{tags} field in the \texttt{meta.*} file for each network. 
\begin{itemize}
\item \texttt{\#acyclic}:  The network is acyclic.  Can only be
  set for directed networks.  If this is not set, a directed
  network must contain at least two pairs of reciprocal edges of
  the form $(u,v)$ and $(v,u)$.  If the network does not contain
  reciprocal edges, but has cycles, the tag
  \texttt{\#nonreciprocal} is used.
\item \texttt{\#aggregatetime}:  The smallest value in the timestamp column
  stands for any earlier time; all timestamp having this smallest value should not be
  considered when performing time-based methods and plots.  In most
  cases, this value is zero, but it can just as well be -1 or the most
  negative integer of the integer type used when creating the dataset. 
\item \texttt{\#clique}:  All possible edges are present, i.e., the
  graph is complete.  This is \emph{not} the opposite of
  \texttt{\#incomplete}.  Which edges are taken into account depends on
  the type of graph, i.e., whether the graph is bipartite, directed,
  allows loops.  This does not take into account edge weights and
  multiplicities, and in fact this tag only really makes sense when
  those are present. 
\item \texttt{\#incomplete}: The network is incomplete, i.e.,
  the present dataset represents a subset of the actual data.  This is
  mostly due to the fact that the data was crawled, or aggregated in
  another way that cannot guarantee that all nodes and links were seen.
  In such graphs, it is not specified \emph{which} nodes and edges are
  missing.  
  This implies that strictly speaking, certain statistics or plots like 
  the degree distribution are not meaningful, since they may
  depend on which parts are missing.  In practice, many datasets
  fall into this category, and they do not necessarily all have this
  tag. 
\item \texttt{\#join}:  The network is the join (in the
  database-theoretical sense) of more
  fundamental networks.  For instance, a co-authorship network
  is a join of the authorship network with itself.  In general, all
  networks that can be described as being a \emph{co-X network}, where
  \emph{X} is some other network, are of this type. 
  Networks that have this tag may have skewed properties, such as skewed
  degree distributions.  In KONECT, we recommend to \emph{not} generate
  the join of a given network, but to publish the underlying network(s)
  itself.  However, many networks are only known as their join, and thus
  are included in KONECT.  (Note that in many cases, the self join has
  been used to make a unipartite network out of a bipartite one, in
  order to apply network analysis methods that otherwise do not apply to
  bipartite networks.)
\item \texttt{\#kcore}: The network contains only nodes with a
  certain minimal degree $k$. In other words, the nodes with
  degree less than a certain number $k$ were removed from the
  dataset.  This changes a network drastically, and is called
  the ``$k$-core'' of a network. This is sometimes done to get
  a less sparse network in applications that do not perform well
  on sparse networks. This tag implies the
  \texttt{\#incomplete} tag.
\item \texttt{\#lcc}:  The dataset actually contains only the
  largest connected component of the actual network.  Implies
  \texttt{\#incomplete}.  This tag is not used when the network
  is connected for other reasons. 
\item \texttt{\#loop}: The network may contain loops, i.e.,
  egdes connecting a vertex to itself.  This tag is only
  allowed for unipartite networks.  When this tag is not
  present, loops are not allowed, and the presence of loops
  will be considered an error by analysis code.
\item \texttt{\#lowmultiplicity}:  Set in networks with multiple
  edges in which the actual maximal edge multiplicity is very
  low.  Used to be able to use the maximal multiplicity as a
  sanity check.  Indicates a dataset error, as edge
  multiplicities usually have a power law-like distribution, and
  thus very high edge multiplicities are usually present. 
\item \texttt{\#missingmultiplicity}:  This tag is used when the
  underlying network had inherent multiple edges, but these are
  not present in the dataset.  For instance, any email network
  that does not contain multiple edges is tagged with this. 
\item \texttt{\#missingorientation}: This tag is used for
  undirected networks which are based on an underlying
  directed network.  For instance, in a citation network, we
  may only know that the documents A and B are linked, but not
  which one cites the other.  In such a case, the network in
  KONECT is undirected, although the underlying network is
  actually directed.
\item \texttt{\#nonreciprocal}:  For directed networks only.
  The network does not contain reciprocal edges.  This is only
  used when the network is non-reciprocal, but does contain
  directed cycles.  If the network is acyclic,
  \texttt{\#acyclic} is used. 
\item \texttt{\#unspecifiedtime}:  The timestamps do not represent Unix
  time, but something else.  The only assumption that can be made is
  that timestamps are monotonously increasing, i.e., that larger values
  denote later times.  These could be simply year numbers, number of
  days since a starting point, or something else.  In networks having
  this tag, the number of unique timestamps may be very low, making
  temporal methods unsuited.  For instance, a network that is known from
  only two snapshots may have the timestamp values 1 and 2. 
\item \texttt{\#path}:  The edges represent paths have have been
  navigated in some form.  Thus, a purely network-analysis approach will
  fail to take into account the paths and only consider the
  adjacencies. 
\item \texttt{\#regenerate}: The network can be regenerated
  periodically and may be updated when a more recent dataset
  becomes available.
\item \texttt{\#skew}:  The graph is directed,
  and can be interpreted such that an inverted edge is the same than a
  negative edge.  This applies for instance to sports results network,
  where a directed edge means A won against B, but could be equivalently
  expressed as a negative edge from B to A.  It also applies to networks
  denoting dominance behavior, in particular between animals. 
\item \texttt{\#tournament}:  The graph is directed and for each
  pair of nodes $\{u,v\}$, either the directed edge $u \rightarrow v$ or
  the directed edge $v \rightarrow u$ exists, but not both.  It
  is an error for a non-directed graph to have this tag.  If
  \texttt{\#tournament} is defined, then
  \texttt{\#nonreciprocal} must also be defined.  Also, the graph must
  not contain loops, and thus
  \texttt{\#loop} must not be defined. 
\item \texttt{\#trianglefree}:  The network is triangle-free, i.e., the
  network does not contain any triangles.  Can only be used on
  unipartite networks.  By default, it is an error for a unipartite
  network to not contain triangles.  This tag allows a network to be
  triangle-free. 
\item \texttt{\#zeroweight}:  Must be set if it is allowed for edge
  weights to be zero. Only used for networks with positive edge
  weights and signed/multisigned networks. 
\end{itemize}

\section{Graph Theory}
\label{sec:definitions}
The areas of graph theory and network analysis are young, and
many concepts within them notoriously lack a single established notation.  The
notation chosen in KONECT represents a compromise between familiarity
with the most common conventions, and the need to use an unambigous
choice of letters and symbols.  This section gives an overview of the
basic definitions used within KONECT, including in the rest of this handbook. 
A general mathematical reference for basic definitions used in graph
theory is given by a book by Bollobás \citeyearpar{b116}.  

To give an example of the inconsistency across different discipline, the
degree of a node (the number of neighbors it has) is usually denoted $d$
in mathematics, but $k$ in certain other fields, in particular physics
and network science.  In KONECT we have chosen to always denote it $d$.
Similar choice have been made for other symbols.  For the sake of
completeness, we give the definition as used in KONECT, then mention
other commonly used symbols; the rest of the handbook and KONECT as a
whole then uses the symbol defined initially. 

\subsection{Graphs}
Graphs will be denoted as $G=(V,E)$, in which $V$ is the set of
vertices, and $E$ is the set of edges. Without loss of
generality, we assume that the vertices $V$ are consecutive natural
numbers starting at one, i.e.,
\begin{align}
  V &= \{ 1, 2, 3, \dotsc, |V| \}.
\end{align}
Edges $e\in E$ will be denoted as sets of two vertices, i.e.,
$e=\{u,v\}$.  We say that two vertices are adjacent if they are
connected by an edge; this will be written as $u \leftrightarrow v$, and is equivalent to $\{u,v\}\in E$. 
For directed networks, $u \rightarrow v$ will denote the existence of a
directed edge from $u$ to $v$, and $u \rightleftarrows v$ will denote
that two directed edges of opposite orientation exist between $u$ and $v$.
We say that an
edge is incident to a vertex if the edge touches the vertex. Strictly
speaking, an edge $e$ is incident to node $u$ when $u\in e$, but this
notation may be confusing and we avoid it. 

We also allow loops, i.e., edges of the form $\{u,u\}=\{u\}$.  Loops
appear for instance in email networks, where it is possible to send an
email to oneself, and therefore an edge may connect a vertex with
itself.  Loops are also called \emph{self loops} in the literature, but
in KONECT, we use the simple term \emph{loop}, as there are no other
loops to consider -- cycles of multiple nodes and edges are called
\emph{cycles}. 
Most networks do not contain loops, and therefore
networks that allow loops are annotated in KONECT with the 
\texttt{\#loop} tag, as described in Section~\ref{sec:format}.  
Loops are special in networks because they often need special treatment:
For instance, when defining the degree of a node, do we count a loop
once (defining the degree as the number of edges that are incident to a
node), twice (definining the degree as the number of half-edges attached
to it) or none at all (counting the number other nodes that are
neighbours).  

Most of the time, we work with only one given graph, and therefore it is
unambigous which node and edge set are meant by $V$ and $E$.  When
ambiguity is possible, we will use the notation
$V[G]$ and $E[G]$ to denote the vertex and edge sets of a graph $G$.
This notation using brackets may occasionally be extended to other graph
characteristics. 

In directed networks, edges are pairs instead of sets, i.e.,
$e=(u,v)$.  In this case, we have $(u,v)\neq(v,u)$ whenever $u \neq v$, 
as both edges connecting two nodes can exist independently of each
other. 
In directed networks, edges are sometimes called
\emph{arcs}, in which case the term \emph{edge} os often reserved for
undirected edges; in KONECT, we use the term \emph{edge} or \emph{directed
  edge} for them, and the term \emph{edge} does not imply
undirectedness.  Note that the term \emph{directed} may apply to both an
individual edge as well as to a whole graph, but in KONECT, graphs never
contain both directed and undirected edges.  
For directed graphs, we use the following terminology:  A directed graph
is \emph{symmetric} if all edges are reciprocated, i.e., if, for every
directed edge $(u,v)$, there is another directed edge $(j,i)$.  A
symmetric directed graph is equivalent to an undirected graph, but we
must take care of a subtelty:  In a symmetric directed graph, the degree
of each node is twice of that in the corresponding undirected graph.
Thus, we cannot just identify both with each other, and must be careful
whether we are talking about an undirected graph, or a symmetric
directed graph.  Directed graphs that are not symmetric will be called
asymmetric. 

In bipartite graphs, we can partition the set of nodes $V$ into two
disjoint sets $V_1$ and $V_2$, which we will call the left and right
set respectively.  Although the assignment of a bipartite network's two
node types to left and right sides is mathematically arbitrary, it is
chosen in KONECT such that the left nodes are \emph{active} and the
right nodes are \emph{passive}, as such a distinction can often be made,
and may provide useful hints to the users of a dataset.  For instance, a rating graph with users
and items will always have users on the left since they are active in
the sense that it is they who give the ratings. 
Such a distinction is sensible in most networks \citep{b732}, as certain
patterns can be observed.  The degree distribution, for instance, is
usually more regular for the \emph{passive} than for the \emph{active} nodes.
The number of left and
right nodes are denoted $n_1 = |V_1|$ and $n_2 = |V_2|$.  As a general
rule, a certain number of quantities used in graph theory can be applied
to the left and to the right node left separately, in which case we will
use the indexes of one and two consistently. 

Additionally, bipartite networks can be interpreted as a way to
represent hypergraphs.  A hypergraph $G=(V,E)$ is similar to a graph,
except that the edges in $E$ are sets that may contain any number of
nodes, and not only two as in ordinary graphs.  These sets in $E$ are
called hyperedges.  Equivalently, a
hypergraph $G=(V,E)$ can be represented as bipartite graph $G=(V \cup E,
F)$, where $\{v,e\} \in F$ whenever $v\in V$, $e \in E$ and $v \in e$.
In other words, nodes of the hypergraph are kept as left nodes of the
bipartite graph, hyperedges become right nodes, and the edges of the
bipartite graph correspond to the inclusion relation between them.
Since this equivalence is isomorphic, KONECT does not include a format
\emph{hypergraph}, and represents all such networks as bipartite ones.
Note that for this isomorphism to be precise, one must allow empty and
singleton hyperedges. 

Networks with multiple edges are written as $G=(V,E)$ just as other networks, with $E$ being
a multiset.  The degree of nodes in such networks takes into account
multiple edges.  Thus, the degree does not equal the number of adjacent
nodes but the number of incident edges.  When $E$ is a multiset, it can
contain the edge $\{u,v\}$ multiple times.  Mathematically, we 
may write $\{u,v\}_1$, $\{u,v\}_2$, etc.\ to distinguish multiple such
edges.   Note however that we will be lax with
this notation.  In expressions valid for all types of networks, we will
use sums such as $\sum_{\{u,v\}\in E}$ and understand that the sum
is over all edges, taking into account multiplicities. 

In positively weighted networks, we define $w$ as the
weight function, returning the edge weight when given an edge. In such
networks, the weights are not taken into account when computing the
degree. 

In a signed network, each edge is assigned a signed weight such as $+1$
or $-1$ \citep{b647}.  In such networks, we define $w$ to be the signed weight
function.  In the general case, we allow arbitrary nonzero real numbers,
representing degrees of positive and negative edges.  Signed
relationships have been considered in both phychology \citep{b862} and
anthropology \citep{b323}.  

In rating networks, we define $r$ to be
the rating function, returning the rating value when given an edge.  Note
that rating values are interpreted to be invariant under shifts, i.e.,
adding a real constant to all ratings in the network must not
change the semantics of the network.  Thus, we will often make use of
the mean rating defined as
\begin{align}
  \mu &= \frac 1 {|E|} \sum_{e\in E} r(e). 
\end{align}

For consistency, we also
define the edge weight function $w$ for unweighted and rating networks: 
\begin{align}
  w(e) &= \left\{ \begin{array}{ll} 
    1 & \text{when $G$ is unweighted} \\
    r(e)-\mu & \text{when $G$ is a rating network} \\
  \end{array} \right. 
\end{align}

We also define a weighting function for node pairs, also denoted
$w$. This function takes into account both the weight of edges and edge
multiplicities. It is defined as $w(u,v)=0$ when the nodes $u$ and $v$ are
not connected and if they are connected as
\begin{align}
  w(u,v) &= \left\{ \begin{array}{ll}
    1 & \text{when $G$ is $-$} \\
    |\{k \mid \{u,v\}_k \in E\}| & \text{when $G$ is $=$} \\
    w(\{u,v\}) & \text{when $G$ is $+$}
    \\
    w(\{u,v\}) & \text{when $G$ is $\pm$} \\
    r(\{u,v\}) - \mu & \text{when $G$ is $*$} \\
    \sum_{\{u,v\}_{k\in E}} [r(\{u,v\}_k) - \mu] & \text{when $G$
      is $_*{}^*$}
    \end{array} \right. 
\end{align}

Dynamic networks are special in that they have a set of events (edge
addition and removal) instead of a set of edges.  In most cases, we will
model dynamic networks as unweighted networks $G=(V,E)$ representing
their state at the latest known timepoint.  For analyses that are
performed over time, we consider the graph at different time points,
with the graph always being an unweighted graph. 

In an unweighted graph $G=(V,E)$, the degree of a vertex is the number
of neighbors of that node
\begin{align}
  d(u) &= \{ v \in V \mid \{u,v\} \in E \}. 
\end{align}
In networks with multiple edges, the degree takes into account multiple
edges, and thus to be precise, it equals the number of incident edges
and not the number of adjacent vertices. 
\begin{align}
  d(u) &= \{ \{u,v\}_k \in E \mid v \in V \}
\end{align}
In directed graphs, the sum is over all of $u$'s neighbors, regardless
of the edge orientation. 
Note that the sum of the degrees of all nodes always equals twice the
number of edges, i.e.,
\begin{align}
  \sum_{v\in V} d(u) &= 2|E|. 
\end{align}

In a directed graph we define the outdegree $d^+$ of a node as the number of
outgoing edges, and the indegree $d^-$ as the number of ingoing edges.
\begin{align}
  d^+(u) &= \{ v \in V \mid (u,v) \in E \} \\
  d^-(u) &= \{ v \in V \mid (v,u) \in E \}
\end{align}
Within source code in KONECT, the outdegree and indegree are also often denoted $d_1(u)$ and
$d_2(u)$, respectively -- the indices $1$ and $2$ used for out-
and indegree respectively can be thought as referring to rows and
columns of a matrix, and are consistent with the usage in bipartite
graphs, where the nodes in the first set of nodes correspond to rows of
matrices, and nodes in the second set to columns. 

The sum of all outdegrees, and likewise the sum of all indegrees always
equals the number of nodes in the network.  
\begin{align}
  \sum_{u\in V} d^+(u) = \sum_{u \in V} d^-(u) = |E|
\end{align}
Thus, the sum of all outdegrees always equals the sum of all indegrees,
and therefore the average outdegree always equals the average indegree.  

We also define the weight of a node, also denoted by the symbol $w$, as
the sum of the absolute weights of incident edges
\begin{align}
  w(u) &= \sum_{ \{u,v\} \in E} |w(\{u,v\})|. 
\end{align}
The weight of a node coincides with the degree of a node in unweighted
networks and networks with multiple edges. 
The weight of a node may also be called its strength \citep{b792}. 

For directed graphs, we can distinguish the outdegree weight and the
indegree weight:
\begin{align}
  w_1(u) &= \sum_{(u,v)\in E} |w((u,v))| \\
  w_2(u) &= \sum_{(v,u)\in E} |w((v,u))| 
\end{align}

\subsection{Graph Transformations}
Sometimes, it is necessary to construct a graph out of another graph.
In the following, we briefly review such constructions.  

Let $G=(V,E,w)$ be any weighted, signed or rating graph, regardless of
edge multiplicities.  Then, $\bar G$ will denote the corresponding
unweighted graph, i.e.,
\begin{align}
  \bar G &= (V,E).
\end{align}
Note that the graph $\bar G$ may still contain multiple edges. 

Let $G=(V,E,w)$ be any graph with multiple edges.  We define the
corresponding unweighted simple graphs as
\begin{align}
  \bar{\bar{G}} = (V, \bar{\bar E}),
\end{align}
where $\bar{\bar E}$ is the set underlying the multiset $E$. For simple
graphs, we define $\bar{\bar G} = G$. 

Let $G=(V,E,w)$ be a signed or rating network.  Then, $|G|$ will denote
the corresponding unsigned graph defined by
\begin{align}
  |G| &= (V,E, w') \\
  w'(e) &= |w(e)|. \nonumber
\end{align}

Let $G=(V,E,w)$ be any network with weight function $w$.  The negative
network to $G$ is then defined as
\begin{align}
  -G &= (V, E, w') \\
  w'(e) &= -w(e). \nonumber
\end{align}
This construction is possible for all types of networks. For unweighted
and positively weighted networks, it leads to signed networks. 

\subsection{Algebraic Graph Theory}
A very useful representation of graph is using matrices. In fact, a
subfield of graph theory, algebraic graph theory, is devoted to
this representation \citep{b118,b646,b902}.  When a graph is represented as a
matrix, operations on graphs can often be expressed as simple algebraic
expressions.  For instance, the number of common friends of
two people in a social network can be expressed as the square of a
matrix. 
This section gives brief definition of the most important graph
matrices.  More properties of these matrices, in particular their
decompositions, as well as other such 
matrices, are given in Section~\ref{sec:matrix}. 

An unweighted graph $G=(V,E)$ can be represented by a $|V|$-by-$|V|$
matrix containing the values 0 and 1, denoting whether a certain edges
between two nodes is present.  This matrix is called the adjacency
matrix of $G$ and will be denoted $\mathbf A$.  Remember that we assume
that the vertices are the natural numbers $1, 2, \dotsc, |V|$.  Then the
entry $\mathbf A_{uv}$ is one when $\{u,v\} \in E$ and zero when not.
This makes $\mathbf A$ square and symmetric for undirected graphs, generally
asymmetric (but still square) for directed graphs.  

For a bipartite graph $G=(V_1 \cup V_2, E)$, the adjacency matrix has
the form 
\begin{align}
  \mathbf A &= \left[ \begin{array}{cc} & \mathbf B \\
      \mathbf B^{\mathrm T} & \end{array} \right].
\end{align}
The matrix $\mathbf B$ is a $|V_1|$-by-$|V_2|$ matrix, and thus
generally rectangular. $\mathbf B$ will be called the biadjacency
matrix. 

In weighted networks, the adjacency matrix takes into account edge
weights.  In networks with multiple edges, the adjacency matrix takes
into account edge multiplicities. Thus, the general definition of the
adjacency matrix is given by
\begin{align}
  \mathbf A_{uv} &= w(u, v). 
\end{align}

The degree matrix $\mathbf D$ is a diagonal $|V|$-by-$|V|$ matrix containing
the absolute weights of all nodes, i.e.,
\begin{align}
  \mathbf D_{uu} &= |w(u)|. 
\end{align}
Note that we define the degree matrix explicitly to contain node weights
instead of degrees, to be consistent with the definition of $\mathbf
A$. 

For directed graphs, we can define the diagonal degree matrix
specifically for outdegrees and indegrees as follows:
\begin{align}
  [\mathbf D_1]_{uu} &= |w_1(u)| \\
  [\mathbf D_2]_{uu} &= |w_2(u)| 
\end{align}
Note that while the outdegree and indegree is denoted $d^+$ and $d^-$, we
avoid superscripts for matrices, as they can be confused with powers or
matrix inverses, etc. 

The normalized adjacency matrix $\mathbf N$ is a $|V|$-by-$|V|$ matrix
given by
\begin{align}
  \mathbf N &= \mathbf D^{-\frac 12} \mathbf A \mathbf D^{-\frac 12}. 
\end{align}

Finally the Laplacian matrix $\mathbf L$ is an $|V|$-by-$|V|$ matrix
defined as
\begin{align}
  \mathbf L = \mathbf D - \mathbf A. 
\end{align}
The term \emph{Laplacian matrix} is also used in a wider sense to refer
to similar matrices, which, like $\mathbf L$, are
positive-semidefinite.  An alternative convention exists in which the
Laplacian matrix is defined as $\mathbf A - \mathbf D$, effectively
negating all eigenvalues.  In KONECT, we use the $\mathbf D - \mathbf A$
convention for all Laplacian-type matrices.  The Laplacian is also
called the \emph{Laplace matrix}. 

Note that in most cases, we work on just a single graph, and it is
implicit that the characteristic matrices apply to this graph.  In a few
cases, we may need to consider the characteristic matrices of multiple
graphs.  In these cases, we will write
\begin{align*}
  \mathbf A[G], \mathbf D[G], \mathbf L[G], \dotsc
\end{align*}
to denote the characteristic matrices of the graph $G$. 

Another matrix that appears in relation to graph analysis is the
incidence matrix.  Unlike almost all other matrices used in KONECT and
in network analysis in general, the incidence matrix is not indexed by
nodes of the graph, but by edges.  The incidence matrix $\mathbf E \in
\{0,1\}^{|V|\times|E|}$ is a nodes-by-edges matrix, and it contains the
value one when a node is incident to an edge, hence the name:
\begin{align*}
  \mathbf E_{u \{u,v\}} &= 1 \;\text{for edges $\{u,v\} \in E$} \\
  \mathbf E_{u \{v,w\}} &= 0 \;\text{when $u \neq v$ and $u \neq w$}
\end{align*}

To be precise, the exact definition of $\mathbf E$ requires one to
define a numbering for the edges, just as a number of the nodes is necessary
to define the adjacency matrix. 

The incidence matrix $\mathbf E$ is highly sparse:  Only a proportion
$2/m$ of all entries are nonzero.  Since the incidence matrix represents
the same information as the adjacency matrix, it is not necessary to represent
it programmatically, except when additional edge attributes are used,
which is not the case in KONECT.  Rather, the matrix is of theoretical
interest.  For instance, we can derive the equality $\mathbf E \mathbf
E^{\mathrm T} = \mathbf D + \mathbf A$, which corresponds to the
\emph{signless Laplacian matrix}, as will be defined in
Section~\ref{sec:signless-laplacian}.  

A variant is the signed incidence matrix $\mathbf E^{\pm}$, which has
the same size and values as $\mathbf E$, with the exception that in each
column, a single value of $1$ is replaced by $-1$.  The choice of which
of the two nonzero values in each column to negate is arbitrary, and
corresponds to chosing an orientation for each edge.  
This matrix appears
for instance in the equality $\mathbf E^{\pm} [\mathbf E^{\pm}]^{\mathrm
  T} = \mathbf D - \mathbf A = \mathbf L$, which proves that the matrix
$\mathbf L$ is positive-semidefinite.  Note that the equality holds
independently of the chosen edge orientations. 

Both variants of the incidence matrix may also be denoted $\mathbf B/\mathbf R$
\cite[e.g.\ in][]{b902} and $\mathbf M$
\cite[e.g.\ in][]{b646} in other literature. 

Finally, the line matrix $\mathbf F \in \{0,1\}^{|E|\times|E|}$ of a graph is the edge-by-edge
matrix given by 
\begin{align}
  \mathbf F &= [\mathbf E^{\pm}]^{\mathrm T} \mathbf E^{\pm}.
\end{align}
This matrix denotes which pairs of edges are incident, i.e., share an
endpoint.  This matrix is sometimes also denoted $\mathbf G$.  It shares
its spectrum with the signless Laplacian matrix $\mathbf K = \mathbf D +
\mathbf A$, and thus, as will be shown in
Section~\ref{sec:signless-laplacian}, it is positive-semidefinite, and
has eigenvalue zero if and only if the graph contains a connected
component that is bipartite. 

The line matrix $\mathbf F$ is also the adjacency matrix of the original
graph's line graph. 

\section{Statistics}
\label{sec:statistics}
A network statistic is a numerical value that characterizes a network.
Examples of network statistics are the number of nodes and the number of
edges in a network, but also more complex measures such as the diameter and the
clustering coefficient.  
Statistics are the basis of most network analysis methods; they can be
used to compare networks, classify networks, detect anomalies in
networks and for many other tasks.  Network statistics are also used to
map a network's structure 
to a simple numerical space, in which many standard statistical
methods can be applied.  Thus, network statistics are essential for the
analysis of almost all network types. 
All statistics described in KONECT are real numbers.
Graph statistics are also called graph invariants and grapg metrics.  In
KONECT we use the term \emph{statistic} exclusively. 

This section gives the definitions for the statistics supported by
KONECT, and briefly reviews their uses.  
All network statistics can be computed using the KONECT Toolbox using
the function \texttt{konect\_statistic()}. Each statistic has an
internal name that must be passed as the first argument to
\texttt{konect\_statistic()}.  The internal names are given in the
margin in this section. 
Additionally, the KONECT Toolbox includes functions named
\texttt{konect\_statistic\_<NAME>()} which compute a single statistic
\texttt{<NAME>}. 

The values of selected statistics are
shown for the KONECT networks on the
website\footnote{\href{http://konect.cc/statistics/}{konect.cc/statistics}}.  

\subsection{Basic Network Statistics}
Some statistics are simple to define, trivial to compute, and 
are reported universally in studies about networks.  These include the
number of nodes, the number of edges, and statistics derived from them
such as the average number of neighbors a node has.  

The size of a network is the number of nodes it contains, and
is almost universally denoted $n$.  The size of a graph is sometimes also
called the order of the graph. 
\begin{align}
  \mathnote{\texttt{size}}
  n &= |V|
\end{align}
In a bipartite graph, the size can be decomposed as $n = n_1 + n_2$ with
$n_1 = |V_1|$ and $n_2=|V_2|$.  The size of a network is not necessarily
a very meaningful number.  For instance, adding a node without edges to
a network will increase the size of the network, but will not change
anything in the network. In the case of an online social
network, this would correspond to creating a user account and not
connecting it to any other users -- this adds an inactive user, which
are often not taken into account.  Therefore, a more representative
measure of the \emph{size} of a network is actually given by the number
of edges, giving the volume of a network.

The volume of a network equals the number of edges and is defined as 
\begin{align}
  \mathnote{\texttt{volume}}
  m &= |E|. 
\end{align}
Note that in mathematical contexts, the number of edges may be called
the \emph{size} of the graph, in which case the number of nodes is
called the \emph{order}.  In this text, we will consistently use
\emph{size} for the number of nodes and \emph{volume} for the number of
edges. 

The volume can be expressed in terms of
the adjacency or biadjacency matrix of the underlying unweighted graph as
\begin{align}
  m &= \left\{ \begin{array}{ll}
    \frac 1 2 \| \mathbf A[\bar G] \|_{\mathrm F} ^2 &
    \text{when $G$ is undirected} \\
    \| \mathbf A[\bar G] \| _{\mathrm F} ^2 &
    \text{when $G$ is directed} \\
    \| \mathbf B[\bar G] \| _{\mathrm F} ^2 &    
    \text{when $G$ is bipartite}
    \end{array} \right.
\end{align}
The number of edges in network is often considered a better measure of
the \emph{size} of a network than the number vertices, since a vertex
unconnected to any other vertices may often be ignored.  On the
practical side, the volume is also a much better indicator of the amount
of memory needed to represent a network.

We will also make use of the number of edges without counting multiple
edges.  We will call this the unique volume of the graph. 
\begin{align}
  \mathnote{\texttt{uniquevolume}}
  \bar{\bar m} &= m[\bar{\bar{G}}]
\end{align}

The weight $w$ of a network is defined as the sum of absolute edge weights.  For
unweighted networks, the weight equals the volume. For rating networks,
remember that the weight is defined as the sum over ratings from which the overall
mean rating has been subtracted, in accordance with the definition of
the adjacency matrix for these networks. 
\begin{align}
  \mathnote{\texttt{weight}}
  w &= \sum_{e \in E} |w(e)|
\end{align}

The number of loops is also a statistic, and is denoted
$l$. \marginpar{\texttt{loops}} 

The average degree is defined as
\begin{align}
  \mathnote{\texttt{avgdegree}}
  d &= \frac 1 {|V|} \sum_{u\in V} d(u) = \frac {2m} n . 
\end{align}
The average degree is sometimes called the \emph{density} or \emph{node density}.  We avoid the term
\emph{density} in KONECT as it is sometimes used for the fill, which
denotes the probability that an edge exists.  The average degree is also
called the expected degree.  
In bipartite networks, we
additionally define the left and right average degree
\begin{align}
  d_1 &= \frac 1 {|V_1|} \sum_{u \in V_1} d(u) = \frac m {n_1} \\
  d_2 &= \frac 1 {|V_2|} \sum_{u \in V_2} d(u) = \frac m {n_2} 
\end{align}
Note that in directed networks, the average outdegree equals the average
indegree, and both are equal to $m/n$. 

The fill of a network is the proportion of edges to the total
number of possible edges. 
The fill is used as a basic parameter in the Erdős--Rényi random graph
model \citep{b569}, where it denotes the probability that an edge is present between
two randomly chosen nodes, and is usually called $p$, which is the
notation we also use in KONECT. 
\begin{align}
  \mathnote{\texttt{fill}}
  p &= \left\{ \begin{array}{ll}
    2m / [n(n-1)] & \text{when $G$ is undirected without loop} \\
    2m / [n(n+1)] & \text{when $G$ is undirected with loops} \\
    m / [n(n-1)] & \text{when $G$ is directed without loops} \\
    m / n^2 & \text{when $G$ is directed with loops} \\
    m / (n_1 n_2) & \text{when $G$ is bipartite} 
  \end{array}\right. 
\end{align}
In the undirected case, the expression is explained by the fact that the
total number of possible edges is $n(n-1)/2$ excluding loops.  
The fill is sometimes also called the density or edge density of the network, in
particular in a mathematical context, or the connectance of the
network\footnote{Used for instance in this blog entry:  \href{https://proopnarine.wordpress.com/2010/02/11/graphs-and-food-webs/}{proopnarine.wordpress.com/2010/02/11/graphs-and-food-webs}}. 

The maximum degree equals the highest degree value attained
by any node.
\begin{align}
  \mathnote{\texttt{maxdegree}}
  d_{\max} &= \max_{u\in V} d(u)
\end{align}
This is can be applied to directed and bipartite networks in the obvious way:
\begin{align*}
  d^+_{\max} &= \max_{u \in V} d^+(u) \\
  d^-_{\max} &= \max_{u \in V} d^-(u) \\
  d_{1 \max} &= \max_{u \in V_1} d(u) \\
  d_{2 \max} &= \max_{v \in V_2} d(u) \\
\end{align*}

The maximum degree can be divided by the average degree to normalize it.
\begin{align}
  \mathnote{\texttt{relmaxdegree}}
  d_{\mathrm{MR}} &= \frac {d_{\max}} {d}
\end{align}

In a directed network, the reciprocity equals the proportion of edges
for which an edge in the opposite direction exists, i.e., that are
reciprocated \citep{b866}.  
\begin{align}
  \mathnote{\texttt{reciprocity}}
  y = \frac 1 m  | \{ (u,v) \in E \mid (v,u) \in E \} | 
\end{align}
The reciprocity has also been noted $r$ \citep[e.g.][]{b867}. 
The reciprocity can give an idea of the type of network.  For instance,
citation networks only contain only few pairs of papers that mutually
cite each other.  On the other hand, an email network will contain many
pairs of people who have sent emails to each other.  Thus, citation
networks typically have low reciprocity, and communnication networks
have high reciprocity. 

In a graph with multiple edges, we may consider the average
multiplicity, i.e., the average number of edges connecting two nodes
that are connected by at least one edge. 
\begin{align}
  \mathnote{\texttt{avgmult}}
  \tilde{m} = \frac m {\bar{\bar m}} 
\end{align}

\subsection{Subgraph Count Statistics}
\label{sec:count-statistics}
The fundamental building block of a network are the edges.  Thus, the
number of edges is a basic statistic of any network.  To understand the
structure of a network, it is however not enough to analyse edges
individually.  Instead, larger patterns such as triangles must be
considered.  These patterns can be counted, and give rise to count
statistics, i.e., statistics that count the number of ocurrences of
specific patterns. 

\begin{figure}
  \centering
  \includegraphics[width=0.20\textwidth]{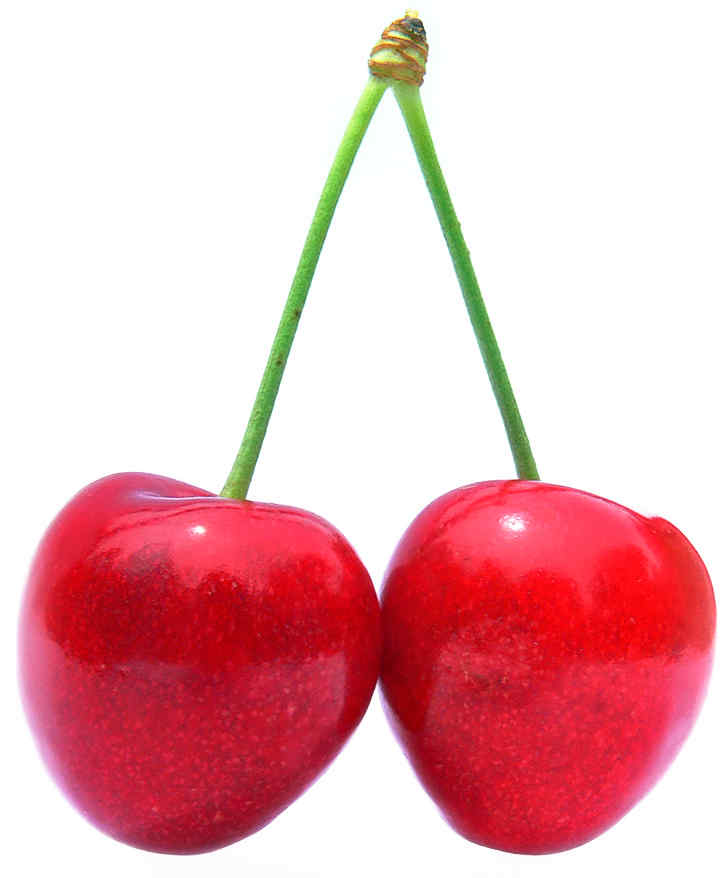}
  \caption{
    A 2-star is a graph consisting of three nodes, two of which are
    connected.  2-stars are occasionally called \emph{cherries} due to
    their resemblance to the fruit.
    \label{fig:cherries}
  }
\end{figure}

Table~\ref{tab:patterns} gives a list of fundamental patterns in
networks, and their corresponding count statistics.

\begin{table}
  \caption{
    \label{tab:patterns}
    Subgraph patterns that occur in networks.  Each pattern can be counted,
    giving rise to a count statistic. 
  }
  \centering
  \makebox[\textwidth]{
  \begin{tabular}{ c l l l }
    \toprule
    \textbf{Pattern} & \textbf{Name(s)} & \textbf{Statistic} &
    \textbf{Internal name} \\

    \midrule
    \begin{tikzpicture}
      [scale=.4,every node/.style={circle,fill=blue!40}]
      \node (n1) at (1,1) {};
    \end{tikzpicture}      
    & Node, 0-star, 0-path, 1-clique
    & $n$ & \texttt{size} \\

    \midrule
    \begin{tikzpicture}
      [scale=.3,every node/.style={circle,fill=blue!40}]
      \node (n1) at (1,1) {};
      \node (n2) at (3,1) {};
      \draw (n1)--(n2);
    \end{tikzpicture}      
    & Edge, 1-star, 1-path, 2-clique
    & $m$ & \texttt{volume} \\

    \midrule
    & Loop 
    & $l$ & \texttt{loops} \\

    \midrule
    \begin{tikzpicture}
      [scale=.3,every node/.style={circle,fill=blue!40}]
      \node (n1) at (2,2.73) {};
      \node (n2) at (1,1) {};
      \node (n3) at (3,1) {};
      \draw (n1)--(n2);
      \draw (n1)--(n3);
    \end{tikzpicture}
    & Wedge, 2-star, 2-path
    & $s$ & \texttt{twostars} \\

    \midrule
    \begin{tikzpicture}
      [scale=.3,every node/.style={circle,fill=blue!40}]
      \node (n1) at (2,2.73) {};
      \node (n2) at (1,1) {};
      \node (n3) at (3,1) {};
      \draw (n1)--(n2);
      \draw (n1)--(n3);
      \draw (n2)--(n3); 
    \end{tikzpicture}
    & Triangle, 3-cycle, 3-clique 
    & $t$ & \texttt{triangles} \\

    \midrule
    \begin{tikzpicture}
      [scale=.3,every node/.style={circle,fill=blue!40}]
      \node (n1) at (1,1) {};
      \node (n2) at (3,1) {};
      \node (n3) at (5,1) {};
      \node (n4) at (3,3) {};
      \draw (n1)--(n2);
      \draw (n2)--(n3);
      \draw (n2)--(n4);
    \end{tikzpicture}
    & Claw, 3-star 
    & $z$ & \texttt{threestars} \\

    \midrule
    \begin{tikzpicture}
      [scale=.3,every node/.style={circle,fill=blue!40}]
      \node (n1) at (1,1) {};
      \node (n2) at (1,3) {};
      \node (n3) at (3,3) {};
      \node (n4) at (3,1) {};
      \draw (n1)--(n2);
      \draw (n2)--(n3);
      \draw (n3)--(n4);
      \draw (n4)--(n1); 
    \end{tikzpicture}
    & Square, 4-cycle 
    & $q$ & \texttt{squares} \\

    \midrule
    \begin{tikzpicture}
      [scale=.3,every node/.style={circle,fill=blue!40}]
      \node (n1) at (2,2) {};
      \node (n2) at (1,1) {};
      \node (n3) at (1,3) {};
      \node (n4) at (3,3) {};
      \node (n5) at (3,1) {};
      \draw (n1)--(n2);
      \draw (n1)--(n3);
      \draw (n1)--(n4);
      \draw (n1)--(n5);
    \end{tikzpicture}
    & Cross, 4-star 
    & $x$ & \texttt{fourstars} \\

    \midrule

    & $k$-Star
    & $S_k$ & \\

    & $k$-Path
    & $P_k$ & \\

    & $k$-Cycle
    & $C_k$ & \\

    & $k$-Clique 
    & $K_k$ & \\
    
    \bottomrule
  \end{tabular}
  }
\end{table}

A star is defined as a graph in which a central node is connected to all
other nodes, and no other edges are present. 
Specifically, a $k$-star is defined as a star in which the central node is
connected to $k$ other nodes.  Thus, a 2-star consists of a node
connected to two other nodes, or equivalently two incident edges, or a
path of length 2.  The specific name for 2-stars is \emph{wedges}.  The number of
wedges can be defined as 
\begin{align}
  \mathnote{\texttt{twostars}}
  s = \sum_{u \in V} {d(u) \choose 2} = \sum_{u \in V} \frac 1 2 d(u) (d(u) - 1),
\end{align}
where $d(u)$ is the degree of node $u$. 
Wedges have many different names:  2-stars, 2-paths,
hairpins \citep[e.g.][]{b853} and cherries.   

Three-stars are defined analogously to two-stars, and their count
denoted $z$.  Three-stars are also called \emph{claws} and
\emph{tripins} \citep[e.g.][]{b853}.   
\begin{align}
  \mathnote{\texttt{threestars}}
  z = \sum_{u \in V} {d(u) \choose 3} = \sum_{u \in V} \frac 1 6 d(u)
  (d(u) - 1) (d(u) - 2)
\end{align}

In the general case, the number of $k$-stars is defined as
\begin{align}
  S_k &= \sum_{u \in V} {d(u) \choose k}
\end{align}

The number of
triangles defined in the following way is independent of the orientation
of edges when the graph is directed.  Loops in the graph, as well as
edge multiplicities, are ignored.
\begin{align}
  \mathnote{\texttt{triangles}}
  t = | \{ \{u, v, w\} \mid u \leftrightarrow v \leftrightarrow w \leftrightarrow u \} | \;/\; 6
\end{align}

A square is a cycle of length four, and the number of squares in a graph
is denoted $q$.
\begin{align}
  \mathnote{\texttt{squares}}
  q = | \{ u, v, w, x \mid u \leftrightarrow v \leftrightarrow w \leftrightarrow x \leftrightarrow u \} | \;/\; 8
\end{align}
The factor 8 ensures that squares are counted regardless of their edge
labeling.  

Multiple edges are ignored in these count statistics, and edges in
patterns are not allowed to overlap. 

Triangles and squares are both cycles~-- which we can generalize to
$k$-cycles, sequences of $k$ distinct vertices that are cyclically
linked by edges.  We denote the number of $k$-cycles by $C_k$. 
For small $k$, we note the following equivalences: 
\begin{align*}
  C_1 &= 0 \\
  C_2 &= m \\
  C_3 &= t \\
  C_4 &= q
\end{align*}
for graphs without loops.  Cycles of length three and four have special
notation:  $C_3 = t$ and $C_4 = q$ and are called triangles and
squares. 

A cycle cannot the same node twice.  Due to this combinatorial
restriction, $C_k$ is quite complex to compute for large $k$.
Therefore, we may use \emph{tours} instead, defined as cyclical lists of
connected vertices in which we allow several vertices to overlap.  The
number of $k$-tours will be denoted $T_k$.  For computational
conveniance, we will define labeled tours, where two tours are not equal
when they are identical up to shifts or inversions.  
We note the following equalities: 
\begin{align}
  T_1 &= 0 \nonumber \\
  T_2 &= 2m \nonumber \\
  T_3 &= 6t \nonumber \\
  \mathnote{\texttt{tour4}}
  T_4 &= 8q + 4s + 2m 
\end{align}
Again, these are true when the graph is loopless.  The last equality
shows that trying to divide the tour count by $2k$ to count them up to
shifts and inversions is a bad idea, since it cannot be implemented by
dividing the present definition by $2k$. 

As mentioned before, counting cycles is a complex problem.  Counting
tours is however much easier.  The number of tours of length $k$ can be
expressed as the trace of a power of the graph's adjacency matrix, and
thus also as a moment of the adjacency matrix's spectrum when $k > 2$.  
\begin{align}
  T_k &= \mathrm{Tr}(\mathbf A^k) = \sum_i \lambda_i[\mathbf A]^k
\end{align}
This remains true when the graph includes loops.  

\subsection{Connectivity Statistics}
Connectivity statistics measure to what extent a network is
connected. 
Two nodes are said to be connected when they are either directly
connected through an edge, or indirectly through a path of several
edges. 
A connected component is a set of vertices all of which are connected,
and unconnected to the other nodes in the network.  
The largest connected component in a network is usually very large and
called the giant connected component. When it contains all nodes, the
network is connected. 

The size of the largest connected component is denoted 
$N$.  
\begin{align}
  \mathnote{\texttt{coco}}
  N &= \max_{F \subseteq \mathcal C} |F|  \\
  \mathcal C &= \{ C \subseteq V \mid \forall u, v \in C:  \exists w_1,
  w_2, \ldots \in V:  u \leftrightarrow w_1 \leftrightarrow w_2 \leftrightarrow \cdots \leftrightarrow v \} \nonumber
\end{align}

In bipartite networks, the number of left and right nodes in the largest
connected components are denoted $N_1$ and $N_2$,
with $N_1 + N_2 = N$. 

The relative size of the largest connected component equals the
size of the largest connected component divided by the size of the
network
\begin{align}
  \mathnote{\texttt{cocorel}}
  N_{\mathrm{rel}} &= \frac N n. 
\end{align}

We also use an inverted variant of the relative size of the largest
connected component, which makes it easier to plot the values on a
logarithmic scale.
\begin{align}
  \mathnote{\texttt{cocorelinv}}
  N_{\mathrm{inv}} &= 1 - \frac N n 
\end{align}

In directed networks, we additionally define the size of the largest
\marginpar{\texttt{cocos}}
strongly connected component $N_{\mathrm s}$.  A strongly
connected component is a 
set of vertices in a directed graph such that any node is reachable from
any other node using a path following only directed edges in the forward
direction.   We always have $N_{\mathrm s} \leq N$. 

\subsection{Distance Statistics}
\label{sec:distance-statistics}
The distance between two nodes in a network is defined as the number of
edges needed to reach one node from another, and serves as the basis for
a class of network statistics.
For instance, the well-known characterisation of networks as having the
\emph{small-world} property by Watts and Strogatz \citeyearpar{b228}
implies that distances between nodes in networks are typically short. 

A path in a network is a sequence of incident edges, or equivalently, a
sequence of nodes $P = (u_0, u_2, \dotsc, u_k)$, such that $(u_i,
u_{i+1})\in E$ for all $i \in \{0, \dotsc, k-1\}$.  The number $k$ is
called the length of the path, and will also be denoted $l(P)$.  A
further restriction can be set on the visited nodes, definining that
each node can only be visited at most once. If the distinction is made,
the term \emph{path} is usually reserved for sequences of non-repeating
nodes, and general sequence of adjacent nodes are then called
\emph{walks}.  We will not make this distinction here.

Paths in networks can be used to model browsing behavior of people in
hyperlink networks, navigation in transport networks, and other types of
movement-like activities in a network.  When considering navigation and
browsing, an important problem is the search for shortest paths.  Since
the length of a path determines the number of steps needed to reach one
node from another, it can be used as a measure of distance between nodes
of a network.  The distance defined in this way may also be called the
shortest-path distance to distinguish it from other distance measures
between nodes of a network.
\begin{align}
  d(u, v) &= \left\{ \begin{array} {ll} \min_{P=(u, \dotsc, v)} l(P) &
    \text{when $u$ and $v$ are connected} \\ \infty & \text{when $u$ and
      $v$ are not connected}
  \end{array} \right. 
\end{align}
In the case that a network is not connected, the distance is defined as
infinite.  In practice, only the largest connected component of a
network may be used, making it unnecessary to deal with infinite values.
The distribution of all $|V|^2$ values $d(u,v)$ for all $u,v\in V$ is
called the distance distribution, and it too characterizes the network.

The eccentricity of a node can then be defined as the maximal distance
from that node to any other node, defining a measure of
\emph{non-centrality}:
\begin{align}
  \epsilon(u) &= \max_{v\in V} d(u,v)
\end{align}

The diameter $\delta$ of a graph equals the longest shortest path in the
network \citep{b779}.  It can be equivalently defined as the largest eccentricity of
all nodes.
\begin{align}
  \mathnote{\texttt{diam}} \delta &= \max_{u \in V} \epsilon(u) =
  \max_{u,v \in V} d(u,v)
\end{align}

Note that the diameter is undefined (or infinite) in unconnected
networks, and thus in numbers reported for actual networks in KONECT we
consider always the diameter of the network's largest connected
component.  Du to the high runtime complexity of computing the diameter,
it may be estimated by various methods, in which case it is noted noted
$\tilde \delta$.

A statistic related to the diameter is the radius, defined as the
smallest eccentricity
\begin{align}
  \mathnote{\texttt{radius}} r &= \min_{u \in V} \epsilon(u) = \min_{u
    \in V} \max_{v \in V} d(u,v)
\end{align}

The diameter is bounded from below by the radius, and from above by
twice the radius.  
\begin{align}
  r \leq \delta \leq 2r
\end{align}
The first inequality follows directly from the definitions of $r$ and
$\delta$ as the minimal and maximal eccentricity.  The second inequality
follows from the fact that between any two nodes, the path joining them
cannot be longer that the path joining them going through a node with
minimal eccentricity, which has length of at most $2r$. 

The radius and the diameter are not very expressive statistics: Adding
or removing an edge will, in many cases, not change their values.  Thus,
a better statistic that reflects the typical distances in a network in
given by the mean and average distance.

The mean path length $\delta_{\mathrm m}$ in a network is defined as as
the mean distance over all node pairs, including the distance between a
node and itself \citep{b779}:
\begin{align}
  \mathnote{\texttt{meandist}} 
  \delta_{\mathrm m} &= \frac 1 {n^2}
  \sum_{u \in V} \sum_{v\in V} d(u,v)
\end{align}
The mean path length defined in this way is undefined when a graph is
disconnected.  Also, the average inverse distance has been used, or
equivalently, the inverse of the harmonic mean of distances
\citep{b877}.  The sum of path lengths has been used on chemistry, where
it is referred to as the Wiener index \citep{b901}.

\marginpar{\texttt{mediandist}}
Likewise, the median path length $\delta_{\mathrm M}$ is the median
length of shortest paths in the network.  In KONECT, both the median and
mean path lengths are computed taking into account node pairs of the
form $(u,u)$.

Both the mean and median path length can be called the
\emph{characteristic path length} of the network.

A related statistic is the 90-percentile effective diameter
$\delta_{0.9}$, which equals the number of edges needed on average to
reach 90\% of all other nodes.

\subsection{Degree Distribution Statistics}
The distribution of degree values $d(u)$ over all nodes $u$ is often
taken to characterize a network.  Thus, a certain number of network
statistics are based solely on this distribution, regardless of overall
network structure.  
The simplest network statistics that depend only on the degree
distribution are the number of nodes $n$, the number of edges $m$ and
the average degree $d=2m/n$, as described previously.  More elaborative
statistics however reveal information about the detailed shape of the
degree distribution. 

The power law exponent is a number that characterizes the degrees of the
nodes in the network.  In many circumstances, networks are modeled to
follow a degree distribution power law, i.e., the number of nodes with
degree $n$ is taken to be proportional to the power $n^{-\gamma}$, for a
constant $\gamma$ larger than one \citep{b439}.  This constant $\gamma$
is called the power law exponent.  Given a network, its degree
distribution can be used to estimate a value $\gamma$.  There are
multiple ways of estimating $\gamma$, and thus a network does not have a
single definite value of it. In KONECT, we estimate $\gamma$ using the
robust method given in \citep[Eq.~5]{b408}
\begin{align}
  \mathnote{\texttt{power}} \gamma &= 1 + n \left( \sum_{u\in V} \ln
  \frac {d(u)} {d_{\min}} \right) ^{-1},
\end{align}
in which $d_{\min}$ is the minimal degree.  This calculation of $\gamma$
takes into account the entire distribution.   Another way to calculate
the power law exponent takes only into account the tail, and it is based
on code by Aaron Clauset.\footnote{\href{http://tuvalu.santafe.edu/~aaronc/powerlaws/}{http://tuvalu.santafe.edu/~aaronc/powerlaws/}}  We call that
variant $\gamma_{\mathrm t}$. \marginpar{\texttt{power2}}

The Gini coefficient is a measure of inequality from economics,
typically applied to distributions of wealth or income.  In KONECT, we
apply it to the degree distribution, as described in
\citep{kunegis:power-law}.  The Gini coefficient can either be defined in
terms of the Lorenz curve, a type of plot that visualizes the inequality
of a distribution, or using the following expression.  Let $d_1 \leq d_2
\leq \dotsb \leq d_{n}$ be the sorted list of degrees in the
network. Then, the Gini coefficient is defined as
\begin{align}
  \mathnote{\texttt{gini}} G &= \frac {2 \sum_{i=1}^{n} i d_i} {n
    \sum_{i-1}^{n} d_i} - \frac {n+1} {n}.
\end{align}
The Gini coefficient takes values between zero and one, with zero
denoting total equality between degrees, and one denoting the dominance
of a single node.

The relative edge distribution entropy is a measure of the equality of
the degree distribution, and equals one when all degrees are equal, and
attains the limit value of zero when all edges attach to a single node
\citep{kunegis:power-law}.  It is defined as
\begin{align}
  \mathnote{\texttt{dentropyn}} H_{\mathrm{er}} &= \frac 1 {\ln n}
  \sum_{u \in V} - \frac{d(u)}{2m} \ln \frac{d(u)}{2m}.
\end{align}

Another statistic for \marginpar{\texttt{own}} measuring the inequality
in the degree distribution is associated with the Lorenz curve (see
Section~\ref{sec:plot:degree}).  Even though it often does appear
indirectly in discussions of skewed distributions, it does not have a
common name.  In KONECT, we call it the balanced inequality ratio. 
It is given by the intersection point
of the Lorenz curve with the antidiagonal given by $y = 1 - x$
\citep{kunegis:power-law}.  By construction, this point equals $(1-P, P)$
for some $0<P<1$, where the value $P$ corresponds exactly to the number
``25\%'' in the statement ``25\% of all users account for 75\% of all
friendship links on Facebook''.  By construction, we can expect $P$ to
be smaller when $G$ is larger, but there is no precise relationship
between the two.

Another statistic, which has been called the ``hubiness'' [sic], is given by
the variance of the degree distribution \cite[e.g. by][]{b883,b884}.

Statistics such as the power law exponent $\gamma$, the Gini coefficient
$G$, the relative edge distribution entropy $H_{\mathrm{er}}$, as well
as the ratio $P$ all correlate with each other, as they measure the
inequality, skewness or ``hubiness'' of the degree distribution. 

The analysis of degrees can be generalized to pairs of nodes:  What is
the distribution of degrees for pairs of connected edges?  In some
networks, high-degree nodes are connected to other high-degree nodes,
while low-degree nodes are connected to low-degree nodes.  This property
is called the degree assortativity \citep{b854}, or, very often, simply the
assortativity.  Inversely, in a network with dissortativity,  
high-degree nodes are typically connected to low-degree and vice versa.
The amount of assortativity can be measured by the Pearson correlation
$\rho$ between the degree of connected nodes.  
\begin{align}
  \mathnote{\texttt{assortativity}} 
  \rho &= \frac
  {\sum_{\{u,v\} \in E} (d(u) - d)(d(v) - d)}
  {\sum_{\{u,v\} \in E} (d(u) - d)^2}
\end{align}
The assortativity is
undefined whenever the Pearson correlation is undefined, for instance,
if all nodes have the same degree (i.e., when the graph is regular), and
when the graph does not contain any edges. 

\marginpar{\texttt{inoutassort}}
In directed networks, we define the in-out assortativity as the Pearson
correlation coefficient between in-degree and the out-degree of nodes,
taking the logarithm of one plus the degree.

\subsection{Preferential Attachment Statistics}
The term \emph{preferential attachment} refers to the observation that
in networks that grow over time, the probability that an edge is added
to a node with $d$ neighbors is proportional to $d$.  This linear
relationship lies at the heart of Barabási and Albert's
\emph{scale-free} network model \citep{b439}, and has been used in a vast
number of subsequent work to model networks, online and offline. The
scale-free network model results in a distribution of degrees, i.e.,
number of neighbors of individual nodes, that follows a power law with
negative exponent. In other words, the number of nodes with degree $d$ is
proportional to $d^{-\gamma}$ in these networks, for a constant
$\gamma>1$.

In basic preferential attachment, the probability that an edge
attached to a vertex $u$ is propertional to its degree $d(u)$.  An
extension of this basic model uses a probability that is a power of the
degree, i.e., $d(u)^\beta$.  The exponent $\beta$ is a positive number,
and can be measured empirically from a dataset
\citep{kunegis:preferential-attachment}.  The value of $\beta$ then
determines the type of preferential attachment:
\begin{enumerate}
\item \textbf{Constant case $\beta=0$}.  
This case is equivalent to a constant probability of attachment, 
and thus this graph growth model results in networks in which each edge
is equally likely and independent from other edges.  This is the
Erdős--Rényi model of random graphs \citep{b569}. 

\item \textbf{Sublinear case $0 < \beta < 1$}. 
In this case, the preferential attachment function is sublinear.  This
model gives rise to a stretched exponential degree distribution
\citep{b764}, whose exact 
expression is complex and given in \citep[Eq. 94]{b773}. 

\item \textbf{Linear case $\beta=1$}.
This is the scale-free network model of Barabási and Albert
\citep{b439}, in which attachment is proportional to the degree. This gives a
power law degree distribution.  

\item \textbf{Superlinear case $\beta > 1$}.
In this case, a single node will acquire 100\% of all edges
asymptotically \citep{b765}. Networks with this behavior will however display
power law degree distributions in the pre-asymptotic regime
\citep{b769}. 
\end{enumerate}

The following minimization problem gives an estimate for the exponent
$\beta$ \citep{kunegis:preferential-attachment}.
\marginpar{\texttt{prefatt}}
\begin{align}
  \min_{\alpha,\beta} \sum_{u\in V} \left( \alpha + \beta \ln[1 + d_1(u)]
  - \ln[\lambda + d_2(u)]  \right)^2
  \label{eq:min}
\end{align}
The resulting value of $\beta$ is the estimated preferential attachment
exponent. 

To measure the error of the fit, the root-mean-square
logarithmic error $\epsilon$ can be defined in the following way: 
\begin{align*}
    \epsilon &= \exp\left\{ \sqrt{ \frac 1 {|V|} \sum_{u \in V}
      \left(\alpha + \beta \ln[1 
        + d_1(u)] - \ln[\lambda + d_2(u)]\right)^2  }  \right\}
\end{align*}
This gives the average factor by which the actual new number of edges
differs from the predicted value, computed logarithmically. 
The value of $\epsilon$ is larger or equal to one by construction. 

\subsection{Clustering Statistics}
\label{sec:statistic:clustering}
The term \emph{clustering} refers to the observation that in almost all
networks, nodes tend to form small groups within which many edges are
present, and such that only few edges connected different clusters with
each other.  In a social network for instance, people form groups in
which almost every member known the other members.  Clustering thus
forms one of the primary characteristics of real-world networks, and
thus many statistics for measuring it have been defined.  
As an example, the well-known characterisation of networks as having the
\emph{small-world} property by Watts and Strogatz \citeyearpar{b228}
uses the clustering coefficient as one network statistic, and states
that it will be particularly small. 
The main
method for measuring clustering numerically is the clustering
coefficient, of which there exist several variants. As a general rule,
the clustering coefficient measures to what extent edges in a network
tend to form triangles. Since it is based on triangles, it can only be
applied to unipartite networks, because bipartite networks do not
contain triangles.

The number of triangles $t$ itself as defined in
Section~\ref{sec:count-statistics} is however not a statistic that can
be used to measure the clustering in a network, since it correlates with
the size and volume of the network.  Instead, the clustering
coefficients in all its variants can be understood as a count of
triangles, normalized in different ways in order to compare several
networks with it.

The local clustering coefficient $c(u)$ of a node $u$ is defined as the
probability that two randomly chosen (but distinct) neighbors of $u$ are
connected.
\begin{align}
  c(u) &= \left\{ \begin{array}{ll} \frac { \{ v, w \in V \mid u \leftrightarrow v
      \leftrightarrow w \leftrightarrow u \} } { \{ v, w \in V \mid u \leftrightarrow v \neq w \leftrightarrow u \}
    } & \text{when } d(u) > 1 \\ 0 & \text{when } d(u) \leq 1
          \end{array} \right. 
\end{align}

The global clustering of a network can be computed in two ways.  The
first way defines it as the probability that two incident edges are
completed by a third edge to form a triangle \citep{b736}. This is also
called the transitivity ratio, or simply the transitivity.
\begin{align}
  \mathnote{\texttt{clusco}} c &= 
  \frac {|\{ u, v, w \in V \mid u \leftrightarrow v \leftrightarrow w \leftrightarrow u \}|} 
        {|\{ u, v, w \in V \mid u \leftrightarrow v \neq w \leftrightarrow u \}|} 
        = \frac {3t} s
\end{align}
This variant of the global clustering coefficient has values between
zero and one, with a value of one denoting that all possible triangles
are formed (i.e., the network consists of disconnected cliques), and
zero when it is triangle free.  Note that the clustering coefficient is
trivially zero for bipartite graphs.  This clustering coefficient is
however not defined when each node has degree zero or one, i.e., when
the graph is a disjoint union of edges and unconnected nodes.  This is
however not a problem in practice.

The second variant variant of the clustering coefficient uses the
average of the local clustering coefficients. This second variant was
historically the first to be defined.  In was defined in
1998 \citep{b228} and precedes the first variant by four years.
\begin{align}
  \mathnote{\texttt{clusco2}} c_2 &= \frac 1 {|V|} \sum_{u \in V} c(u)
\end{align}
This second variant of the global clustering coefficient is zero when a
graph is triangle-free, and one when the graph is a disjoint union of
cliques of size at least three.  This variant of the global clustering
coefficient is defined for all graphs, except for the empty graph, i.e.,
the graph with zero nodes.  A slightly different definition of the
second variant computes the average only over nodes with a degree of at
least two, as seen for instance in \citep{b845}.

Because of the arbitrary decision to define $c(u)$ as zero when the
degree of $c$ is zero or one, we recommend to use the first variant of
the clustering coefficient.  In the following, the extensions to the
clustering coefficient we present are all based on the first variant,
$c$.

For directed graphs, the directed clustering $c^{\pm}$ can be considered.  It is
\marginpar{\texttt{cluscoasym}}
defined in analogy to $c$, but only considering directed 2-paths closed
by a directed edge that has the same orientation as the path. 

For signed graphs, we may define the clustering coefficient to take into
account the sign of edges.  The signed clustering coefficient is based
on balance theory \citep{kunegis:slashdot-zoo}.  In a signed network,
edges can be positive or negative.  For instance in a signed social
network, positive edges represent friendship, while negative edges
represent enmity.  In such networks, balance theory stipulates than
triangles tend to be balanced, i.e., that three people are either all
friends, or two of them are friends with each other, and enemies with
the third.  On the other hand, a triangle with two positive and one
negative edge, or a triangle with three negative edges is unbalanced.
In other words, we can define the sign of a triangle as the product of
the three edge signs, which then leads to the stipulation that triangles
tend to have positive weight.  To extend the clustering coefficient to
signed networks, we thus distinguis between balanced and unbalanced
triangles, in a way that positive triangles contribute positively to the
signed clustering coefficient, and negative triangles contribute
negatively to it.  For a triangle $\{u,v,w\}$, let
$\sigma(u,v,w)=w(u,v)w(v,w)w(w,u)$ be the sign of the triangle, then the
following definition captures the idea:
\begin{align}
  c_{\mathrm s} &= \frac {\sum_{u,v,w\in V} \sigma(u,v,w)} {|\{ u, v, w
    \in V \mid u \leftrightarrow v \neq w \leftrightarrow u \}|}
\end{align}
Here, the sum is over all triangles $\{u,v,w\}$, but can also be taken
over all triples of vertices, since $w(u,v)=0$ when $\{u,v\}$ is not an
edge.

The signed clustering coefficient is bounded by the clustering
coefficient:
\begin{align}
  | c_{\mathrm s} | \leq c
\end{align}

The relative signed clustering coefficient can then be defined as
\begin{align}
  c_{\mathrm r} = \frac {c_{\mathrm s}} c = \frac {\sum_{u,v,w\in V}
    \sigma(u,v,w)} {|\{ u, v, w \in V \mid u \leftrightarrow v \leftrightarrow w \leftrightarrow u \}|}
\end{align}
which also equals the proportion of all triangles that are balanced,
minus the proportion of edges that are unbalanced.

\subsection{Algebraic Statistics}
Algebraic statistics are based on a network's characteristic matrices.
They are motivated by the broader field of spectral graph theory, which
characterizes graphs using the spectra of these matrices \citep{b285}.
This section only describes algebraic statistics that do not fall under
another category, such as bipartivity statistics, or signed graph
statistics. 

In the following we will denote by $\lambda_k[\mathbf X]$ the
$k$\textsuperscript{th} dominant eigenvalue of the matrix $\mathbf X$.
For the adjacency matrix $\mathbf A$, the dominant eigenvalues are the
largest absolute ones; for the Laplacian $\mathbf L$ they are the
smallest ones.

Also, the matrix $\mathbf L$ will only be considered for the network's
largest connected component.

The spectral norm of a network equals the spectral norm (i.e., the
largest absolute eigenvalue) of the network's adjacency matrix
\begin{align}
  \mathnote{\texttt{snorm}} 
  \alpha &= 
  \left\| \mathbf A \right\|_2 = 
  | \lambda_1[\mathbf A] | .
\end{align}
The spectral norm can be understood as an alternative measure of the
size of a network.

For directed graphs, we consider the largest singular value of the
\marginpar{\texttt{opnorm}}
graph's (generally) asymmetric adjacency matrix.  This is called the
operator 2-norm or the Ky Fan 1-norm.  We denote it $\nu$.

Also for directed graphs only, the largest absolute eigenvalue of the
(generally asymmetric) adjacency matrix $\mathbf A$ can be considered.  
\marginpar{\texttt{maxdiag}} 
We denote it $\pi$ and call it the cyclic eigenvalue in KONECT.  It is zero if and
only if the graph is acyclic. 

The algebraic connectivity equals the second smallest nonzero eigenvalue
of $\mathbf L$ \citep{b652}
\begin{align}
  \mathnote{\texttt{alcon}} a &= \lambda_2[\mathbf L].
\end{align}
The algebraic connectivity is zero when the network is disconnected~--
this is one reason why we restrict the matrix $\mathbf L$ to each
network's giant connected component.  The algebraic connectivity is
larger the better the network's largest connected component is
connected.

The largest absolute eigenvalue of $\mathbf W$ is a network statistic.
\marginpar{\texttt{seidelnorm}}

\subsection{Bipartivity Statistics}
Some unipartite networks are almost bipartite.  Almost-bipartite
networks include networks of sexual contact \citep{b719} and ratings in
online dating sites \citep{b311,kunegis:split-complex-dating}.  Other,
more subtle cases, involve online social networks.  For instance, the
follower graph of the microblogging service Twitter is by construction
unipartite, but has been observed to reflect, to a large extent, the
usage of Twitter as a news service \citep{b545}. This is reflected in the
fact that it is possible to indentify two kinds of users: Those who
primarily get followed and those who primarily follow.  Thus, the
Twitter follower graph is almost bipartite.  Other social networks do
not necessarily have a near-bipartite structure, but the question might
be interesting to ask to what extent a network is bipartite.  To answer
this question, measures of bipartivity have been developed.

Instead of defining measures of bipartivity, we will instead consider
measures of non-bipartivity, as these can be defined in a way that they
equal zero when the graph is bipartite.  Given an (a priori) unipartite
graph, a measure of non-bipartivity characterizes the extent to which it
fails to be bipartite.  These measures are defined for all networks, but
are trivially zero for bipartite networks.  For non-bipartite networks,
they are larger than zero.

A first measure of bipartivity consists in counting the minimum number
of \emph{frustrated edges} \citep{b531}. Given a bipartition of vertices
$V=V_1\cup V_2$, a frustrated edge is an edge connecting two nodes in
$V_1$ or two nodes in $V_2$.  Let $f$ be the minimal number of
frustrated edges in any bipartition of $V$, or, put differently, the
minimum number of edges that have to be removed from the graph to make
it bipartite.  Then, a measure of non-bipartivity is given by
\begin{align}
  \mathnote{\texttt{frustration}} F &= \frac f {|E|}.
\end{align}
This statistic is always in the range $[0, 1/2]$.  It attains the value
zero if and only if $G$ is bipartite.

The minimal number of frustrated edges $f$ can be approximated by
algebraic graph theory.  First, we represent a bipartition $V = V_1 \cup
V_2$ by its characteristic vector $\mathbf x \in \mathbb R^{|V|}$
defined as
\begin{align*}
  \mathbf x_u &= \left\{ \begin{array}{ll} +1/2 & \text{when $u \in
      V_1$} \\ -1/2 & \text{when $u \in V_2$}
    \end{array} \right. 
\end{align*}
Note that the number of edges connecting the sets $V_1$ and $V_2$ is
then given by
\begin{align*}
\left\{ \{u,v\} \mid u \in V_1, v \in V_2 \right\} = \frac 12 \mathbf
x^{\mathrm T} \mathbf K[\bar G] \mathbf x &= \frac 12 \sum_{(u,v) \in E}
(\mathbf x_u + \mathbf x_v)^2,
\end{align*}
where $\mathbf K[\bar G] = \mathbf D[\bar G] + \mathbf A[\bar G]$ is the
signless Laplacian matrix of the underlying unweighted graph (see
Section~\ref{sec:signless-laplacian}).  Thus, the 
minimal number of frustrated edges $f$ is given by
\begin{align*}
  f &= \min_{\mathbf x \in \{\pm 1/2\}^{|V|}} \frac 12 \mathbf
  x^{\mathrm T} \mathbf K[\bar G] \mathbf x.
\end{align*}
By relaxing the condition $\mathbf x \in \{\pm 1/2\}^{|V|}$, we can
express $f$ in function of $\mathbf K[\bar G]$'s minimal eigenvalue,
using the fact that the norm of all vectors $\mathbf x \in \{\pm
1/2\}^{|V|}$ equals $\sqrt{|V|/4}$, and the property that the minimal
eigenvalue of a matrix equals its minimal Rayleigh quotient.
\begin{align*}
  \frac {2f} {|V|/4} &\approx \min_{\mathbf x \neq \mathbf 0} \frac
        {\mathbf x^{\mathrm T} \mathbf K[\bar G] \mathbf x} {\left\|
          \mathbf x \right\|^2} = \lambda_{\min}[\mathbf K[\bar G]]
\end{align*}
We can thus approximate the previous measure of non-bipartivity by
\begin{align}
  \mathnote{\texttt{anticonflict}} b_{\mathrm K} &= \frac {|V|} {8 |E[\bar
      G]|} \lambda_{\min}[\mathbf K[\bar G]]
\end{align}
The eigenvalue itself,
\begin{align}
  \mathnote{\texttt{nonbipal}}
  \chi &= \lambda_{\min}[\mathbf K[\bar G]],
\end{align}
as the algebraic non-bipartivity of the graph. 

The eigenvalue $\lambda_{\min}[\mathbf K[\bar G]]$ can also be
interpreted as the algebraic conflict in $G$ interpreted as a signed
graph in which all edges have negative weight.

A further measure of bipartivity exploits the fact that the adjacency
matrix $\mathbf A$ of a bipartite graph has eigenvalues symmetric around
zero, i.e., all eigenvalues of a bipartite graph come in pairs $\pm
\lambda$. Thus, the ratio of the smallest and largest eigenvalues can be
used as a measure of non-bipartivity
\begin{align}
  \mathnote{\texttt{nonbip}} b_{\mathrm A} &= 1 - \left| \frac
             {\lambda_{\min}[\mathbf A[\bar G]]} {\lambda_{\max}[\mathbf
                 A[\bar G]]} \right|,
\end{align}
where $\lambda_{\min}$ and $\lambda_{\max}$ are the smallest and largest
eigenvalue of the given matrix, and $\bar G$ is the unweighted graph
underlying $G$.  Since the largest eigenvalue always has a larger
absolute value than the smallest eigenvalue (due to the
Perron--Frobenius theorem, and from the nonnegativity of $\mathbf A[\bar
  G]$), it follows that this measure of non-bipartivity is always in the
interval $[0,1)$, with zero denoting a bipartite network.  The value one
is excluded in non-empty loopless graphs.  This can be seen by
considering the trace of the graph's adjacency matrix, which equals the
sum of its diagonal values (and therefore equals the number of loops),
and at the same time equals the sum of eigenvalues of the adjacency
matrix. 

Another spectral measure of non-bipartivity is based on considering the
smallest eigenvalue of the matrix $\mathbf N[\bar G]$.  This eigenvalue
is $-1$ exactly when $G$ is bipartite.  Thus, this value minus one is a
measure of non-bipartivity. Equivalently, it equals two minus the
largest eigenvalue of the normalized Laplacian matrix $\mathbf Z$.
\begin{align}
  \mathnote{\texttt{nonbipn}} b_{\mathrm N} &= \lambda_{\min}[\mathbf
    N[\bar G]] + 1 = 2 - \lambda_{\max}[\mathbf Z[\bar G]]
\end{align}

\subsection{Signed Network Statistics}
In networks that allow negative edges such as signed networks and rating
networks, we may be interested in the proportion of edges that are
actually negative.  We call this the \emph{negativity} of the network. 
\begin{align}
  \mathnote{\texttt{negativity}}
  \zeta &= \frac { | \{ e \in E \mid w(e) < 0 \} | } {m}
\end{align}
The negativity is denoted $q$ in \citep{b868}. 

In directed signed networks, we can additionally compute the dyadic
conflict, i.e., the propostion of node pairs connected by two oppositely
oriented edges of different, compared to the total number of pairs of
nodes connected by two edges of opposite orientation. 
\begin{align}
  \mathnote{\texttt{dconflict}}
  \eta &= \frac 
       {|\{ u,v \mid u \rightleftarrows v, w(u,v) = - w(v,u) \}|} 
       {|\{ u, v \mid u \rightleftarrows v \}|}
\end{align}

Furthermore, the triadic conflict can be defined as the proportion of
triangles that are in conflict, i.e., that are unbalanced. 
\begin{align}
  \mathnote{\texttt{tconflict}}  
  \tau &= \frac 
  {|\{ u, v, w \mid w(u, v) w(v, w) w(w, u) < 0\}|}
  {|\{ u, v, w \mid u \sim v \sim w \sim u \}|}
\end{align}
This is also known as the triangle index.
It is also related to the relative signed clustering coefficient by
\begin{align}
\tau &= 2 c_{\mathrm r} - 1.
\end{align}

The smallest eigenvalue of the adjacency matrix $\mathbf L$ can
be larger than zero.  (In other networks, it is always zero.) The
algebraic conflict equals this smallest eigenvalue
\begin{align}
  \mathnote{\texttt{conflict}} 
  \xi &= \lambda_1[\mathbf L].
\end{align}
The algebraic conflict measures the amount of conflict in the network,
i.e., the tendency of the network to contain cycles with an odd number
of negatively weighted edges.  It is applied in KONECT only to the
network's largest connected component. 
A normalized version of the algebraic conflict is an approximation to
the number of edges whose sign must be switched for the graph to become
balanced.  We call it the spectral signed frustration $\phi$, defined as
\begin{align}
  \mathnote{\texttt{fconflict}}
  \phi &= \frac{n \xi}{8 m} = \frac{n \lambda_1[\mathbf L]}{8m}.
\end{align}
The normalization factor can be derived in analogous way as for the
corresponding measure of bipartivity that is based on the signless
Laplacian matrix. 

\subsection{Miscellaneous}
This section lists miscellaneous statistics that do not fit any of the
other sections

\subsubsection{Controllability}
A less-known way to assess the structure of a network consists in
measuring how well it can be controlled.  For instance, assume that we
want to influence opinions in a social network, but are only able to
directly influence $k$ persons in the network. Assuming that opinions
will spread through the network, how big has $k$ to be in order for us
to be able to influence all nodes in a network, in a way that any
arbitrary opinion can be given to any node?  A solution to this problem
is given in \citep{b673}, in which such \emph{driver nodes} are
identified and, surprisingly, they are not necessarily the nodes with
highest degree.  In fact, the authors of that article state that driver
nodes tend to \emph{avoid the hubs} of the network. 

The resulting computational model uses differential equations to model
diffusion and can be reduced to finding a maximal matching in the
bipartite double cover of the network \citep{b673}. The maximal matching
in a bipartite graph can be computed efficiently because of Kőnig's
theorem. It states that perfect matchings and minimal vertex covers have
equal size in bipartite graphs, and thus the corresponding integer
program formulations are equivalent to their relaxations.

The number of driver nodes $C$ needed to control a graph $G=(V,E)$
equals $|V|$ minus the size of the maximum
matching in the bipartite cover of the network, which equals the
maximal directed 2-matching in the network.
The bipartite double cover of a network $G=(V,E)$ is constructed by
mapping each vertex $u\in V$ to two vertices $(u,1)$ and $(u,2)$, and
mapping each edge $\{u,v\}$ to the edges $\{(u,1),(v,2)\}$ and
$\{(u,2),(v,1)\}$.  Equivalently this corresponds
to the size of the maximal directed 2-matching.  A 2-matching is a set
of edges such that each vertex is incident to at most two edges.  A
directed 2-matching is a set of directed edges, such that each vertex is
incident to at most one ingoing and one outgoing edge.  Here, we
interpret an undirected graph as a directed graph where each edge
corresponds to two directed edges:
\begin{align*}
  \mathnote{\texttt{controllability}}
  C = |V| - & \max |M| \\
  & \text{s.t.} \quad \left| \left\{ (i,j) \in M \mid i = u, \{i,j\}\in E\right\} \right| \leq 1 
  & \text{for all } u \in V, \\
  & \phantom{\text{s.t.}} \quad \left| \left\{ (i,j) \in M \mid j = u, \{i,j\}\in E  \right\} \right| \leq 1
  & \text{for all } u \in V. 
\end{align*}
The result is the number $C$ of vertices needed to control a given
network.  
A maximal matching in a bipartite graph can be found in runtime
$O(|V|^{1/2}|E|)$ \citep{b673}, and thus can be computed efficiently even for large
networks. 

We also define the relative controllability as
\begin{align*}
  \mathnote{\texttt{controllabilityn}}
  C_{\mathrm r} &= \frac C n 
\end{align*}

This statistic has been used by the authors of KONECT in
\citep{kunegis:network-rank}. 

\section{Matrices and Decompositions}
\label{sec:matrix}
In order to analyse graphs, algebraic graph theory is a common
approach.  In algebraic graph theory, a graph with $n$ vertices is
represented by an $n\times n$ matrix called the adjacency matrix, from
which other matrices can be derived.  The defined matrices can then be
decomposed to reveal properties of the graph. 
In this section, we review characteristic graph matrices, their
decompositions, and their uses.  Since most decompositions are based on
a specific matrix, this section also serves as a survey of
characteristic graph matrices.  

As a general rule, matrices are noted with bold uppercase letters.  The
letters shown in this section for each characteristic matrix are used
systematically within KONECT, but not necessarily in the literature.
Among established matrix names, there is the adjacency matrix ($\mathbf
A$) and the Laplacian ($\mathbf L$), but we must note that even the
adjacency matrix is not always called $\mathbf A$ in the literature.
Each matrix resp.\ decomposition has an internal name in KONECT, which
is given in the margin.  Using the KONECT Toolbox, the corresponding
matrices can be generated using the \texttt{konect\_matrix()} function,
and the corresponding (rank reduced) matrix decomposition can be
computed with the function \texttt{konect\_decomposition()}. 

The first subsection contains decompositions that apply to simple
graphs.  Subsequent subsections contain matrices that apply to specific
types of graphs: signed, bipartite, and directed graphs.  Almost all
decompositions given in the first subsection can also be applied to
non-simple graphs, in which cases the KONECT implementation will ignore
the extra information.  Exceptions to this are when a decomposition
given in a subsequent subsection shares an internal name with one given
in the first subsection.  In that case, KONECT will do the obvious
generalization when a signed or directed graph is given. 

An unfinished proposal for a more precise naming scheme is given in 
tables \ref{tab:matcross} and~\ref{tab:matov}, including
the internal names of corresponding decompositions and eigenvalue
statistics. 

\begin{table}
  \caption{
    \label{tab:matcross} Matrices that characterize networks.  All shown
    cases include weighted and signed graphs. 
    All matrices shown here are written with a diacritic or a subscript.
    When used without, $\mathbf{\bar X}$ is meant for undirected graphs,
    and $\vec{\mathbf X}$ for undirected graphs; hence the need to use
    diacritics when covering both types of networks. 
  }
  \makebox[\textwidth]{
  \begin{tabular}{ l l c@{\;}l c@{\;}l c }
    \toprule
    & \textbf{Name} 					& \multicolumn{2}{l}{\textbf{Plain}}      & \multicolumn{2}{l}{\textbf{Normalized}} & \textbf{Stochastic} \\
    \midrule
    \multirow{6}{*}{\textbf{Undirected}} 
    & Adjacency matrix    				& $\mathbf{\bar A}$         && $\mathbf{\bar N}$ && $\mathbf{\bar P}$ \\
    & Laplacian 					& $\mathbf{\bar L}$         && $\mathbf{\bar Z}$ && $\mathbf{\bar S}$ \\
    & Signless Laplacian				& $\mathbf{\bar K}$         && $\mathbf{\bar C}$ && $\mathbf{\bar J}$ \\
    \cmidrule(lr){2-7}
    & Repulsive adjacency matrix\textsuperscript{1} 	& $\mathbf{\hat A}$   && $\mathbf{\hat N}$ && $\mathbf{\hat P}$ \\
    & Repulsive Laplacian$\dagger$\textsuperscript{2}	& $\mathbf{\hat L}$   && $\mathbf{\hat Z}$ && $\mathbf{\hat S}$\\
    & Repulsive signless Laplacian$\dagger$\textsuperscript{2}& $\mathbf{\hat K}$   && $\mathbf{\hat C}$ && $\mathbf{\hat J}$\\
    \midrule
    \multirow{15}{*}{\textbf{Directed}} 
    & Symmetric adjacency matrix			& $\mathbf{\bar A}$         && $\mathbf{\bar N}$   && $\mathbf{\bar P}$ \\
    & Symmetric Laplacian		  		& $\mathbf{\bar L}$         && $\mathbf{\bar Z}$   && $\mathbf{\bar S}$ \\
    & Symmetric signless Laplacian			& $\mathbf{\bar K}$         && $\mathbf{\bar C}$   && $\mathbf{\bar J}$ \\
    \cmidrule(lr){2-7}
    & Hermitian adjacency matrix  			& $\mathbf{\check A}$&$=\mathbf H$ & $\mathbf{\check N}$ \\
    & Hermitian Laplacian         			& $\mathbf{\check L}$&$=\mathbf T$ & $?$ \\
    & Hermitian signless Laplacian                      & & &\\
    \cmidrule(lr){2-7}
    & Skew-Hermitian adjacency matrix       		& $\mathbf{\acute A}$&$=i\mathbf Y$ & $\mathbf{\acute N}$ \\
    & Skew-Hermitian Laplacian    			& $\mathbf{\acute L}$ && $?$ \\
    & Skew-Hermitian signless Laplacian 		& & &\\
    \cmidrule(lr){2-7}
    & Magnetic adjacency matrix 			& $\mathbf A_{\theta}$  && $?$ \\
    & Magnetic Laplacian				& $\mathbf L_{\theta}$  && \\
    & Magnetic signless Laplacian			& $\mathbf K_{\theta}$  && \\
    \cmidrule(lr){2-7}
    & Unidirectional adjacency matrix$\dagger$		& $\vec{\mathbf A}$   && $\vec{\mathbf N}$ && $\vec{\mathbf P}$ \\
    & Unidirectional Laplacian$\dagger$			& $\vec{\mathbf L}$   && $\vec{\mathbf Z}$ && $\vec{\mathbf S}$ \\
    & Unidirectional signless Laplacian$\dagger$	& $\vec{\mathbf K}$   && $\vec{\mathbf C}$ && $\vec{\mathbf J}$ \\
    \midrule
    \multirow{1}{*}{\textbf{Bipartite}} 
    & Biadjacency matrix*				& $\mathbf{\grave A}$&$=\mathbf B$ & $\mathbf{\grave N}$&$=\mathbf M$ & $\times$ \\
    \bottomrule
  \end{tabular}
  }
  * Non-square. As a general rule, bipartite graphs are a special case
  of unipartite graphs and thus all variants apply; we only show the two
  cases here in which the matrix has the bipartite form $[\mathbf 0
    \mathbf X; \mathbf X^{\mathrm T} \mathbf 0]$. \\
  $\dagger$ Non-normal.  These matrices allow both the singular value
  decomposition, as well as considering their eigenvalues.  \\
  \textsuperscript{1} Accent usually omitted because the matrix is the natural extension of the non-accented matrix \\
  \textsuperscript{2} Identical to the non-accented matrix for unsigned networks 
\end{table}

\begin{table}
  \caption{
    \label{tab:matov}
    The nine base matrices of KONECT for undirected, unipartite graphs,
    including unweighted, weighted, and signed cases.  
    The transpose of the stochastic matrices are not shown; they are
    left-stochastic and are analogous to the shown right-stochastic
    variants. 
    Question marks indicate aspects we have not yet investigated. 
  }
  \makebox[\textwidth]{
    \scalebox{0.8}{
  \begin{tabular}{ l c@{\;}l l l l l }
    \toprule
    & \multicolumn{2}{l}{\textbf{Expression}} & \textbf{Int.} &
    \textbf{Name} & \textbf{Prop.} & \textbf{Eigenvalues} \\
    \midrule \midrule

    \ref{sec:matrix.A} & $\mathbf A$ && \texttt{sym} & Adjacency m. &
    Sym 		& $\ldots, \alpha$ \texttt{snorm} \\
    \ref{sec:matrix.N} & $\mathbf N$&$ = \mathbf D^{-\frac 12}\mathbf A\mathbf D^{-\frac 12}$ & \texttt{sym-n} & Norm.\ adjacency m. &
    Sym 		& $-1+?, \ldots, 1-?, 1-?$  \\
    \ref{sec:matrix.P} & $\mathbf P$&$ = \mathbf D^{-1}\mathbf A$ & \texttt{stoch1} & Stoch.\ adjacency m. &
    Right-stoch\textsuperscript{1} 	& $\lambda[\mathbf N]$ \\

    \midrule

    \ref{sec:matrix.L} & $\mathbf L$&$ = \mathbf D - \mathbf A$ & \texttt{lap} & Laplacian &
    Sym 		& $0\leq \xi$ \texttt{conflict}, $a$ \texttt{alcon}, $\ldots, ?$ \\ 
    \ref{sec:matrix.Z} & $\mathbf Z$&$ = \mathbf I - \mathbf D^{-\frac 12}\mathbf A\mathbf D^{-\frac 12}$ & \texttt{*} $\mathbf N$ & Norm.\ Laplacian &
    Sym 		& $1-\lambda[\mathbf N]$ \\
    \ref{sec:matrix.S} & $\mathbf S$&$ = \mathbf I - \mathbf D^{-1}\mathbf A$ & \texttt{*} $\mathbf P$ & Stoch.\ Laplacian &
    Row sum = 0\textsuperscript{1}    	& $1-\lambda[\mathbf N]$ \\

    \midrule

    \ref{sec:matrix.K} & $\mathbf K$&$ = \mathbf D + \mathbf A$ & \texttt{lapq} & Signless Laplacian &
    Sym      		& $0\leq \chi$ \texttt{nonbipal}$, ?, \ldots, ?$ \\
    \ref{sec:matrix.C} & $\mathbf C$&$ = \mathbf I + \mathbf D^{-\frac 12}\mathbf A\mathbf D^{-\frac 12}$ & \texttt{*} $\mathbf N$ & Norm.\ signless Laplacian &
    Sym 		& $1+\lambda[\mathbf N]$ \\
    \ref{sec:matrix.J} & $\mathbf J$&$ = \mathbf I + \mathbf D^{-1}\mathbf A$ & \texttt{*} $\mathbf P$ & Stoch.\ signless Laplacian &
    Row sum = 2\textsuperscript{1} & $1+\lambda[\mathbf N]$ \\
    
    \midrule \midrule

    \ref{sec:matrix.repulsive} & $\mathbf{\hat A}$ &$=\mathbf A$& \texttt{*} $\mathbf A$ & Repulsive adjacency m. &
    Sym 		& $\lambda[\mathbf A]$ \\
    \ref{sec:matrix.repulsive} & $\mathbf{\hat N}$&$= \mathbf N$ & \texttt{*} $\mathbf N$ & Repulsive norm.\ adjacency m. &
    Sym 		& $\lambda[\mathbf N]$  \\
    \ref{sec:matrix.repulsive} & $\mathbf{\hat P}$&$ = \mathbf P$ & \texttt{*} $\mathbf P$ & Repulsive stoch.\ adjacency m. &
    Right-stoch\textsuperscript{1} 	& $\lambda[\mathbf N]$ \\

    \midrule

    \ref{sec:matrix.repulsive} & $\mathbf{\hat L}$&$=\mathbf{\hat D}-\mathbf A$ &
    \texttt{?} & Repulsive Laplacian & Sym & ? \\
    \ref{sec:matrix.repulsive} & $\mathbf{\hat Z}$&$=\mathbf D^{-\frac 12}(\mathbf{\hat D}-\mathbf A)\mathbf D^{-\frac 12}$ &
    \texttt{?} & Repulsive norm.\ Laplacian & Sym & ? \\
    \ref{sec:matrix.repulsive} & $\mathbf{\hat S}$&$=\mathbf D^{-1}(\mathbf{\hat D}-\mathbf A)$ &
    \texttt{*} $\mathbf{\hat Z}$ & Repulsive stoch.\ Laplacian & & $\lambda[\mathbf{\hat Z}]$ \\
    
    \midrule
    \ref{sec:matrix.repulsive} & $\mathbf{\hat K}$&$=\mathbf{\hat D}+\mathbf A$ &
    \texttt{?} & Repulsive signless Laplacian & Sym & ? \\
    \ref{sec:matrix.repulsive} & $\mathbf{\hat C}$&$=\mathbf D^{-\frac 12}(\mathbf{\hat D}+\mathbf A)\mathbf D^{-\frac 12}$ &
    \texttt{?} & Repulsive norm.\ signless Laplacian & Sym & ? \\
    \ref{sec:matrix.repulsive} & $\mathbf{\hat J}$&$=\mathbf D^{-1}(\mathbf{\hat D}+\mathbf A)$ &
    \texttt{*} $\mathbf{\hat C}$ & Repulsive stoch.\ signless Laplacian & & $\lambda[\mathbf{\hat C}]$ \\

    \midrule \midrule

    \ref{sec:matrix.A} & $\mathbf{\bar A}$&$=\vec{\mathbf A}+\vec{\mathbf A}^{\mathrm T}$ & \texttt{sym} & Symmetric adjacency m. &
    Sym 		& $\ldots, \alpha$ \texttt{snorm} \\
    \ref{sec:matrix.N} & $\mathbf{\bar N}$&$=\mathbf D^{-\frac 12}(\vec{\mathbf A}+\vec{\mathbf A}^{\mathrm T})\mathbf D^{-\frac 12}$ & \texttt{sym-n} & Symmetric norm.\ adjacency m. &
    Sym 		& $-1+?, \ldots, 1-?, 1-?$  \\
    \ref{sec:matrix.P} & $\mathbf{\bar P}$&$=\mathbf D^{-1}(\vec{\mathbf A}+\vec{\mathbf A}^{\mathrm T})$ & \texttt{stoch1} & Symmetric stoch.\ adjacency m. &
    Right-stoch\textsuperscript{1} 	& $\lambda[\mathbf{\bar N}]$ \\

    \midrule

    \ref{sec:matrix.L} & $\mathbf{\bar L}$&$ = \mathbf D - (\vec{\mathbf A}+\vec{\mathbf A}^{\mathrm T})$ & \texttt{lap} & Symmetric Laplacian &
    Sym 		& $0\leq \xi$ \texttt{conflict}, $a$ \texttt{alcon}, $\ldots, ?$ \\ 
    \ref{sec:matrix.Z} & $\mathbf{\bar Z}$&$ = \mathbf I - \mathbf D^{-\frac 12}(\vec{\mathbf A}+\vec{\mathbf A}^{\mathrm T})\mathbf D^{-\frac 12}$ & \texttt{*} $\mathbf{\bar N}$ & Symmetric norm.\ Laplacian &
    Sym 		& $1-\lambda[\mathbf{\bar N}]$ \\
    \ref{sec:matrix.S} & $\mathbf{\bar S}$&$ = \mathbf I - \mathbf D^{-1}(\vec{\mathbf A}+\vec{\mathbf A}^{\mathrm T})$ & \texttt{*} $\mathbf{\bar P}$ & Symmetric stoch.\ Laplacian &
    Row sum = 0\textsuperscript{1}    	& $1-\lambda[\mathbf{\bar N}]$ \\

    \midrule

    \ref{sec:matrix.K} & $\mathbf{\bar K}$&$ = \mathbf D + \vec{\mathbf A}+\vec{\mathbf A}^{\mathrm T}$ & \texttt{lapq} & Symmetric signless Laplacian &
    Sym      		& $0\leq \chi$ \texttt{nonbipal}$, ?, \ldots, ?$ \\
    \ref{sec:matrix.C} & $\mathbf{\bar C}$&$ = \mathbf I + \mathbf D^{-\frac 12}(\vec{\mathbf A}+\vec{\mathbf A}^{\mathrm T})\mathbf D^{-\frac 12}$ & \texttt{*} $\mathbf{\bar N}$ & Symmetric norm.\ signless Laplacian &
    Sym 		& $1+\lambda[\mathbf{\bar N}]$ \\
    \ref{sec:matrix.J} & $\mathbf{\bar J}$&$ = \mathbf I + \mathbf D^{-1}(\vec{\mathbf A}+\vec{\mathbf A}^{\mathrm T})$ & \texttt{*} $\mathbf{\bar P}$ & Symmetric stoch.\ signless Laplacian &
    Row sum = 2\textsuperscript{1} & $1+\lambda[\mathbf{\bar N}]$ \\

    \midrule \midrule
    
    \ref{sec:matrix.skew} & $\mathbf{\acute A}$&$=i\mathbf Y=i(\vec{\mathbf A}-\vec{\mathbf A}^{\mathrm T})$& \texttt{skewi} & Skew-Hermitian Adjacency m. &
    Skew-Herm. 		& ? \\
    \ref{sec:matrix.skewn} & $\mathbf{\acute N}$&$=i\mathbf D^{-\frac 12}(\vec{\mathbf A}-\vec{\mathbf A}^{\mathrm T})\mathbf D^{-\frac 12}$ & \texttt{skewin} & Skew-Hermitian norm.\ adjacency m. &
    Skew-Herm., ? 		& ?  \\
    ? & $\mathbf{\acute P}$&$ = i\mathbf D^{-1}(\vec{\mathbf A}-\vec{\mathbf A}^{\mathrm T})$ & \texttt{?} & Skew-Hermitian stoch.\ adjacency m. &
    ? 	& $\lambda[\mathbf{\acute N}]$ \\

    \midrule \midrule
    
    \ref{sec:matrix.diag}, \ref{sec:matrix.svd} & $\vec{\mathbf A}$ && \texttt{diag;svd} & Unidirectional adjacency m. &
    &  $\ldots, \pi$ \texttt{maxdiag}; ? \\
    \ref{sec:decomposition.svd-n}, \ref{sec:matrix.diag-n} & $\vec{\mathbf N}$&$=\mathbf D_1^{-\frac 12}\vec{\mathbf A}\mathbf D_2^{-\frac 12}$ & \texttt{diag-n;svd-n} & Unidirectional norm.\ adjacency m. &
    & ?; ?  \\
    ?;? & $\vec{\mathbf P}$&$ = \mathbf D_1^{-1}\vec{\mathbf A}$ & \texttt{stoch1;?} & Unidirectional stoch.\ adjacency m. &
    Right-stoch\textsuperscript{1} 	& ? \\

    \midrule

    \ref{sec:decomposition.lapdiag2};? & $\vec{\mathbf L}$&$ = \mathbf D_1 - \vec{\mathbf A}$ & \texttt{lapdiag2;?} & Unidirectional Laplacian &
    & ? \\ 

    \bottomrule
  \end{tabular}
    }
  }
  \texttt{*} These decompositions can be derived by a transformation of eigenvalues of the given decomposition, and thus no internal name is used, and the decompositions are not explicitly computed in KONECT.  \\ 
  \textsuperscript{1} Only valid for unsigned graphs \\
  When two decomposition names are given, the matrix is non-normal and
  the first represents eigenvectors and -values and the second is the
  singular value decomposition.
\end{table}

\subsection{Decompositions of Undirected Graphs}
This section covers the case of unipartite unsigned undirected graphs.
Weighted graphs and graphs with multiple edges are included, as their
treatment is usually simple. 
We define nine basic matrices in this section, which can be thought of
spanning the two dimensions of matrix type (adjacency/Laplacian/signless
Laplacian) and normalization type (none, symmetric normalization,
stochastic).  These matrices have the names $\mathbf A$, $\mathbf L$,
$\mathbf K$, $\mathbf N$, $\mathbf Z$, $\mathbf C$, $\mathbf P$,
$\mathbf S$ and $\mathbf J$.  
An overview is given in Table~\ref{tab:basic-nine}
These matrices serve as the ``basic nine'' matrices
in KONECT, and derivatives with similar names are used for other types
of graphs.  Of these, $\mathbf A$, $\mathbf L$, $\mathbf N$ and $\mathbf
P$ are
ubiquitous in network analysis; $\mathbf K$ is
used from time to time; $\mathbf Z$, $\mathbf C$, $\mathbf S$,
and $\mathbf J$ can be derived by shifts of eigenvalues from $\mathbf N$
and $\mathbf P$, and thus don't usually need to be considered
separately.  $\mathbf N$ and $\mathbf P$ share eigenvalues, but not
eigenvectors.  Only the names $\mathbf A$ and $\mathbf L$ can be
expected to be understood without explanation -- the other names are
specific to KONECT, although the notation $\mathbf N$, $\mathbf K$ and
$\mathbf P$ are used in the literature from time to time. 

\begin{table}
  \caption{
    \label{tab:basic-nine}
    The basic nine matrices of KONECT.
  }
  \centering
  \begin{tabular}{ l @{\qquad\qquad} c @{\qquad\quad} c c }
    \toprule
    & \textbf{Plain} & \textbf{Normalized} & \textbf{Stochastic} \\
    \midrule
    Adjacency matrix & $\mathbf A$ & $\mathbf N$ & $\mathbf P$ \\
    Laplacian        & $\mathbf L$ & $\mathbf Z$ & $\mathbf S$ \\
    Signless Laplacian & $\mathbf K$ & $\mathbf C$ & $\mathbf J$ \\
    \bottomrule 
  \end{tabular}
\end{table}

In KONECT, these decompositions can be applied to directed graphs as well, in
which case edge directions are ignored.  A directed graph that contains
reciprocal edges (the majority) will thus result in an undirected graph
with multiple parallel edges, even if the original directed graph did
not contain parallel directed edges.  As a result, one must be careful:
Expressions involving the number of edges $m$ in this chapter refer
to the number of unique edges, which gives a different result for
directed and undirected graphs.  In the same way, the matrix $\mathbf A$
in this section refers to the symmetric adjacency matrix; in the case of
directed graphs, this thus equals the matrix normally written as
$\mathbf A + \mathbf A^{\mathrm T}$. 

\subsubsection{Symmetric Adjacency Matrix ($\mathbf A$)}
\label{sec:matrix.A}
The symmetric adjacency matrix $\mathbf A$ is the most basic graph
characteristic matrix.  It is a symmetric $n \times n$ matrix defined as
$\mathbf A_{uv}=1$ when the nodes $u$ and $v$ are connected, and
$\mathbf A_{uv}=0$ when $u$ and $v$ are not connected. 

The eigenvalue decomposition of the matrix $\mathbf A$ for undirected
graphs is widely used to analyse graphs: 
\begin{align}
  \mathnote{\texttt{sym}}
  \mathbf A &= \mathbf U \mathbf \Lambda \mathbf U^{\mathrm T}
\end{align}
$\mathbf \Lambda$ is an $n \times n$ real diagonal matrix containing the
eigenvalues of $\mathbf A$, i.e., $\mathbf \Lambda_{ii} =
\lambda_i[\mathbf A]$.  
$\mathbf U$ is an $n \times n$ orthogonal matrix having the
corresponding eigenvectors as columns. 

The eigenvalues of $\mathbf A$ are widely used to characterize graphs,
and often, the expression \emph{eigenvalue of a graph} is used without
further qualification to refer to the eigenvalues of $\mathbf A$.  An
example is given in \citep{b911}.

The largest absolute eigenvalue of $\mathbf A$ is the networks spectral
norm $\alpha$, i.e.,
\begin{align*}
  \alpha = \max_i |\mathbf \Lambda_{ii}| = \left\| \mathbf A \right\|_2.
\end{align*}

The sum of all eigenvalues $\lambda_i$ equal the trace of $\mathbf A$,
i.e., the sum of its diagonal elements.  The sum of the eigenvalues of
$\mathbf A$ thus equals the number of loops in the graphs. 
In particular, when a graph
has no loops, then the sum of the eigenvalues of its adjacency matrix is
zero.  

Higher moments the eigenvalues of $\mathbf A$ give the number of tours
in the graph.  Remember that a tour of length $k$ is defined as a
sequence of $k$ connected nodes, such that the first and the last node
are connected, such that two tours are considered as distinct when they
have a different starting node or orientation. 
The sum of $k$\textsuperscript{th} powers of the
eigenvalues of $\mathbf A$ then equals the number of $k$-tours $T_k$. 
We thus have in a loopless graph, that the traces of powers of $\mathbf
A$ are related to the number of edges $m$, the number of triangles $t$,
the number of squares $q$ and the number of wedges $s$ by:
\begin{align*}
  \mathrm{Tr}(\mathbf A) &= 0 \\
  \mathrm{Tr}(\mathbf A^2) &= 2m \\
  \mathrm{Tr}(\mathbf A^3) &= 6t \\
  \mathrm{Tr}(\mathbf A^4) &= 8q + 4s + 2m
\end{align*}
The traces of $\mathbf A$ can also be expressed as sums of powers
(moments) of the eigenvalues of $\mathbf A$:
\begin{align*}
  \mathrm{Tr}(\mathbf A^k) &= \sum_{i=1}^n \lambda_i^k
\end{align*}

The spectrum of $\mathbf A$ can also be characterized in terms of graph
bipartivity.  When the graph is bipartite, then all eigenvalues come in
pairs $\{\pm\lambda\}$, i.e., they are distributed around zero
symmetrically.  When the graph is not bipartite, then their distribution
is not symmetric.  It follows that when the graph is bipartite, the
smallest and largest eigenvalues have the same absolute value. 

The sum of the absolute values of $\mathbf A$ is called the graph
energy \citep{b910}. 

\subsubsection{Normalized Adjacency Matrix ($\mathbf N$)}
\label{sec:matrix.N}
The normalized adjacency matrix $\mathbf N$ of an undirected graph is
defined as 
\begin{align}
  \mathnote{\texttt{sym-n}}
  \mathbf N &= \mathbf D^{-\frac 1 2} \mathbf A \mathbf D^{-\frac 1 2},
\end{align}
where we remind the reader that the diagonal matrix $\mathbf D$
contains the node degrees, i.e., $\mathbf D_{uu} = d(u)$. 
The matrix $\mathbf N$ is symmetric and its eigenvalue decomposition can
be considered:
\begin{align}
  \mathbf N = \mathbf U \mathbf \Lambda \mathbf U^{\mathrm T}
\end{align}
The eigenvalues $\lambda_i$ of $\mathbf N$ can be used to characterize
the graph, in analogy with those of the nonnormalized adjacency
matrix.
The spectrum of $\mathbf N$ is also called the weighted spectral
distribution \citep{b864}.
All eigenvalues of $\mathbf N$ are contained in the range
$[-1,+1]$.  When the graph is unsigned, the largest eigenvalue is one.
In addition, the eigenvalue one has multiplicity one if the graph is
connected and unsigned.  It follows that for general unsigned graphs,
the multiplicity of the eigenvalue one equals the number of connected
components of the graph. 

Minus one is the smallest eigenvalue of $\mathbf N$ if and only iff the graph is
bipartite.  As with the nonnormalized adjacency matrix, the eigenvalues
of $\mathbf N$ are distributed symmetrically around zero if and only if
the graph is bipartite. 

When the graph is connected, 
the eigenvector corresponding to eigenvalue one has entries proportional
to the square root of node degrees, i.e.,
\begin{align}
  \mathbf U_{u1} &= \sqrt{\frac{d(u)}{2m}}.
\end{align}
Note that this equivalence only holds for undirected graphs.  For
directed graphs, there is no such equivalence. 

\subsubsection{Stochastic Adjacency Matrix ($\mathbf P$)}
\label{sec:matrix.P}
The consideration of random walks on a graph leads to the definition of
the stochastic adjacency matrix $\mathbf P$.  Imagine a random walker on
the nodes of a graph, who can walk from node to node by following
edges.  If, at each edge, the probability that the random walker will go
to each neighboring node with equal probability, then the random walk
can be described be the transition probability matrix defined as
\begin{align}
  \mathnote{\texttt{stoch1}}
  \mathbf P &= \mathbf D^{-1} \mathbf A.
\end{align}
This matrix is called the stochastic adjacency matrix.  It is asymmetric,
even when the graph is undirected, except when the graph is regular,
i.e., when all degrees are the same.  
Thus, its eigenvalue decomposition
is not always defined, and in any case may not involve orthogonal
matrices. 

The sum of rows of $\mathbf P$ sum to one, making this matrix
right-stochastic. 
For directed graphs we may distinguish the right-stochastic (or
row-stochastic) matrix 
$\mathbf P = \mathbf D^{-1} \mathbf A$ and the left-stochastic (or column-stochastic)
matrix $\mathbf A \mathbf D^{-1}$.  Note the subtle terminology here:
$\mathbf D^{-1} \mathbf A$ is left-normalized but right-stochastic. 

The matrix $\mathbf P$ is related to the normalized adjacency matrix $\mathbf N$ by
\begin{align}
  \mathbf P &= \mathbf D^{-\frac 1 2} \mathbf N \mathbf D^{\frac 1 2}
\end{align}
and therefore both matrices have the same set of eigenvalues.
Thus, the eigenvalues of $\mathbf P$ are all real, even though
$\mathbf P$ is asymmetric, and they are contained in the range $[-1,+1]$. 
Also, the relationship between $\mathbf P$ and $\mathbf N$ implies that that
eigenvectors of $\mathbf P$ are related to those of 
$\mathbf N$ by factors of the diagonal elements of $\mathbf D^{\frac 1 2}$,
i.e., the square roots of node degrees. 
Since $\mathbf P$ is asymmetric, its left eigenvectors differ from its
right eigenvectors.  When the graph is undirected, the left eigenvector
corresponding to the eigenvalue one has entries proportional to the
degree of nodes, 
while the right eigenvector corresponding to the eigenvalue one 
is the constant vector.  This is consistent with the fact that for a
random walk on an undirected graph, the stationary distribution of nodes
is proportional to the node degrees. 

The alternative matrix $\mathbf A \mathbf D^{-1}$ can also be
considered. \marginpar{\texttt{stoch2}} It is left-stochastic, and
can be derived by considering random walks that tranverse edges in a
backward direction. 

The matrix $\mathbf P$ is the state transition matrix of a random walk
on the graph, and thus its largest eigenvector is one if the graph is
(strongly) connected.  
The matrix $\mathbf P$ is also related to the PageRank matrix $\mathbf
G$ (``Google matrix''), which
equals 
\begin{align}
  \mathbf G &= (1-\alpha) \mathbf P + \alpha\mathbf 1
\end{align}
where $0 < \alpha < 1$ is a damping factor (the teleportation
probability), and $\mathbf 1$ is the matrix containing all ones.  The
left 
eigenvalues of the PageRank matrix give the PageRank values, and thus
we see that (ignoring the teleportation term), the PageRank of nodes in
an undirected network equals the degrees of the nodes. 

The matrix $\mathbf P$ is also related to random walks with restarts on
the graph, i.e., random walks that have a certain probability $0 <
\alpha < 1$
to return to an initial node at each step, instead of taking an edge at
random.  For any two nodes $u$ and $v$, the number
\begin{align}
  \left[\alpha  (\mathbf I - (1-\alpha) \mathbf P^{\mathrm T})^{-1}\right]_{uv}
\end{align}
gives the asymptotic probability that a random walk with restart starting at node $u$
finds itself at node $v$.  

The matrix $\mathbf P$ is further related to the mean first-passage time
(MFPT) on the network \citep{b880}.

\subsubsection{Laplacian Matrix ($\mathbf L$)}
\label{sec:matrix.L}
The Laplacian matrix of an undirected graph is defined as
\begin{align}
  \mathnote{\texttt{lap}}
  \mathbf L &= \mathbf D - \mathbf A,
\end{align}
i.e., the diagonal degree matrix from which we subtract the adjacency
matrix.  The Laplacian matrix is also called the Kirchhoff matrix. 
The Laplacian matrix is the discrete analogue of the Laplace operator
ubiquitous in physics and other areas employing multivariate calculus.
This operator is also called the Laplacian, and denoted $\Delta$ or
$\nabla^2$.  In that context, the Laplacian matrix is also called the
discrete Laplace operator. 
Due to this equivalence, the Laplacian matrix $\mathbf L$
has many uses in network analysis. 
The Laplacian matrix is also denoted with other letters, such as for
instance $\mathbf Q$ as used by \cite{b902}.  It may also be called the
admittance matrix.  

We consider the eigenvalue decomposition of the Laplacian:
\begin{align}
  \mathbf L &= \mathbf U \mathbf \Lambda \mathbf U^{\mathrm T}
\end{align}
The Laplacian matrix of positive-semidefinite, i.e., all eigenvalues are
nonnegative.  
When the graph is unsigned, the smallest eigenvalue is zero and its
multiplicity equals the number of connected components in the graph. 

The second-smallest eigenvalue is called the algebraic connectivity of
the graph, and is denoted $a = \lambda_2[\mathbf L]$ \citep{b652}.  If the graph is
unconnected, that value is zero, i.e., an unconnected graph has an
algebraic connectivity of zero. 

When the graph is connected, the eigenvector corresponding to eigenvalue
zero is a constant vector, i.e., a vector with all entries equal. The
eigenvector corresponding the the second-smallest eigenvalue is called
the Fiedler vector, and can be used to cluster nodes in the
graph. Together with further eigenvectors, it can be used to draw
graphs \citep[see][]{kunegis:signed-kernels}. 

The Laplacian matrix $\mathbf L$ can be expressed in terms of the signed
incidence matrix $\mathbf E^{\pm}$:
\begin{align}
  \mathbf L &= \mathbf E^{\pm} [\mathbf E^{\pm}]^{\mathrm T}
\end{align}
From this equality follows that $\mathbf L$ is positive-semidefinite.   
Note that while the matrix $\mathbf E^{\pm}$ is called the signed
incidence matrix, it is not related to signed graphs.  In fact, the
equality only holds for unsigned, unweighted graphs. 

The Laplacian matrix can be used to compute the \emph{effective
  resistance} in an electrical network \citep{b101}.  If a given graph is interpreted
as an electrical network in which each edge is a resistor, then the
graph as a whole acts as a resistor between any two nodes $u$ and $v$.
The value of the resistance between two nodes $u$ and $v$ can be
expressed using the Moore--Penrose inverse $\mathbf \Gamma = \mathbf
L^+$ of the Laplacian $\mathbf L$:
\begin{align}
  r(u,v) &= \mathbf \Gamma_{uu} + \mathbf \Gamma_{vv} 
  - \mathbf \Gamma_{uv} - \mathbf \Gamma_{vu} \\
  &= (x^u - x^v)^{\mathrm T} \mathbf \Gamma (x^u - x^v)
\end{align}
Here, $x^u$ is a vector of size $|V|$ with $(x^u)_u=1$ and $(x^u)_v=0$
whenever $u \neq v$.  The Moore--Penrose pseudoinverse can be expressed
using the eigenvalue decomposition $\mathbf L = \mathbf U \mathbf
\Lambda \mathbf U^{\mathrm T}$ as
\begin{align}
  \mathbf \Gamma &= \mathbf L^+ = \mathbf U \mathbf \Lambda^+ \mathbf U^{\mathrm T},
\end{align}
in which the Moore--Penrose pseudodoinverse of the diagonal matrix
$\mathbf \Lambda$ is also diagonal and given by $(\mathbf
\Lambda^+)_{ii}=(\mathbf \Lambda_{ii})^{-1}$ when $\mathbf \Lambda_{ii} \neq
0$, and $(\mathbf L^+)_{ii}=0$ otherwise. 

\subsubsection{Normalized Laplacian Matrix ($\mathbf Z$)}
\label{sec:matrix.Z}
The Laplacian matrix too, can be normalized.  It turns out that the
normalized Laplacian and the normalized adjacency matrix
are tighly related
to each other:  They share the same set of eigenvectors, and their
eigenvalues are reflections of each other.  

The normalized Laplacian matrix of an undirected graph is defined as 
\begin{align}
  \mathbf Z &= \mathbf D^{-\frac 1 2} \mathbf L \mathbf D^{-\frac 1 2}.
\end{align}
As opposed to $\mathbf A$, $\mathbf L$ and $\mathbf N$, there is no
standardized notation of the normalized Laplacian.  The notation
$\mathbf Z$ is specific to KONECT, and was chosen as the letter Z
resembles a turned letter N, and the matrices represented by those
letters share eigenvectors and have flipped eigenvalues. 
The matrix $\mathbf Z$ is also called the \emph{random walk Laplacian}. 

The normalized Laplacian is related to the normalized adjacency matrix
by 
\begin{align}
  \mathbf Z &= \mathbf I - \mathbf N = \mathbf I - \mathbf D^{-\frac 1 2}
  \mathbf A \mathbf D^{-\frac 1 2}, 
\end{align}
as can be derived directly from their definitions. 
It follows that $\mathbf Z$ and $\mathbf N$ have the same set of
eigenvectors, and that their eigenvalues are related by the
transformation $1 - \lambda$. Thus, the properties of $\mathbf Z$ can be
derived from those of $\mathbf N$. For instance, all eigenvalues of
$\mathbf Z$ are contained in the range $[0, 2]$, and the multiplicity of
the eigenvalue zero equals the number of connected components (when the
graph is unsigned). If the undirected graph is connected, the eigenvector of
eigenvalue zero contains entries proportional to the square root of the
node degrees. 

In KONECT, the decomposition of the normalized Laplacian is not
included, since it can be derived from that of the normalized adjacency
matrix. 

\subsubsection{Stochastic Laplacian Matrix ($\mathbf S$)}
\label{sec:matrix.S}
A further variant of the Laplacian exists, based on the stochastic
adjacency matrix:
\begin{align}
  \mathbf S &= \mathbf I - \mathbf P = \mathbf I - \mathbf D^{-1}
  \mathbf A = \mathbf I - \mathbf D^{-\frac 1 2} \mathbf N \mathbf D^{\frac 1 2} =
  \mathbf D^{-\frac 1 2} \mathbf Z \mathbf D^{\frac 1 2}
  = \mathbf D^{-1} \mathbf L
\end{align}
This matrix shares much properties with $\mathbf P$ and thus with
$\mathbf N$ and $\mathbf Z$.  The eigenvalues of $\mathbf S$ are
contained in the interval $[0, 2]$.  The eigenvalue zero has a
multiplicity equal to the number of connected components of the graph,
and when the graph is connected its corresponding right eigenvector is the
constant vector, while its corresponding left eigenvector is
proportional to the node degrees.  For connected graphs, the largest
eigenvalue of $\mathbf S$ is two if and only if the graph is bipartite.
In the general case, the eigenvalue two has a multiplicity equal to the
number of connected components that are bipartite. 

\subsubsection{Signless Laplacian ($\mathbf K$)}
\label{sec:signless-laplacian}
\label{sec:matrix.K}
\label{sec:matrix.C}
\label{sec:matrix.J}
The signless Laplacian of a graph is defined as the Laplacian of the
corresponding graph in which all edges are interpreted as negative \citep{b900}.  It
thus equals
\begin{align}
  \mathnote{\texttt{lapq}}
  \mathbf K &= \mathbf D + \mathbf A.
\end{align}
The signless Laplacian is also denoted $\mathbf Q$. 
It corresponds to the ordinary Laplacian
$\mathbf L$ of the graph with inverted edge weights, i.e., $\mathbf K[G] =
\mathbf L[-G]$. 

This matrix is positive-semidefinite, and its smallest eigenvalue is
zero if and only if at least one connected component of the graph is bipartite.  Thus, $\mathbf K$ is used
in measures of bipartivity \citep[see e.g.][]{kunegis:bipartivity}.   We
call this smallest eigenvalue the algebraic non-bipartivity $\chi =
\lambda_{\min}[\mathbf K]$. 

In unsigned graphs, the multiplicity of the eigenvalue zero equals the
number of connected components that are bipartite.  
In signed graphs, the multiplicity of the eigenvalue zero equals the
number of connected components that are balanced after all their edges
have been negated (i.e., switch between positive and negative). 

In connected graphs (that are not necessarily bipartite), the smallest
eigenvalue of $\mathbf K$ is a measure of the non-bipartivity of the
graph. 

The signless Laplacian exists in the following two normalized variants:
\begin{align}
  \mathbf C &= \mathbf I + \mathbf D^{-\frac 12}\mathbf A\mathbf D^{-\frac 12} \\
  \mathbf J &= \mathbf I + \mathbf D^{-1} \mathbf A
\end{align}

\subsubsection{Seidel Adjacency Matrix ($\mathbf W$)}
The Seidel adjacency matrix $\mathbf W$ is a symmetric $n \times n$ matrix given by 
\begin{align}
  \mathnote{\texttt{seidel}}
  \mathbf W_{uv} &= 
  \left\{ \begin{array}{ll}
    0  & \text{when $u=v$} \\
    -1 & \text{when $u \neq v$ and $u \leftrightarrow v$} \\
    +1 & \text{when $u \neq v$ and $u \not\leftrightarrow v$} \\
  \end{array} \right.
\end{align}
The notation $\mathbf W$ is not standardized. 
This type of matrix originates in \citep{b906}; see \citep{b907} for
another reference.  This matrix is also called the $(-1,1,0)$-adjacency
matrix.  
In that context, the ordinary adjacency matrix $\mathbf A$ may be called
the $(0,1)$-adjacency matrix. 
The matrix is also written as $\mathbf A^*$ \citep[e.g.\ in][]{b908} and
$\mathbf S$ \citep[e.g.\ in][]{b909}. 
Symmetric matrices such as $\mathbf W$ which have a zero diagonal and $\pm 1$ off
the diagonal are also studied outside of graph theory, see
e.g.\ \cite{b912}. 

The matrix can also be expressed as
\begin{align}
  \mathbf W &= \mathbf 1 - \mathbf I - 2 \mathbf A,
\end{align}
where $\mathbf 1$ is the matrix completely filled with ones, and
$\mathbf A$ is the symmetric adjacency matrix of the graph.

The matrix is not sparse, but due to its simple definition, its
eigenvalue decomposition can be computed using sparse matrix methods. 
In the general case, the matrix has both positive and negative
eigenvalues. 
The largest absolute eigenvalue of $\mathbf W$ is a network statistic.
The sum of the absolute values of all eigenvalues of $\mathbf W$ has
been called the Seidel energy of the graph \citep{b909}, in analogy with
the sum of absolute eigenvalues of $\mathbf A$, called the energy. 

The Seidel adjacency matrix has a particular significance for regular
graphs \citep{b908}.  In particular, the number of eigenvalues of
$\mathbf A$ and $\mathbf W$ whose eigenvector is not orthogonal to the
all-ones vector has been considered.  This number is one for regular
graphs, and larger otherwise. 

Relations between the spectrum of $\mathbf W$ and the spectrum of
$\mathbf A$ are given by \cite{b908}.

\subsection{Decompositions of Signed Graphs}
In this subsection, we review matrices and decompositions that apply to
signed graphs.  

In KONECT, in cases where the matrices are generalizations of a matrix and
decomposition for unsigned graphs, they share the internal name of the
decomposition with the unsigned case.  On matrix at least is however
specific to signed graphs. 

\subsubsection{Signed Adjacency Matrix}
In undirected signed graphs, the adjacency matrix contains the values
$\pm 1$.  In general, it can be used in the same fashion as the
adjacency matrix of an unsigned graph, with the property that when edge
weights get multiplied with each other, the multipliciation follows the
multiplication rules for signed multiplication, i.e., 
\begin{align*}
  +1 \times +1 &= +1 \\
  +1 \times -1 &= -1 \\
  -1 \times +1 &= -1 \\
  -1 \times -1 &= +1 
\end{align*}
As an example of this, powers of the adjacency matrix of a signed graph
will not contain counts of paths, but counts of paths weighted by the
sign of each path, i.e., denoting whether each path contains an even or
odd number of edges.  As another example, the trace of the cube of
$\mathbf A$ will equal six times the number of balanced triangles minus
the number of unbalanced triangles. 

\subsubsection{Signed Laplacian Matrix}
\label{sec:decomposition:lap:signed}
In a signed graph, the Laplacian matrix is defined as in an unsigned
graph as
\begin{align*}
  \mathbf L &= \mathbf D - \mathbf A.
\end{align*}
Note that the values in $\mathbf A$ are signed, but those in $\mathbf D$
are not.  Thus, the properties of $\mathbf L$ differ from the unsigned
case.  In particular, it is no longer true that the smallest eigenvalue
of $\mathbf L$ is always zero.  While $\mathbf L$ is still
positive-semidefinite in the signed case, 
the smallest eigenvalue of $\mathbf L$ may be larger than zero,
rendering $\mathbf L$ positive-definite. 
The smallest eigenvalue of $\mathbf L$ is signed graphs is called 
the algebraic conflict $\xi$.  In connected signed graphs, it is zero if and only if the graph is balanced,
i.e., when the nodes can be divided into two groups such that all
positive edges connect nodes within the same group, and all negative
edges connect nodes of different groups.  Equivalently, $\xi$ is larger
than zero if and only if each connected component contains at least one
cycle with an odd number of negative edges. 
In signed graphs consisting of multiple connected components, the
smallest eigenvalue of $\mathbf L$ is zero if and only if at least one
connected component is balanced.  

If a graph $G$ is balanced, then the eigenvalue decomposition of
$\mathbf L[G]$ can be derived from that of underlying unsigned graph 
\citep{kunegis:signed-kernels}.
If $G$ is a balanced signed graph, and $|G|$ is underlying unsigned
graph, then the eigenvalue decompositions of both graphs' Laplacian
matrices can be considered:
\begin{align*}
  \mathbf L[G] &= \mathbf U \mathbf \Lambda \mathbf U^{\mathrm T} \\
  \mathbf L[|G|] &= \mathbf V \mathbf \Lambda \mathbf V^{\mathrm T} 
\end{align*}
In that case, both Laplacian matrices have the same eigenvalues, and
$\mathbf U$ and $\mathbf V$ differ only by negation of rows
corresponding to nodes in one of the two clusters. 

In signed graphs, the equivalent resistance can be defined by
defining that an edge carrying a negative resistance value acts like the
corresponding positive resistance in series with a component that
negates potentials \citep{kunegis:netflix-srd}.  The result 
gives a \emph{signed resistance distance kernel} based on the signed
Laplacian:
\begin{align}
  r(u,v) &= \mathbf \Gamma_{uu} + \mathbf \Gamma_{vv} 
  - \mathbf \Gamma_{uv} - \mathbf \Gamma_{vu} 
\end{align}
As in the unsigned case, the matrix $\mathbf \Gamma=\mathbf L^+$ is the
Moore-Penrose pseudoinverse of $\mathbf L$.  See also Section~5.8.1 in
the author's Phd thesis \citep{kunegis:phd}.  Note that if all connected
components in the graph are unbalanced, then the Moore--Penrose
pseudoinverse reduces to the ordinary matrix inverse, i.e., $\mathbf
L^+=\mathbf L^{-1}$, due to the fact that $\mathbf L$ does not have
eigenvalues of zero for such graphs. 

\subsubsection{Repulsive Matrices}
\label{sec:matrix.repulsive}
The Laplacian matrix $\mathbf L = \mathbf D - \mathbf A$ as defined
earlier is based on the the degree matrix $\mathbf D$, whose diagonal
elements equal the sum of absolute weights of all edges incident to each
node.  
Taking the absolute value in that definition is justified for
many applications related to signed graphs, and results in a positive-semidefinite matrix, whose
smallest eigenvalue is zero if and only if there is at least one
balanced component in the signed network.  Alternatively, using the sum
of edge weights instead results in a different matrix, which may have
negative eigenvalues:
\begin{align}
  \mathbf{\hat L} &= \mathbf{\hat D} - \mathbf A,
\end{align}
where the diagonal matrix $\mathbf{\hat D}$ is defined as
\begin{align*}
  [\mathbf{\hat D}]_{uu} &= \sum_{v \sim u} \mathbf A_{uv}.
\end{align*}
$\mathbf{\hat L}$ has also been called simply the \emph{signed Laplacian
  matrix}, but should not be confused with the matrix $\mathbf L =
\mathbf D - \mathbf A$ as defined in
Section~\ref{sec:decomposition:lap:signed}. 
The matrix $\mathbf{\hat L}$ appears in various contexts; see e.g.\ the
examples given by \cite{b875}.  We call it the \emph{repulsive
  Laplacian} due to its applications; the name is not standard.  The
matrix $\mathbf{\hat D}$ as defined above will be called the
\emph{repulsive degree matrix}. 

The matrix $\mathbf{\hat D}$ may have zero entries on the diagonal even
when no node has degree zero -- this happens when a node has an equal
number of positive and negative edges.  As a result, $\mathbf{\hat D}$
must be considered non-invertible.  Thus, variants of the repulsive
Laplacian are defined using the ordinary degree matrix for
normalization.

Other matrices that can be labeled \emph{repulsive} are given in the
following. 
When applied to ajacency matrices the hat may be regarded as a no-op.
Thus, $\mathbf{\hat A}=\mathbf A$, $\mathbf{\hat N}=\mathbf N$, and $\mathbf{\hat P}=\mathbf P$.
The repulsive Laplacian matrices are 
\begin{align}
  \mathbf{\hat L} &= \mathbf{\hat D} - \mathbf A \\
  \mathbf{\hat Z} &= \mathbf D^{-\frac 12} (\mathbf{\hat D} - \mathbf A) \mathbf D^{-\frac 12} \\
  \mathbf{\hat S} &= \mathbf D^{-1} \mathbf{\hat D} - \mathbf P = \mathbf D^{-1}(\mathbf{\hat D} - \mathbf A) \\
  \mathbf{\hat K} &= \mathbf{\hat D} + \mathbf A \\
  \mathbf{\hat C} &= \mathbf D^{-\frac 12} (\mathbf{\hat D} + \mathbf A) \mathbf D^{-\frac 12} \\
  \mathbf{\hat J} &= \mathbf D^{-1} \mathbf{\hat D} + \mathbf P = \mathbf D^{-1}(\mathbf{\hat D} + \mathbf A) 
\end{align}
The use of the repulsive matrices is rare, and we have not seen uses of
any beyond $\mathbf{\hat L}$ in the wild.  

\subsection{Decompositions of Bipartite Graphs}
Bipartite graphs can be considered a subset of unipartite graphs, and
thus the methods described previously apply to them.  However, this
treatment ignores the special structure, and instead, methods specific
to bipartite graphs can be used.  Indeed, bipartite graphs have
adjacency matrices of the form  
\begin{align}
  \mathbf A &= \left[ \begin{array}{cc} \mathbf 0^{\phantom {\mathrm T}}
      & \mathbf B \\ \mathbf 
      B^{\mathrm T} & \mathbf 0 \end{array} \right], 
\end{align}
where $\mathbf B$ is called the biadjacency matrix of the graph.  This form can be
exploited to reduce the eigenvalue decomposition of $\mathbf A$ to the equivalent
singular value decomposition of $\mathbf B$~\citep{b614}. 
Given the singular value
decomposition $\mathbf B = \mathbf U \mathbf \Sigma \mathbf
V^{\mathrm T}$, the eigenvalue decomposition of $\mathbf A$ is given by
\begin{align}
  \mathbf A &=
  \left[ \begin{array}{cc} \mathbf {\bar U} & {\phantom -}\mathbf
      {\bar U} \\ \mathbf {\bar V} & -\mathbf {\bar V} \end{array} \right] 
  \left[ \begin{array}{cc} +\mathbf \Sigma & \mathbf 0 \\ \mathbf 0 &
      -\mathbf \Sigma \end{array} \right] 
  \left[ \begin{array}{cc} \mathbf {\bar U} & {\phantom -}\mathbf
      {\bar U} \\ \mathbf {\bar V} & -\mathbf {\bar V} \end{array}
    \right]^{\mathrm T}  
\end{align}
with $\mathbf {\bar U} =\mathbf U/\sqrt 2$ and $\mathbf {\bar V} =
\mathbf V/\sqrt 2$.
In this decomposition, 
each singular value $\sigma$ corresponds to the eigenvalue pair $\{\pm
\sigma\}$.  
Odd powers of $\mathbf A$ then have the form
\begin{align*}
  \mathbf A^{2k+1} &= \left[ \begin{array}{cc} 
      \mathbf 0 & (\mathbf B \mathbf B^{\mathrm T})^k \mathbf B \\
      (\mathbf B^{\mathrm T} \mathbf B)^k \mathbf B^{\mathrm T} &
      \mathbf 0 
    \end{array} \right],
\end{align*}
where the alternating power $(\mathbf B \mathbf B^{\mathrm T})^k \mathbf
B$ can be explained by the fact that in the bipartite network, a path
will follow edges from one vertex set to the other in alternating
directions, corresponding to the alternating transpositions of $\mathbf
B$. 

The same is true in the normalized case:  The eigenvalue decomposition
of the normalized adjacency matrix $\mathbf N$ can be computed using the
singular value decomposition of the normalized biadjacency matrix
$\mathbf M$. 

However, this technique cannot be extended to other matrices such as the Laplacian
$\mathbf L$, since that matrix contains nonzero entries in its
``block diagonal'' entries. 
I.e., there is no corresponding expression relating the decomposition
of the bipartite Laplacian matrix $\mathbf L = [\mathbf D_1 \; {-\mathbf B}; -\mathbf
  B^{\mathrm T} \; \mathbf D_2]$ and the decomposition of the biadjacency
matrix $\mathbf B$. 

\subsection{Decompositions of Directed Graphs}
In directed graphs, the adjacency matrix is itself
asymmetric, and there is no special half-adjacency matrix.  Since the
adjacency matrix is symmetric, decompositions are more complex.  For
instance, the adjacency is not normal in the general case, and therefore
there is no simply defined eigenvalue decomposition anymore. 

For directed graphs, there is an ambiguity about the meaning of the
matrix $\mathbf A$:  It may refer to the (generally asymmetric)
adjacency matrix, or to its symmetrized variants.  As a general rule in
KONECT, in contexts in which all graphs (directed and undirected) are
considered, $\mathbf A$ is always symmetric.  To avoid ambiguities in
this section, we use $\vec{\mathbf A}$ for the (generally asymmetric)
adjacency matrix, and $\mathbf{\bar A}=\vec{\mathbf A}+\mathbf
A^{\mathrm T}$ for the symmetric variant.  When a bare $\mathbf A$ is
used, it is equal to $\vec{\mathbf A}$. 

\subsubsection{Singular Value Decomposition}
\label{sec:matrix.svd}
The singular value decomposition is defined for any matrix, including
those that are not symmetric, and even those that are not quadratic.
Thus, it can be applied to the adjacency matrix of directed graphs. 
\begin{align}
  \mathnote{\texttt{svd}}
  \mathbf A &= \mathbf U \mathbf \Sigma \mathbf V^{\mathrm T}
\end{align}
The matrices $\mathbf U$ and $\mathbf V$ are each orthogonal, but they
are not equal.  They contain the left and right singular vectors as
columns.  The matrix $\mathbf \Sigma$ is diagonal, and contains the
singular values, which are all nonnegative. 

This decomposition corresponds to the eigenvalue decomposition of the
directed graph's bipartite double cover. 
It also corresponds to HITS \citep[Hyperlink-Induced Topic Search, ][]{b27}.

The largest singular value, i.e., $\Sigma_{11}$, equals the graphs
operator 2-norm $\nu$. 

\subsubsection{Normalized Adjacency Matrix}
\label{sec:decomposition.svd-n}
The adjacency matrix can be normalized for directed network, in the same
way as for undirected networks.
\begin{align}
  \mathbf N &= \mathbf D_1^{-\frac 1 2} \mathbf A \mathbf D_2^{-\frac 1 2}
\end{align}
Here, $\mathbf D_1^{-\frac 1 2}$ and $\mathbf D_2^{-\frac 1 2}$
are the diagonal matrices of out- and indegrees. 

The normalized adjacency matrix $\mathbf N$ can be used in the singular
value decomposition, too:
\begin{align}
  \mathnote{\texttt{svd-n}}
  \mathbf N &= \mathbf U \mathbf \Sigma \mathbf V^{\mathrm T}
\end{align}

\subsubsection{Eigenvectors and Eigenvalues}
\label{sec:matrix.diag}
While the eigenvalue decomposition is not defined in the general case
for an asymmetric matrix $\mathbf A$, its eigenvectors and eigenvalues
are well-defined, if we distinguish between left and right
eigenvectors.

\marginpar{\texttt{diag}}
Thus, we define the method \texttt{diag}, which is a decomposition in
the KONECT sense, but not in the strict mathematical sense. 
$\mathbf U$ and $\mathbf V$ are then defined as the matrices containing
the left and right eigenvectors of $\mathbf A$, and $\mathbf \Lambda$ is
the diagonal matrix of corresponding eigenvalues. 

Note that while left and right eigenvectors differ, their eigenvalues
are identical. 

The largest absolute eigenvalue $\pi=|\Lambda_{11}|$ is zero if and only
if the directed graph is acyclic.  We call it the cyclic eigenvalue in
KONECT.  
In the general case, the eigenvalues of the asymmetric $\mathbf A$ may
be complex; if they are, they come in complex conjugate pairs. (This is
a consequence of the fact that the matrix $\mathbf A$ itself is real.)

\subsubsection{Eigenvectors and Eigenvalues (Normalized)}
\label{sec:matrix.diag-n}
This corresponds to finding the eigenvectors of the asymmetric matrix
$\mathbf N$. 
\marginpar{\texttt{diag-n}}
The same comments as for the non-normalized case apply. 

\subsubsection{Stochastic Adjacency Matrix ($\mathbf P$)}
The matrix $\mathbf P$ can be defined in the directed case in the same
way as for undirected graphs.  Properties given in Section
\ref{sec:matrix.P} remain valid. 

\subsubsection{Skew Adjacency Matrix ($\mathbf Y$)}
\label{sec:matrix.skew}
While an asymmetric matrix $\mathbf A$ may be transformed to $\mathbf A
+ \mathbf A^{\mathrm T}$ to give a symmetric matrix, we can also use
$\mathbf Y = \mathbf A - \mathbf A^{\mathrm T}$ to get a skew-symmetric matrix.  A
skew symmetric matrix $\mathbf X$ is a matrix such that $\mathbf X =
-\mathbf X^{\mathrm T}$.

One motivation for considering the matrix $\mathbf A - \mathbf
A^{\mathrm T}$ is simply that this matrix is skew-symmetric, which makes
it a normal matrix, and thus the eigenvalue decomposition is
well-defined.  Beyond this, there is also an argument for why $\mathbf A
- \mathbf A^{\mathrm T}$ in particular may be used -- if it was only for
deriving a normal matrix from $\mathbf A$, then after all $\mathbf A +
\mathbf A^{\mathrm T}$ would be perfectly adequate:  When $\mathbf A +
\mathbf A^{\mathrm T}$ is used, the underlying assumption is that the
direction of edges is unimportant and can be ignored.  This is usually a
good assumption in directed networks in which edges are very often
reciprocated anyway, i.e., in highly symmetric graphs.  Examples of such
graphs are communication networks such as email networks.  In other
directed networks however, the semantics of edge directions are
different.  For instance, a sports result network in which nodes are
teams and a directed edge denotes that a team has won against another
team:  In such networks, we may interpret ``A has won against B'' as
equivalent to ``B has lost against A''.  Thus, we can consider such a
network as a signed directed network, making $\mathbf A - \mathbf
A^{\mathrm T}$ the natural way to \emph{normalize} the network.  This
has the advantage of taking into account who has won each game, as
opposed to just considering the ``who has played whom'' network that is
implicitly given by the symmetrical $\mathbf A + \mathbf A^{\mathrm
  T}$.  This explains that the matrix $\mathbf Y$ is used in situations
such as sports results which are inherently skew symmetric, i.e., where
an edge in one direction is equivalent to a negated edge in the other
direction.  Another example of such networks are dominance networks
between animals, for instance cattle
(\href{http://konect.cc/networks/moreno_cattle/}{\textsf{MA}}).  Such
networks are indicated by the tag \texttt{\#skew}, as defined in
Section~\ref{sec:tags}. 

Skew symmetric matrices have well-defined eigenvalue decompositions,
which are however complex, as both the eigenvectors and eigenvalues will
be complex numbers.  The eigenvectors and eigenvalues however follow a
specific pattern, that we can exploit to represent such a decomposition
using only real numbers:
\begin{align}
  \mathbf Y &= \mathbf A - \mathbf A^{\mathrm T} = \mathbf Q \mathbf R \mathbf Q^{\mathrm T}
\end{align}
such that
\begin{align}
  \mathbf Q &= \frac 1 {\sqrt{2}} \left[
    \begin{array}{cc}
      \mathbf U + i \mathbf V \\
      \mathbf U - i \mathbf V 
    \end{array}
    \right] \\
  \mathbf R &= \left[ \begin{array}{cc}
      i \mathbf D & \mathbf 0 \\
      \mathbf 0 & -i \mathbf D
    \end{array} \right]
\end{align}
where $\mathbf U$, $\mathbf V$ and $\mathbf D$ are real matrices.  In
fact, this decomposition can be equivalently written as
\begin{align}
  \mathnote{\texttt{skew}}
  \mathbf Y = \mathbf A - \mathbf A^{\mathrm T} &=
  \mathbf U \mathbf D \mathbf V^{\mathrm T} - \mathbf V \mathbf D \mathbf U^{\mathrm T}.
\end{align}
This decomposition is equivalent to that given by 
\cite{b869}.  It shows that the skew-symmetric matrix $\mathbf Y$ has
eigenvalues that are purely imaginary, come  
in pairs $\{\pm i \lambda\}$ that are the negative of each other (or
equivalently, the complex conjugate), and their corresponding
eigenvectors are also complex conjugates of each other.

Note also that the number of columns of both $\mathbf U$ and $\mathbf V$
is at most $\lfloor n \rfloor$, and thus, if $n$ is odd, the
skew-symmetric matrix $\mathbf Y$ has the
eigenvalue zero, which is the complex conjugate of itself.  Also, the
expression $\mathbf U \mathbf D \mathbf V^{\mathrm T}$ is \emph{not} the
singular value decomposition of $\mathbf A$, even if the form of the
decomposition is the same.  
In particular, the matrix $\mathbf U \mathbf
D \mathbf V^{\mathrm T}$ has at most rank $\lfloor n/2 \rfloor$, while
$\mathbf A$ itself may have rank up to $n$. 

In some cases, we may be interested in the matrix
\begin{align}
  \mathnote{\texttt{skewi}}
  i \mathbf Y &= i ( \mathbf A - \mathbf A^{\mathrm T} ),
\end{align}
which has purely real eigenvalues distributed symmetrically around
zero. 

\subsubsection{Normalized Skew Adjacency Matrix}
\label{sec:matrix.skewn}
This corresponds to the matrix
\begin{align}
  \mathnote{\texttt{skewn}}
  \mathbf N - \mathbf N^{\mathrm T} = 
  \mathbf D_1^{-\frac 1 2} \mathbf A \mathbf D_2^{-\frac 1 2} - 
  \mathbf D_2^{-\frac 1 2} \mathbf A^{\mathrm T} \mathbf D_1^{-\frac 1 2} 
\end{align}

\subsubsection{Hermitian Adjacency Matrix ($\mathbf H$)}
In certain contexts, for instance when constructing the Hamiltonian of a
system, it is necessary to specify Hermitian matrices, i.e.,
diagonalizable matrices that have only real eigenvalues.  To take into
account all connectivity information of a directed graph in a
directed graph, the symmetric and skew-symmetric adjacency matrices can
be combined in the following way:
\begin{align}
  \mathnote{\texttt{herm}}
  \mathbf H &= \frac 1 {\sqrt 2} \left[
  \mathbf A + \mathbf A^{\mathrm T} + i(\mathbf A - \mathbf A^{\mathrm T})
  \right]. 
\end{align}
The matrix $\mathbf H$ is Hermitian, i.e., it equals the complex
conjugate of its transpose, $\mathbf H = \mathbf {\bar H}^{\mathrm T} =
\mathbf H^{\dagger}$.
Furthermore, the Hermitian adjacency matrix can be defined by
\begin{align}
  \mathbf H_{uv} &= \exp\left\{i \frac \pi 4 \mathbf Y_{uv}\right\},
\end{align}
which additionally justifies the factor $1/\sqrt 2$ in the initial
definition of $\mathbf H$, as $e^{i \frac \pi 4} = 1/\sqrt 2 + i/\sqrt
2$.  Using this form, the Hermitian adjacency matrix can be generalized
to the matrix $\mathbf H_\theta$ where $0 \leq \theta \leq \pi/2$ is a
real parameter, giving 
\begin{align}
  [\mathbf H_{\theta}]_{uv} &= \exp\left\{i \theta \mathbf Y_{uv} \right\},  
\end{align}
from which the following special cases can be recovered:
\begin{align}
  \mathbf H_0 &= \mathbf A + \mathbf A^{\mathrm T}, \\
  \mathbf H_{\pi/4} &= \mathbf H, \\
  \mathbf H_{\pi/2} &= i \mathbf Y.
\end{align}
These matrices appear in the modeling of quantum walks, as used for
instance by \cite{toedtli:ctqw}.  

\subsubsection{Normalized Hermitian Matrix}
The normalized Hermitian matrix is
\begin{align}
  \mathnote{\texttt{hermn}}
  \frac 1{\sqrt 2} \mathbf D^{-\frac 1 2} \mathbf H \mathbf D^{-\frac 1 2}. 
\end{align}
(We currently do not have a symbol for this matrix.)

\subsubsection{Directed Laplacian (Chung)}
This variant of a Laplacian matrix for directed graphs is given by
\cite{b264}.  It is itself symmetric, but takes into account edge
directions. \marginpar{\texttt{lapd}}

A normalized variant is also possible. \marginpar{\texttt{lapd-n}} 

\subsubsection{Magnetic Laplacian ($\mathbf T$)}
The matrix
\begin{align}
  \mathbf T_{\theta} &= \mathbf D - \mathbf H_{\theta} 
\end{align}
is the magnetic Laplacian \citep{b903}.  It is Hermitian and positive-semidefinite.   

For the case $\theta = 0$, this reduces to the ordinary undirected
Laplacian, effectively ignoring the edge directions.
\begin{align}
  \mathbf T_{0} &= \mathbf D - \mathbf H_{0} = \mathbf D - (\mathbf A + \mathbf A^{\mathrm T}) = \mathbf L
\end{align}

In the case $\theta \neq 0$, the magnetic Laplacian $T_{\theta}$ is
distinct from the ordinary Laplacian, even if the directed graph is
symmetric, i.e., when all edges are reciprocated. 

For the case $\theta = \pi/4$, this can be called the Hermitian
Laplacian, and is simply denoted $\mathbf T$, in the same way that
$\mathbf H_{\pi/4}$ is denoted $\mathbf H$:
\begin{align}
  \mathnote{\texttt{lapherm}}
  \mathbf T &= \mathbf T_{\pi/4} = \mathbf D - \mathbf H
\end{align}
Note that we define a bare $\mathbf T$ to be $\mathbf T_{\pi/4}$ rather
than $\mathbf T_0$ -- this is the reason why use the distinct symbol
$\mathbf T$, as otherwise we could just use $\mathbf L_{\theta}$ and define
$\mathbf L_0 = \mathbf L$. 

For $\theta = \pi/2$ this is called the skew Laplacian:
\begin{align}
  \mathnote{\texttt{lapskew}}
  \mathbf T_{\pi/2} &= \mathbf D - i (\mathbf A - \mathbf A^{\mathrm T}) 
  = \mathbf D - i \mathbf Y
\end{align}

For any value of $\theta$, the matrix $\mathbf T_{\theta}$ is Hermitian,
and therefore has real spectrum.  As is the ordinary Laplacian matrix,
the magnetic Laplacian matrix is positive-semidefinite, i.e., all
eigenvalues are nonnegative, for all values of $\theta$. 

\subsubsection{Unidirectional Laplacian}
\label{sec:decomposition.lapdiag2}
This is the generally non-normal matrix
\begin{align}
  \mathnote{\texttt{lapdiag2}}
  \vec{\mathbf L} &= \mathbf D_1 - \vec{\mathbf A}, 
\end{align}
where $\mathbf D_1$ is the diagonal outdegree matrix, and $\vec{\mathbf A}$
is the generally asymmetric adjacency matrix of the given directed
graph.  

This matrix is in general non-normal and as such its eigenvalue
decomposition cannot be considered.  However, the eigenvalues and
eigenvectors are well-defined.  The eigenvalues have nonnegative real
part, and are either real, or come in complex conjugate pairs. 

\subsubsection{Decomposition into Directed Components}
DEDICOM (decomposition into directed components) refers to a class of
matrix decompositions for directed networks of the form
\begin{align}
  \mathbf A = \mathbf U \mathbf X \mathbf U^{\mathrm T},
\end{align}
in which $\mathbf U$ is orthogonal and $\mathbf X$ is asymmetric, and
thus non-diagonal.  These represent a family of decompositions for which
there is no single best one, and in particular the problem of finding
the best rank-$k$ such decompositions is not given by the truncation to
rank $k$ of the solution to the full-rank problem.  An overview of
algorithms is given in \citep{kunegis:directed-decomposition}. 

\subsubsection{Impractical Decompositions}
The decompositions presented in this section apply to the (generally) asymmetric
adjacency matrix $\mathbf A$ of directed graphs.  
They all have practical problems that forbids their application to
almost all networks in KONECT: 
Some are fully dense in the sense that
they need $O(n^2)$ memory to be represented.  Others are numerically
unstable. 
Thus, these decomposition can only be used with the
smallest of networks.  As a result, they are not of practical interest
in KONECT, but rather of theoretical importance. 

As a general rule, implementations of these decompositions only exist
with full rank, and no low-rank approximation is possible, as is the
case with the eigenvalue and singular value decompositions. 

A useful overview of rare and esoteric matrix decompositions and
decomposition-like methods for generally asymmetric matrices is given by
\cite{b136} in the context of computing the matrix exponential. 

The \textbf{Schur decomposition}
\marginpar{\texttt{schur}} reduces a matrix to cycle-free form
\begin{align}
  \mathbf A = \mathbf U \mathbf R \mathbf U^{\mathrm T},
\end{align}
with $\mathbf U$ orthogonal and $\mathbf R$ triangular.  
This is the preferred method used in \citep{b136}. 
The Octave/Matlab function \texttt{schur()} only takes dense a matrix and
computes a full decomposition. 

The \textbf{orthogonal Hessenberg decomposition} \citep{b139} is defined
as
\begin{align}
  \mathbf A = \mathbf U \mathbf H \mathbf U^{\mathrm T},
\end{align}
where $\mathbf U$ is orthogonal and $\mathbf H$ is an upper Hessenberg
matrix, i.e., an upper triangular matrix in which additionally the
subdiagonal contains non-zeroes.  This corresponds to the function
\texttt{OTHES} in EISPACK, and to \texttt{hess()} in Octave/Matlab.  It
is only implemented for dense matrices. 

The \textbf{Jordan decomposition} is defined as
\begin{align}
  \textbf{U} \textbf{F} \textbf{U}^{-1}.
\end{align}
This decomposition is not stable, i.e., it is not robust against
multiple eigenvalues. 
An implementation exists in Matlab only in the symbolic toolbox; due to its
instability, an implementation for floating point values would not
work. 

The \textbf{companion decomposition} or Frobenius normal form
\citep[p.\ 19]{b136} is given by
\begin{align}
  \mathbf A = \mathbf Y \mathbf C \mathbf Y^{\mathrm T},
\end{align} 
in which $\mathbf C$ is a companion matrix.

\section{Plots}
\label{sec:plots}
Plots are drawn to visualize a certain aspect of a dataset. These plots
can be used to compare several network visually, or to illustrate the
definition of a certain numerical statistic.

As a running example, we show the plots for the Wikipedia elections
network
(\href{http://konect.cc/networks/elec/}{\textsf{EL}}), or for larger
networks in some cases.  Plots
for all networks (in which computation was feasible) are shown on the
KONECT
website\footnote{\href{http://konect.cc/plots/}{konect.cc/plots}}. The
KONECT Toolbox contains Matlab code for generating these plot types.

\subsection{Layout}
\begin{figure}
  \centering
  \includegraphics[width=\wPlot]{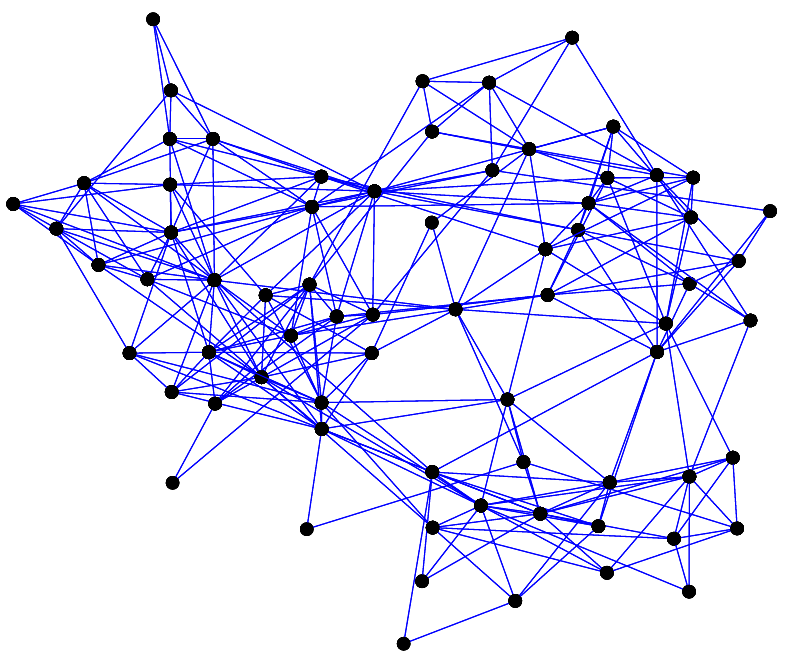}
  \caption{
    \label{fig:fruchterman-reingold}
    The layout of the highschool social network given by \cite{konect:coleman}
    (\href{http://konect.cc/networks/moreno_highschool/}{\textsf{MH}}),
    generated using the algorithm of \cite{b870}.  This network has 70 vertices,
    and for such small networks, drawing a graph leads to sensible
    plot.  For large graphs, graph drawing usually lead to
    ``hairball''-like plots and are thus less useful.
  }
\end{figure}

Layout plots show the nodes and edges of a graph in a way that makes
features if the graph visible.  Usually, this only makes sense for small
graphs.\footnote{See
  \href{https://networkscience.wordpress.com/2016/06/22/no-hairball-the-graph-drawing-experiment/}{networkscience.wordpress.com/2016/06/22/no-hairball-the-graph-drawing-experiment}
  for an explanation.}
In KONECT, we use the algorithm of \cite{b870}.  An
example is shown in Figure~\ref{fig:fruchterman-reingold}.

\subsection{Temporal Distribution}
\begin{figure}
  \centering 
  \subfigure[Temporal distribution]{
    \includegraphics[width=\wPlot]{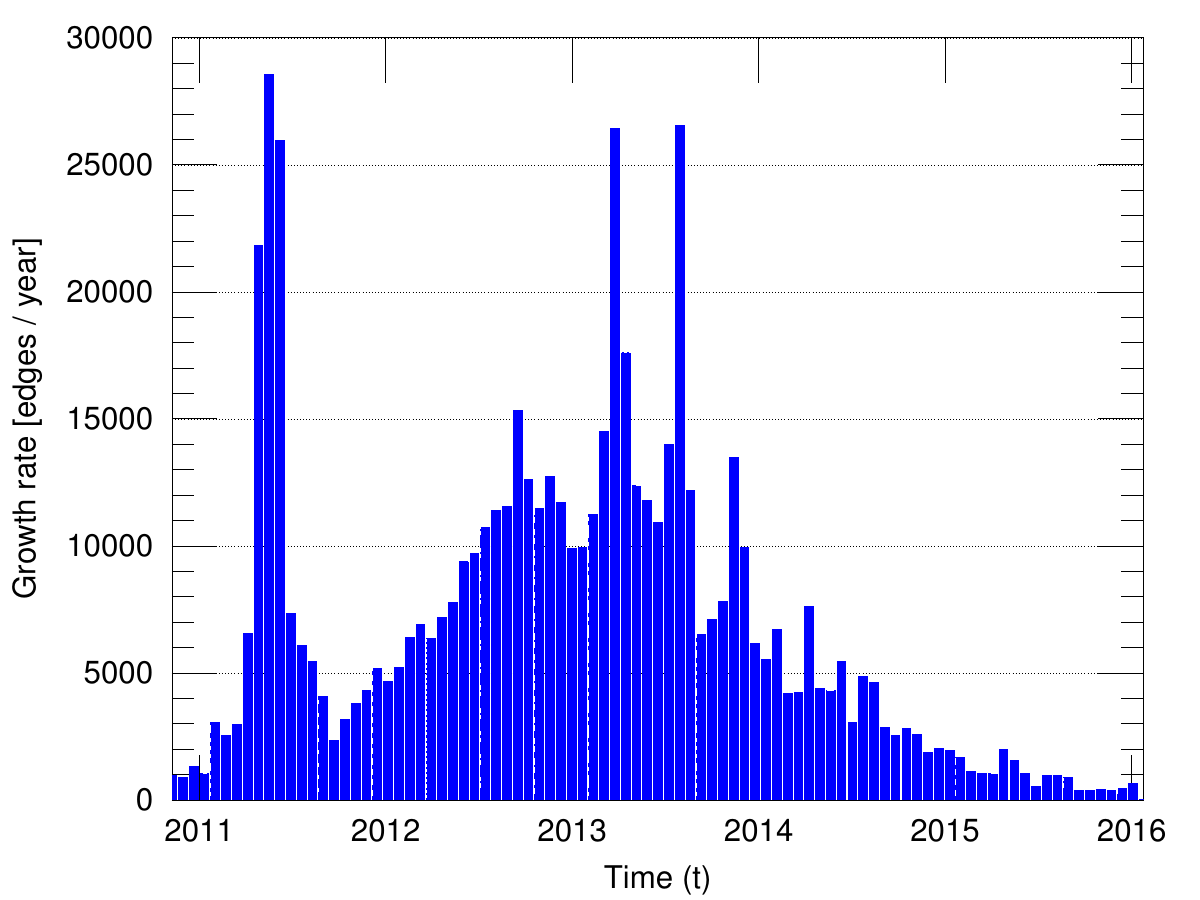}}
  \subfigure[Cumulative temporal distribution]{
    \includegraphics[width=\wPlot]{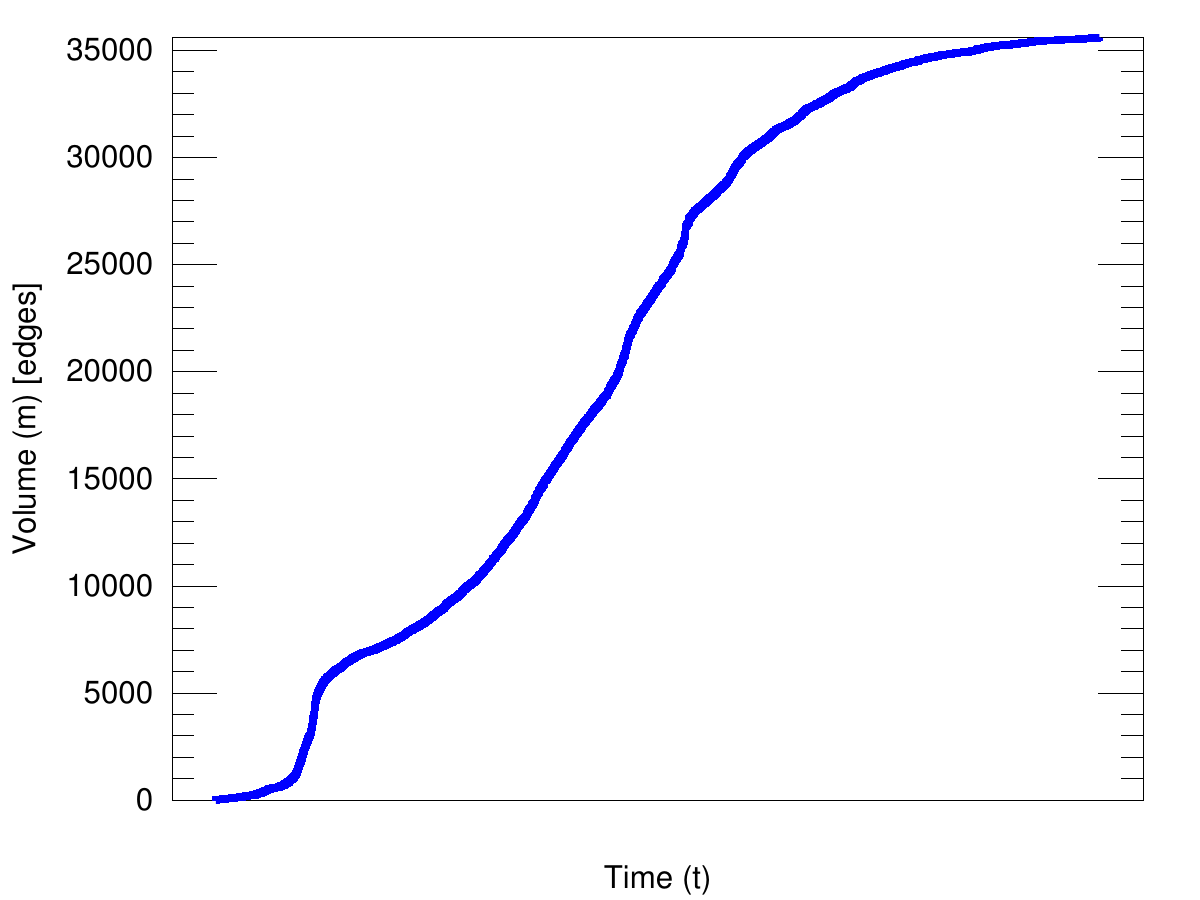}}
  \caption{ 
    \label{plot:time_histogram}
    The temporal distribution and cumulative temporal distribution of
    the Bitcoin OTC network
    (\href{http://konect.cc/networks/soc-sign-bitcoinotc/}{\textsf{BO}}).  
  }
\end{figure}

The temporal distributions shows the distribution of edge creation
times.  It is only defined for networks with known edge creation times.
The X axis is the time, and the Y axis is the number of edges added
during each time interval.  We compute two plots:  the distribution, and
the cumulative distribution.  Examples are shown for the BItcoin OTC
network (\href{http://konect.cc/networks/soc-sign-bitcoinotc/}{\textsf{BO}}) in Figure~\ref{plot:time_histogram}. 

\subsection{Edge Weight and Multiplicity Distribution}
The edge weight and multiplicity distribution plots show the
distribution of edge weights and of edge multiplicities, respectively.
They are not generated for unweighted networks.  The X axis shows values
of the edge weights or multiplicities, and the Y axis shows frequencies.
Edge multiplicity distributions are plotted on doubly logarithmic
scales.

\begin{figure}
  \centering 
  \subfigure[Edge weight distribution]{
    \includegraphics[width=\wPlot]{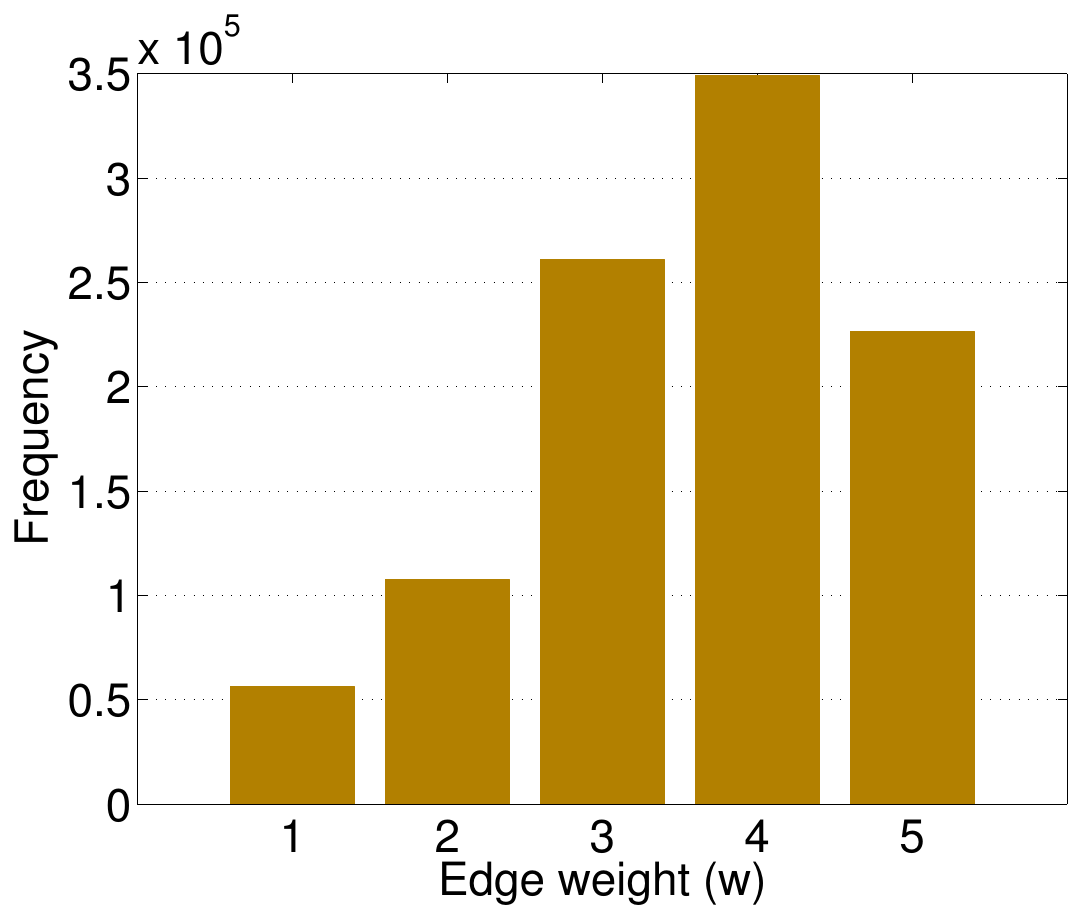}}
  \subfigure[Edge multiplicity distribution]{
    \includegraphics[width=\wPlot]{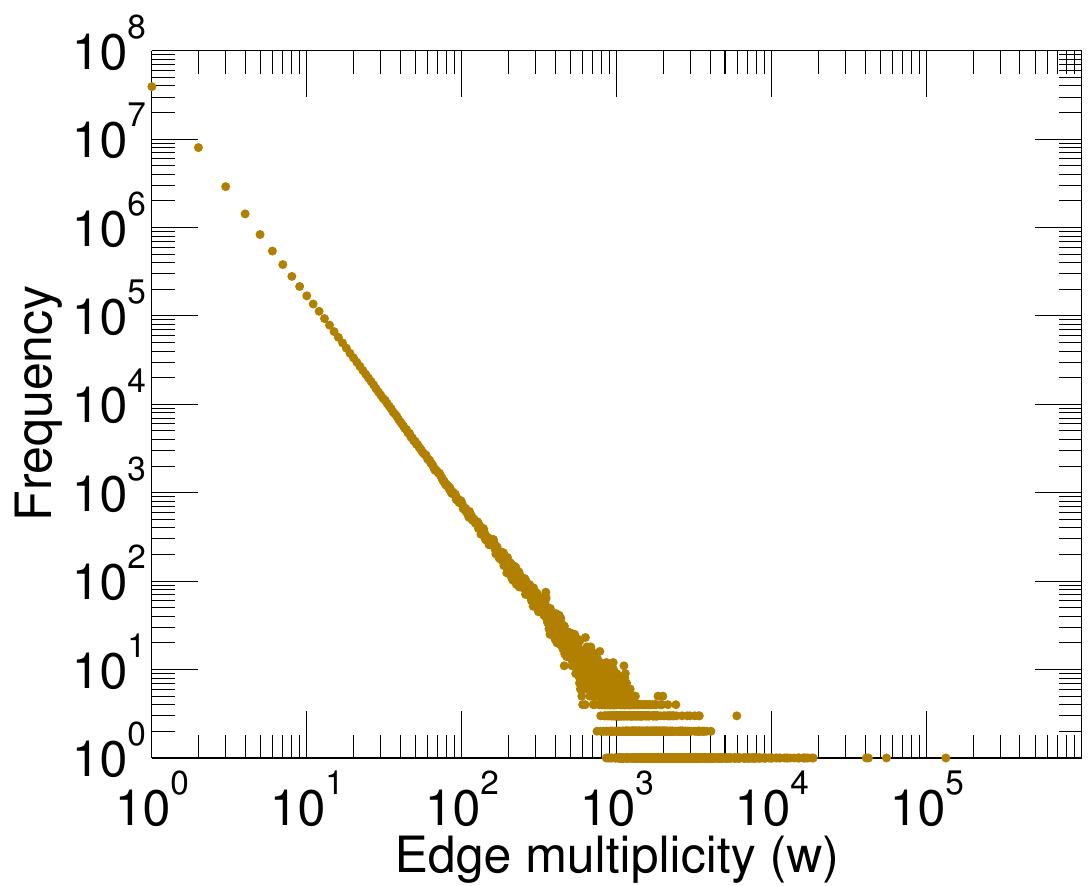}}
  \caption{ 
    \label{fig:weight}
    The distribution of (a)~edge weights for the MovieLens
    rating network
    (\href{http://konect.cc/networks/movielens-1m/}{\textsf{M2}})
    and (b)~edge multiplicities for the German Wikipedia edit network
    (\href{http://konect.cc/networks/edit-dewiki/}{\textsf{de}}).
  }
\end{figure}

\subsection{Degree Distribution} 
\label{sec:plot:degree}
The distribution of degree values $d(u)$ over all vertices $u$
characterizes the network as a whole, and is often used to visualize a
network. In particular, a power law is often assumed, stating that the
number of nodes with $n$ neighbors is proportional to $n^{-\gamma}$, for
a constant $\gamma$ \citep{b439}. This assumption can be inspected
visually by plotting the degree distribution on a doubly logarithmic
scale, on which a power law renders as a straight line.  KONECT supports
two different plots: The degree distribution, and the cumulative degree
distribution. The degree distribution shows the number of nodes with
degree $n$, in function of $n$.  The cumulative degree distribution
shows the probability that the degree of a node picked at random is
larger than $n$, in function of $n$. Both plots use a doubly logarithmic
scale.

Another visualization of the degree distribution supported by KONECT is
in the form of the Lorenz curve, a type of plot to measure inequality
originally used in economics (not shown).

\begin{figure}
  \centering 
  \subfigure[Degree distribution]{
    \includegraphics[width=\wPlot]{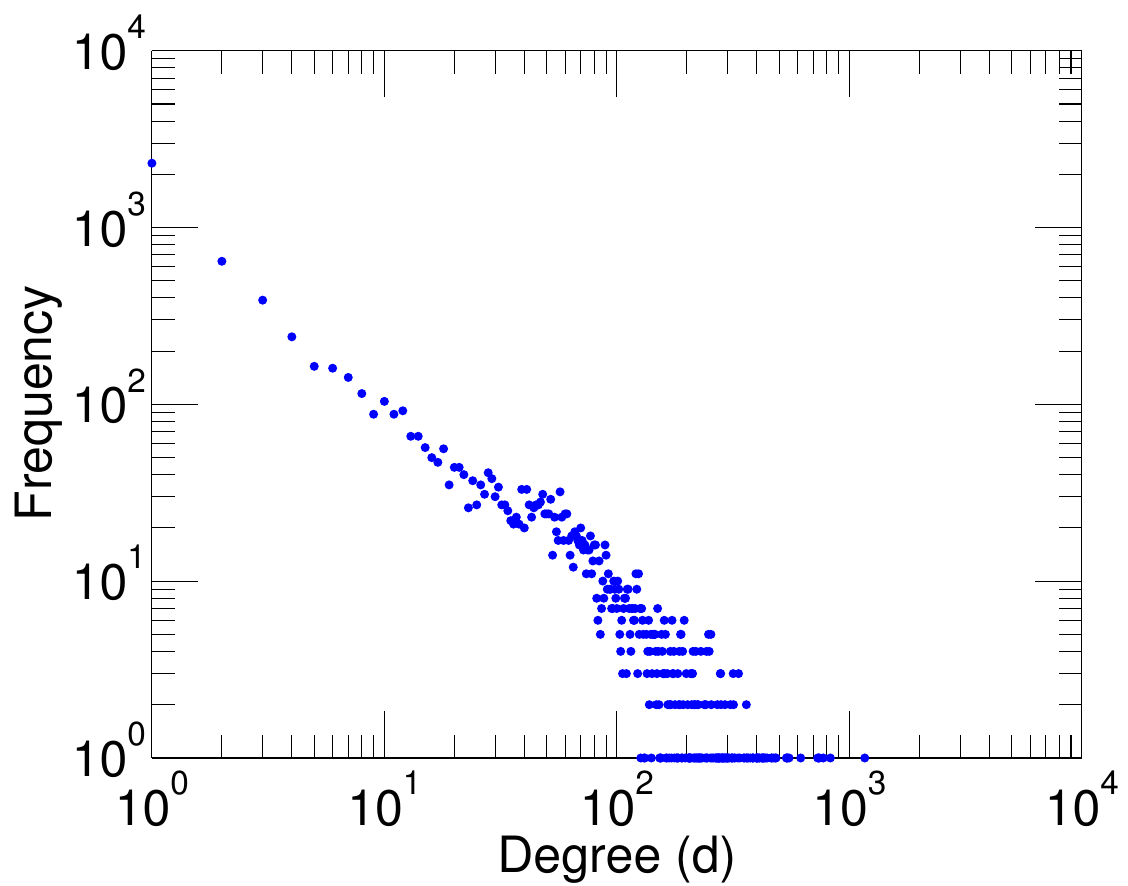}}
  \subfigure[Cumulative degree distribution]{
    \includegraphics[width=\wPlot]{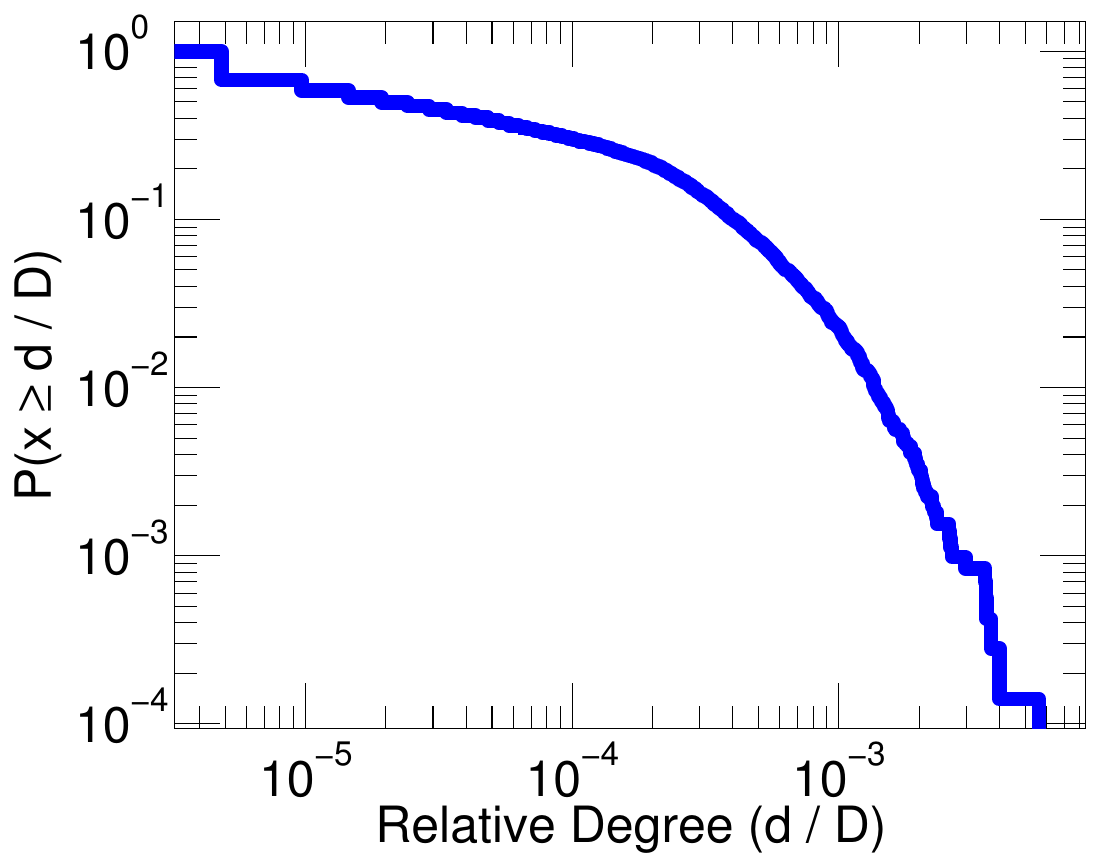}}
  \caption{ 
    \label{fig:degree}
    The degree distribution and cumulative degree distribution
    for the Wikipedia election network
    (\href{http://konect.cc/networks/elec/}{\textsf{EL}}).
  }
\end{figure}

The Lorenz curve is a tool originally from economics that visualizes
statements of the form ``X\% of nodes with smallest degree account for
Y\% of edges''.  The set of values $(X,Y)$ thus defined is the Lorenz
curve. In a network the Lorenz curve is a straight diagonal line when
all nodes have the same degree, and curved
otherwise \citep{kunegis:power-law}.  The area between the Lorenz curve
and the diagonal is half the Gini coefficient (see above).

\begin{figure}
  \centering 
  \includegraphics[width=\wPlot]{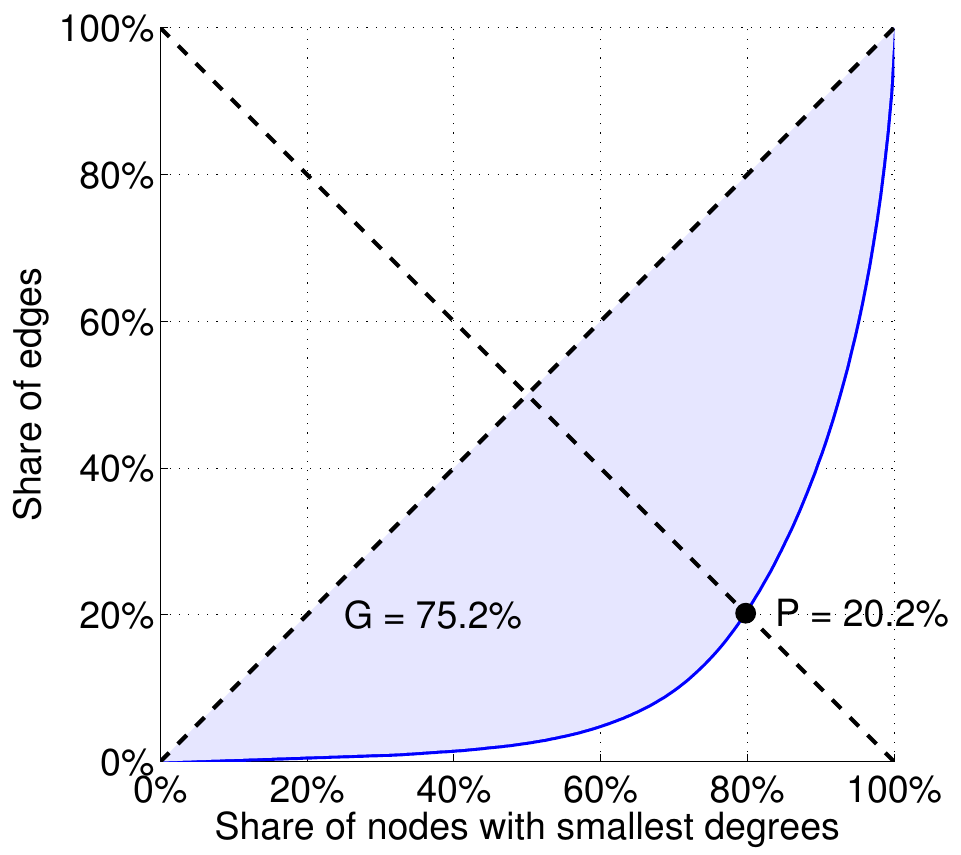}
  \caption{ 
    The Lorenz curve for the Wikipedia election network
    (\href{http://konect.cc/networks/elec/}{\textsf{EL}}).
    \label{fig:lorenz}
  }
\end{figure}

\subsection{Out/indegree Comparison}
The out/indegree comparison plots show the joint distribution of
outdegrees and indegrees of all nodes of directed graphs.  The plot
shows, for one directed network, each node as a point, which the
outdegree on the X axis and the indegree on the Y axis.  

\begin{figure}
  \centering
  \includegraphics[width=\wPlot]{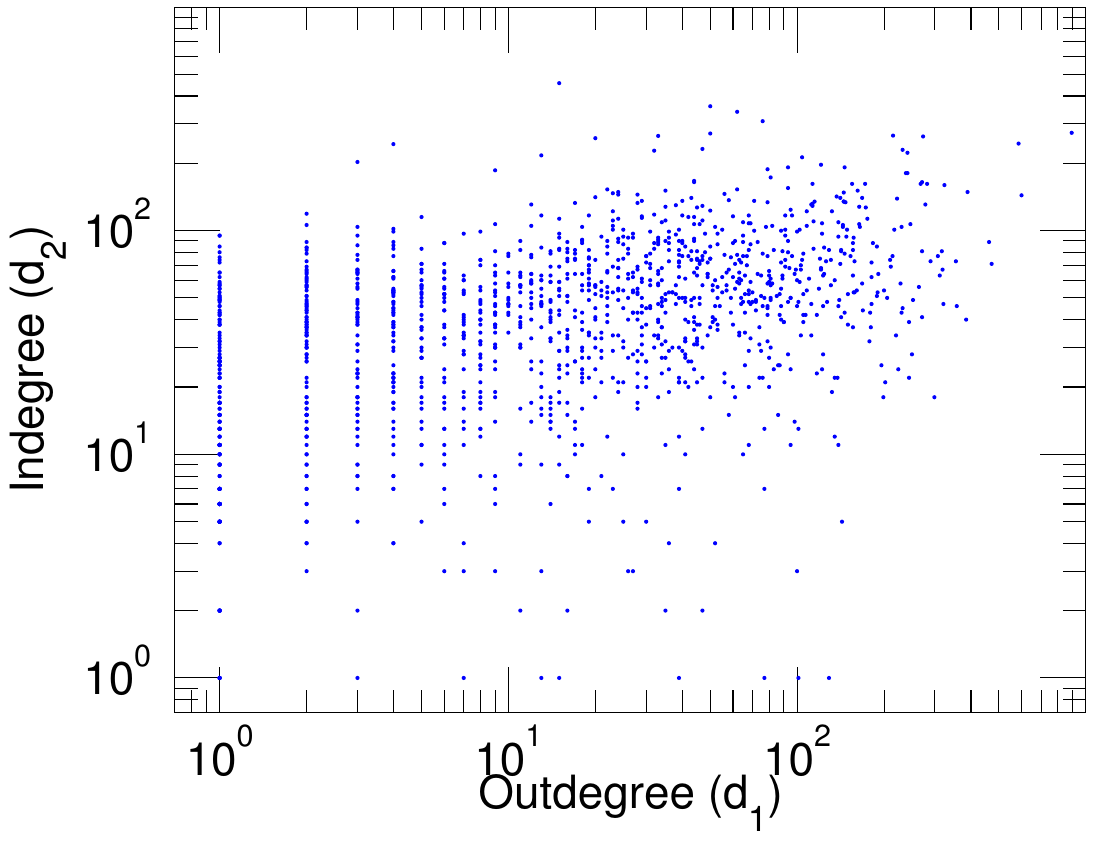}
  \caption{
    \label{fig:outin}
    The out/indegree comparison plot of the Wikipedia election network 
    (\href{http://konect.cc/networks/elec/}{\textsf{EL}}).
  }
\end{figure}

An example is shown in Figure~\ref{fig:outin} for the Wikipedia
elections network. 

\subsection{Assortativity Plot}
In some networks, nodes with high degree are more often connected with
other nodes of high degree, while nodes of low degree are more often
connected with other nodes of low degree.  This property is called
assortativity, i.e., such networks are said to be assortativity.  On the
other hand, some networks, are dissortative, i.e., in them nodes of high
degree are more often connected to nodes of low degree and vice versa.
In addition to the assortativity $\rho$ defined as the Pearson
correlation coefficient between the degrees of connected nodes, the
assortativity or dissortativity of networks may be analyse by plotting
all nodes of a network by their degree and the average degree of their
neighbors.  Thus, the assortativity plot of a network shows all nodes of
a network with the degree on the X axis, and the average degree of their
neighbors on the Y axis. 

\begin{figure}
  \centering
  \includegraphics[width=\wPlot]{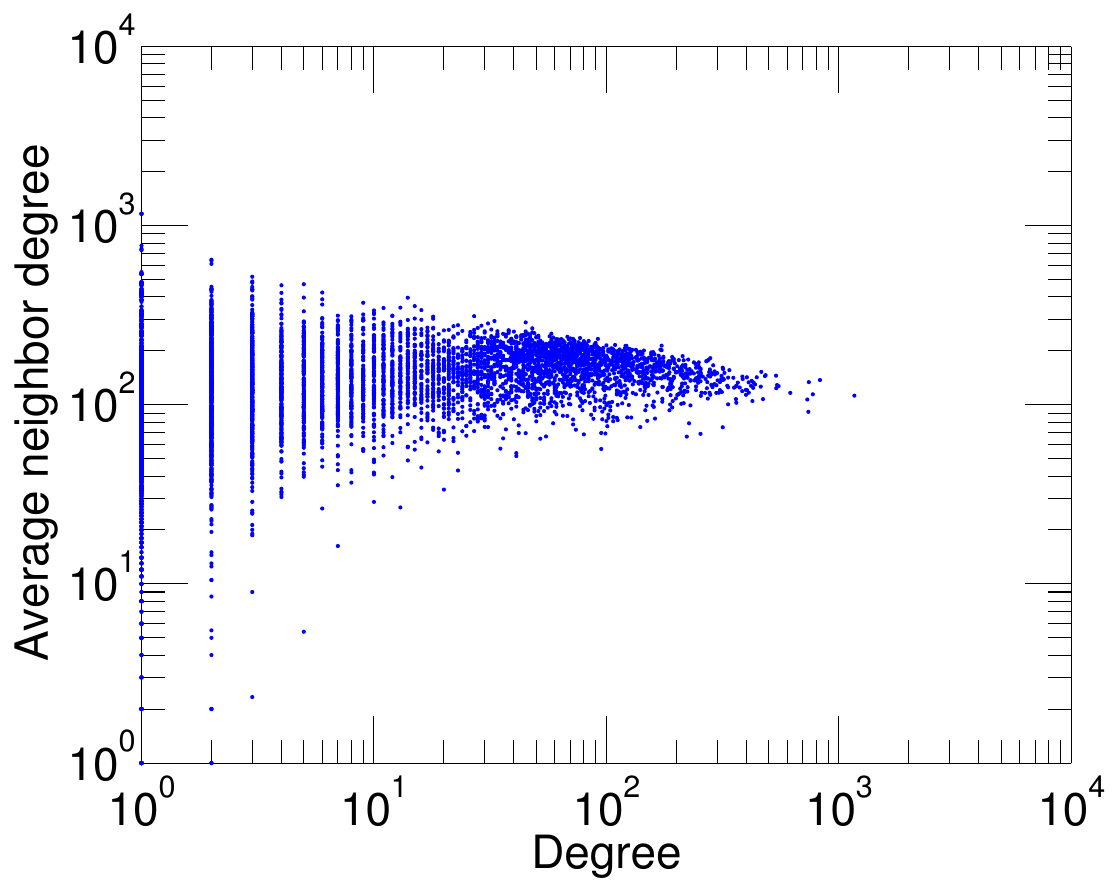}
  \caption{
    The assortativity plot of the Wikipedia election network 
    (\href{http://konect.cc/networks/elec/}{\textsf{EL}}).
    \label{fig:assortativity}
  }
\end{figure}

An example of the assortativity plot is shown for the Wikipedia
elections network in Figure~\ref{fig:assortativity}. 

\subsection{Clustering Coefficient Distribution}
In Section~\ref{sec:statistic:clustering}, we defined the clustering
coefficient of a node in a graph as the propotion of that node's
neighbors that are connected, and proceeded to define the clustering
coefficient as the corresponding measure applied to the whole network.
In some case however, we may be interested in the distribution of the
clustering coefficient over the nodes in the network.  For instance, a
network could have some very clustered parts, and some less clustered
parts, while another network could have many nodes with a similar,
average clustering coefficient.  Thus, we may want to consider the
distribution of clustering coefficient.  This distribution can be
plotted as a cumulated plot.

\begin{figure}
  \centering
  \includegraphics[width=\wPlot]{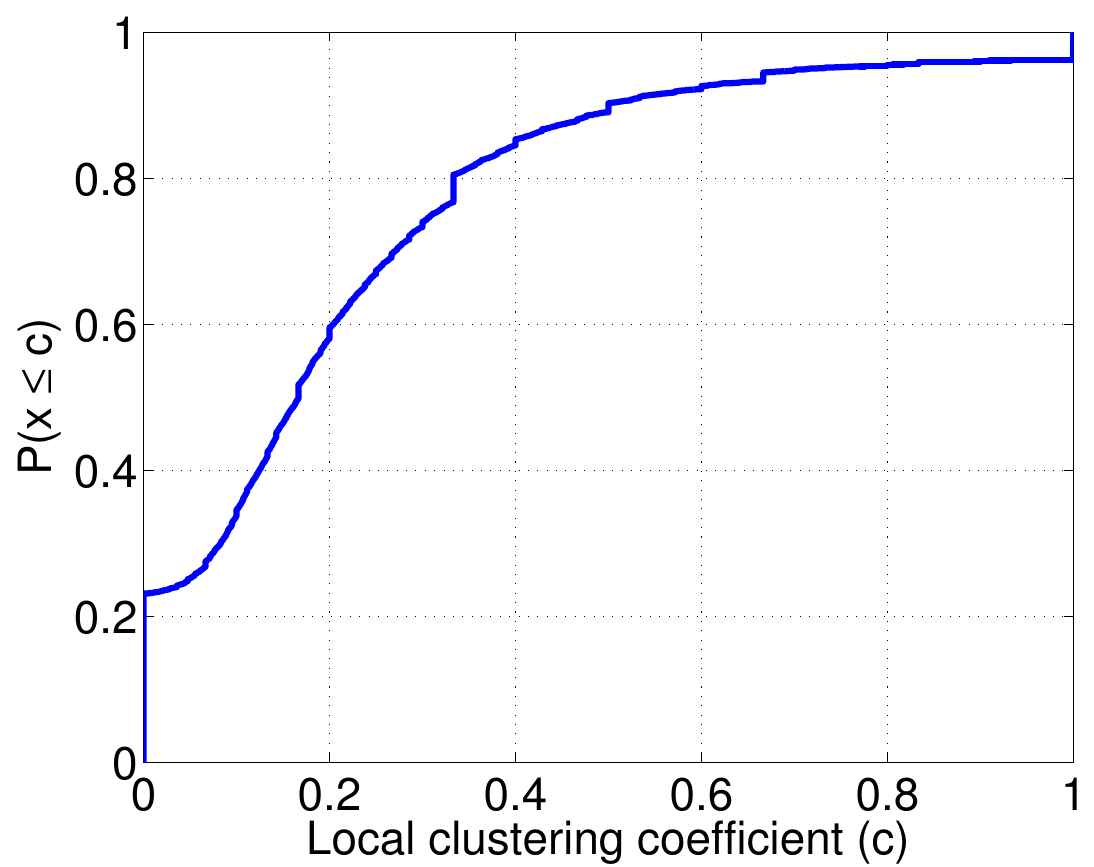}
  \caption{ The clustering coefficient distribution for Facebook link
    network~(\href{http://konect.cc/networks/facebook-wosn-links/}{\textsf{Ol}}).
  }
\end{figure}

\subsection{Spectral Plot}
The eigenvalues of a network's characteristic matrices $\mathbf A$,
$\mathbf N$ and $\mathbf L$ are often used to characterize the network
as a whole.  KONECT supports computing and visualizing the spectrum
(i.e., the set of eigenvalues) of a network in multiple ways.  Two types
of plots are supported: Those showing the top-$k$ eigenvalues computed
exactly, and those showing the overall distribution of eigenvalues,
computed approximately. The eigenvalues of $\mathbf A$ are positive and
negative reals, the eigenvalues of $\mathbf N$ are in the range
$[-1,+1]$, and the eigenvalues of $\mathbf L$ are all nonnegative.  For
$\mathbf A$ and $\mathbf N$, the largest absolute eigenvalues are used,
while for $\mathbf L$ the smallest eigenvalues are used.  The number of
eigenvalue shown $k$ depends on the network, and is chosen by KONECT
such as to result in reasonable runtimes for the decomposition
algorithms.

\begin{figure}
  \centering \subfigure[Top-$k$ eigenvalues of $\mathbf A$]{
    \includegraphics[width=\wPlot]{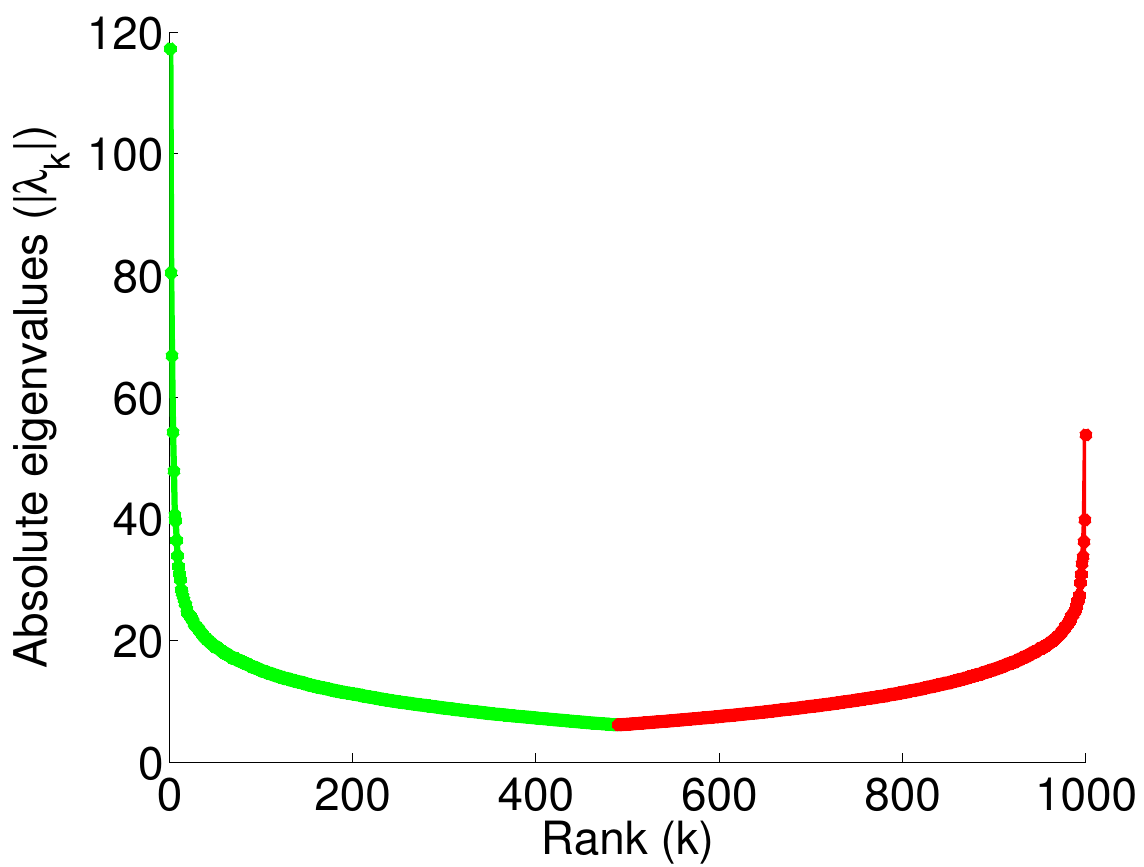}}
  \subfigure[Cumulative eigenvalue distribution of $\mathbf N$]{
    \includegraphics[width=\wPlot]{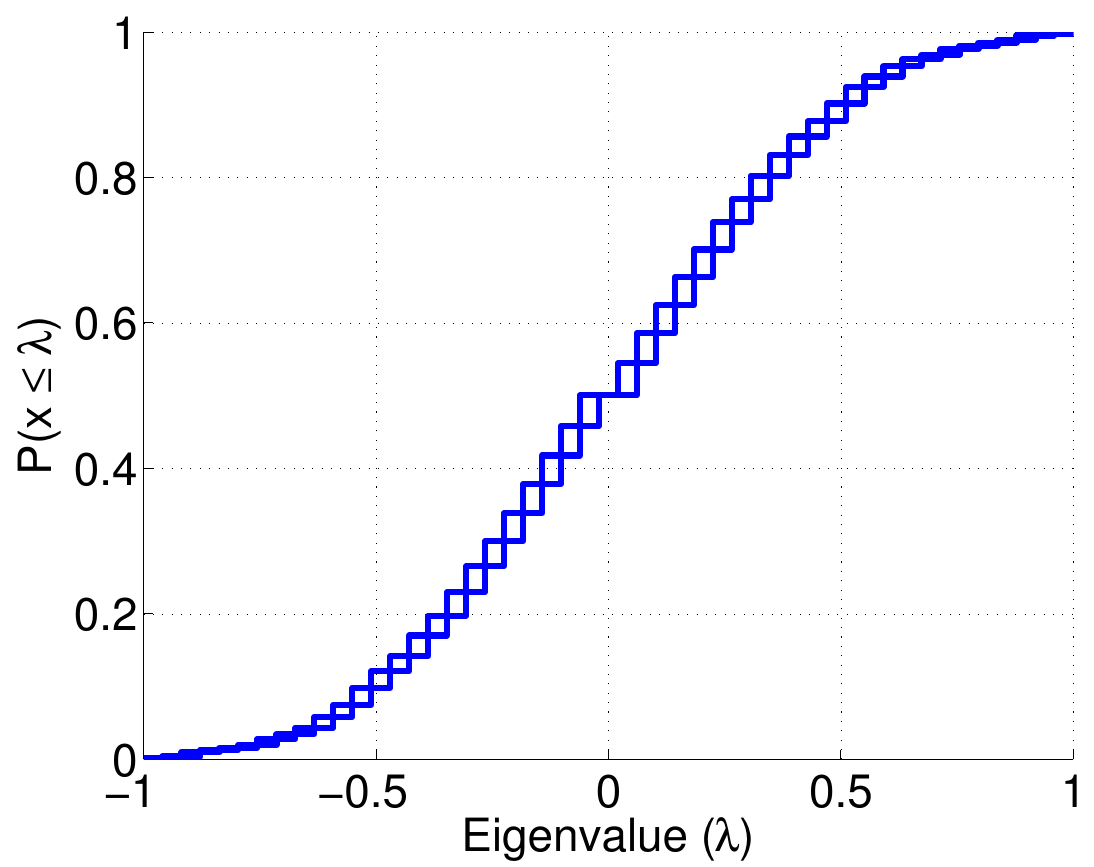}}
  \caption{ The top-$k$ eigenvalues of $\mathbf A$ and the cumulative
    spectral distribution of $\mathbf N$ for the Wikipedia election
    network
    (\href{http://konect.cc/networks/elec/}{\textsf{EL}}).
    In the first plot~(a), positive eigenvalues are shown in green and
    negative ones in red.
    \label{fig:spectrum}
  }
\end{figure}

Two plots are generated: the non-cumulative eigenvalue distribution, and
the cumulative eigenvalue distribution.  For the non-cumulative
distribution, the absolute $\lambda_i$ are shown in function of $i$ for
$1 \leq i \leq k$.  The sign of eigenvalues (positive and negative) is
shown by the color of the points (green and red).  For the cumulated
eigenvalue plots, the range of all eigenvalues is computed, divided into
49 bins (an odd number to avoid a bin limit at zero for the matrix
$\mathbf N$), and then the number of eigenvalues in each bin is
computed.  The result is plotted as a cumulated distribution plot, with
boxes indicating the uncertainty of the computation, due to the fact
that eigenvalues are not computed exactly, but only in bins.

\subsection{Complex Eigenvalues Plot}
The adjacency matrix of an undirected graph is symmetric and therefore
its eigenvalues are real.  For directed graphs however, the adjacency
matrix $\mathbf A$ is asymmetric, and in the general case its
eigenvalues are complex.  We thus plot, for directed graphs, the top-$k$
complex eigenvalues by absolute value of the adjacency matrix $\mathbf
A$.

\begin{figure}
  \centering \subfigure[Wikipedia elections]{
    \includegraphics[width=\wTwo]{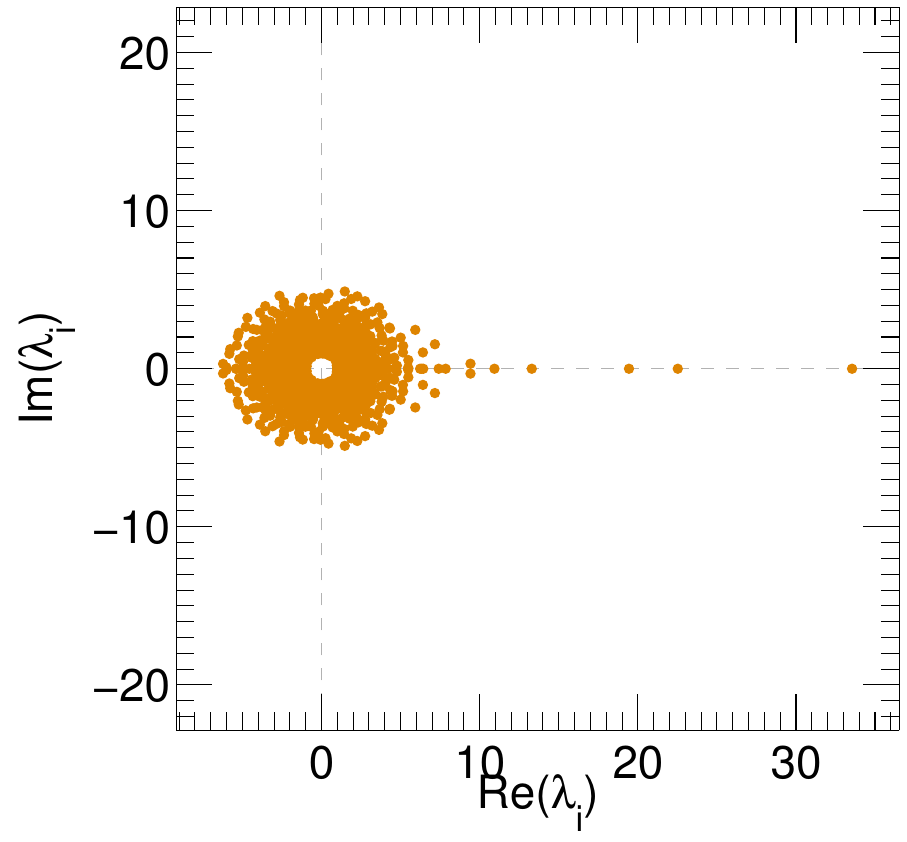}}
  \subfigure[UC Irvine messages]{
    \includegraphics[width=\wTwo]{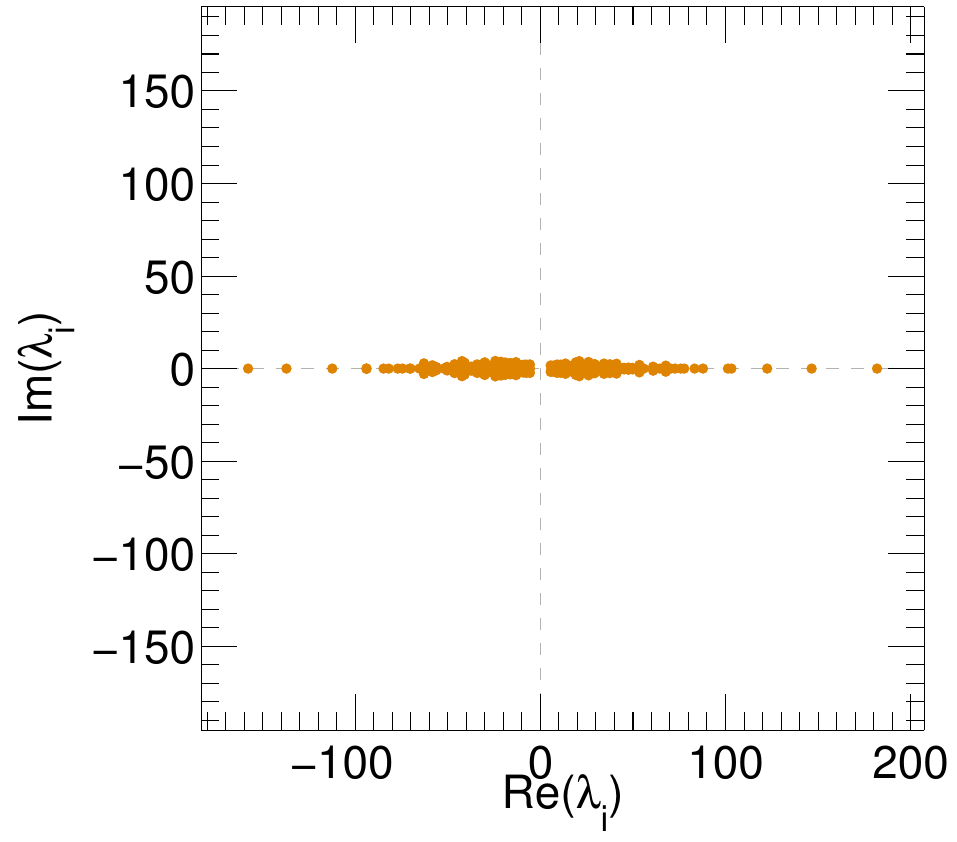}}
  \caption{ The top-$k$ complex eigenvalues $\lambda_i$ of the
    asymmetric adjacency matrix $\mathbf A$ of the directed Wikipedia
    election~(\href{http://konect.cc/networks/elec/}{EL}) and
    UC Irvine
    messages~(\href{http://konect.cc/networks/opsahl-ucsocial/}{UC})
    networks.
    \label{fig:complex}
  }
\end{figure}

Three properties can be read off the complex eigenvalues: whether a graph
is nearly acyclic, whether a graph is nearly symmetric, and whether a
graph is nearly bipartite.  If a  
directed graph is acyclic, its adjacency matrix is nilpotent and
therefore all its eigenvalues are zero. The complex eigenvalue plot can
therefore serve as a test for networks that are nearly acyclic: the
smaller the absolute value of the complex eigenvalues of a directed
graph, the nearer it is to being acyclic.  When a directed network is
symmetric, i.e., all directed edges come in pairs connecting two nodes
in opposite direction, then the adjacency matrix $\mathbf A$ is
symmetric and therefore all its eigenvalues are complex. Thus, a nearly
symmetric directed network has complex eigenvalues that are near the
real line.  
Finally, the eigenvalues of a bipartite graph are symmetric around the
imaginary axis.  In other words, if $a+bi$ is an eigenvalue, then so
is $-a+bi$ when the graph is bipartite.  Thus, the amount of symmetric
along the imaginary axis is an indicator for bipartivity. Note that
bipartivity here takes into account edge directions:  There must be two
groups such that all (or most) directed edges go from the first group to
second. 
Figure~\ref{fig:complex} shows two examples of such plots.

\subsection{Distance Distribution Plot}
Distance statistics can be visualized in the distance distribution plot.
The distance distribution plot shows, for each integer $k$, the number
of node pairs at distance $k$ from each other, divided by the total
number of node pairs.  The distance distribution plot is also called the
\emph{hop plot}.  
The distance distribution plot can be used to
read off the diameter, the median path length, and the 90-percentile
effective diameter (see Section~\ref{sec:distance-statistics}).  For
temporal networks, the distance distribution plot can be shown over
time.

The non-temporal distance distribution plot shows the cumulated distance
distribution function between all node pairs $(u,v)$ in the network,
including pairs of the form $(u,u)$, whose distance is zero.

The temporal distance distribution plot shows the same data in function
of time, with time on the X axis, and each colored curve representing
one distance value.

\begin{figure}
  \centering \subfigure[Distance distribution plot]{
    \includegraphics[width=\wPlot]{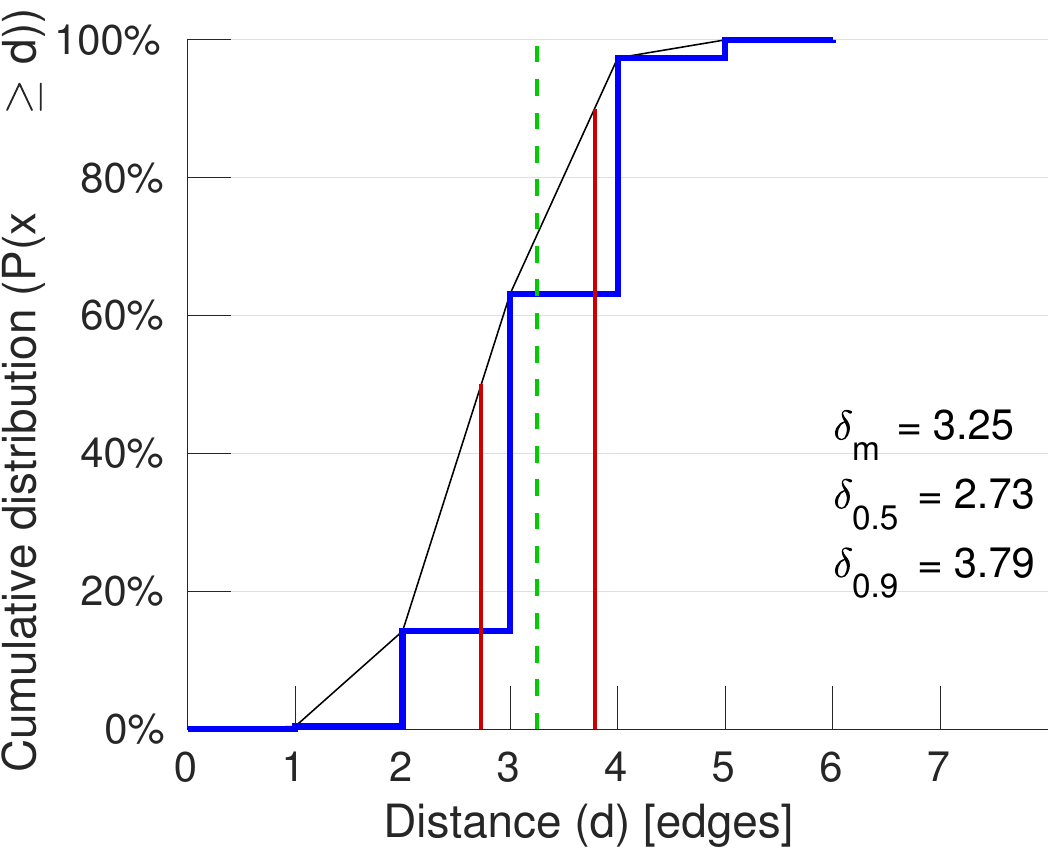}}
  \subfigure[Temporal distance distribution plot]{
    \includegraphics[width=\wPlot]{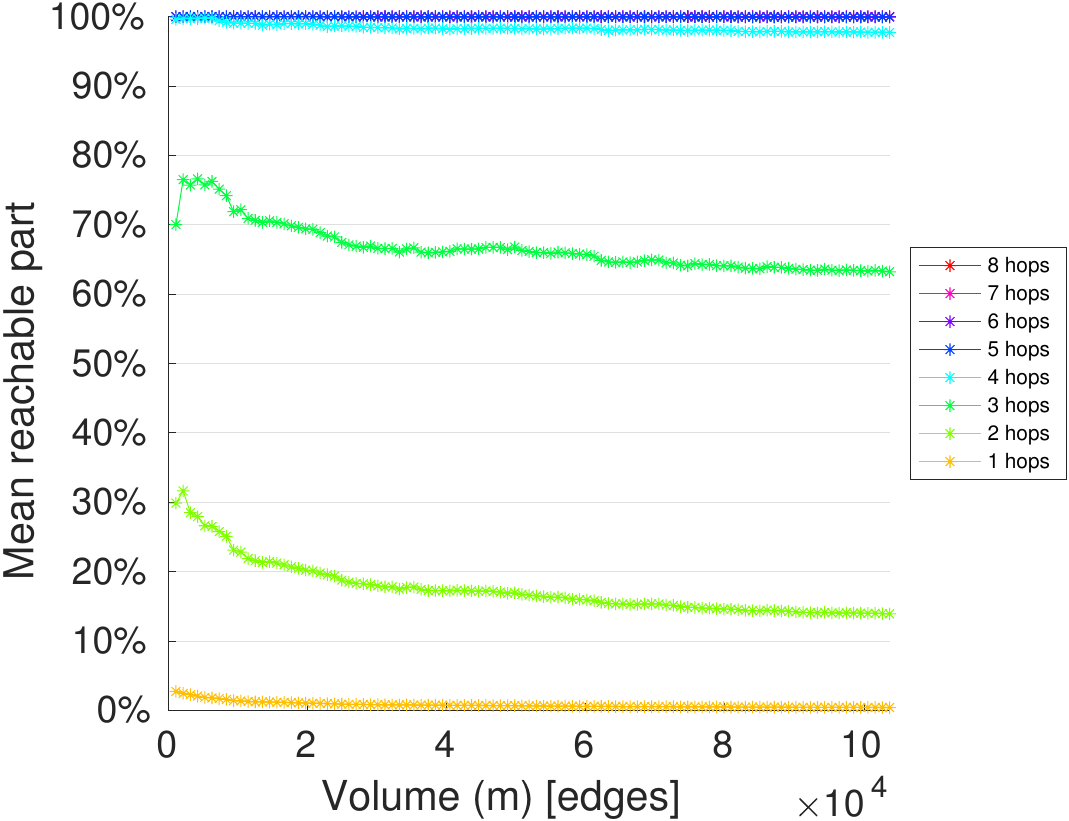}}
  \caption{ The distance distribution plot and temporal distance
    distribution plot of the Wikipedia election network
    (\href{http://konect.cc/networks/elec/}{\textsf{EL}}).
    \label{fig:hopdistr}
  }
\end{figure}

\subsection{Graph Drawings}
A graph drawing is a representation of a graph, showing its vertices and
egdes laid out in two (or three) dimensions in order for the graph
structure to become visible.  Graph drawings are easy to produce when a
graph is small, and become harder to generate and less useful when a
graph is larger.

Given a graph, a graph drawing can be specified by the placement of its
vertices in the plane.  To determine such a placement is a non-trivial
problem, for which many algortihms exist, depending on the required
properties of the drawing. For instance, each vertex should be placed
near to its neighbors, vertices should not be drawn to near to each
other, and edges should, if possible, not cross each other.  It is clear
that it is impossible to fulfill all these requirements at once, and
thus no best graph drawing exists.

In KONECT, we show drawings of small graphs only, such that vertices and
edges remain visible. The graph drawings in KONECT are spectral graph
drawings, i.e., they are based on the eigenvectors of characteric graph
matrices.  In particular, KONECT included graph drawings based on the
adjacency matrix $\mathbf A$, the normalized adjacency matrix $\mathbf
N$ and the Laplacian matrix $\mathbf L$ \citep{b405}. Let $\mathbf x$ and $\mathbf y$
be the two chosen eigenvector of each matrix, then the coordinate of the
node $u\in V$ is given by $\mathbf x_u$ and $\mathbf y_u$.

\begin{figure}
  \centering \subfigure[Adjacency matrix~$\mathbf A$]{
    \includegraphics[width=\wFour]{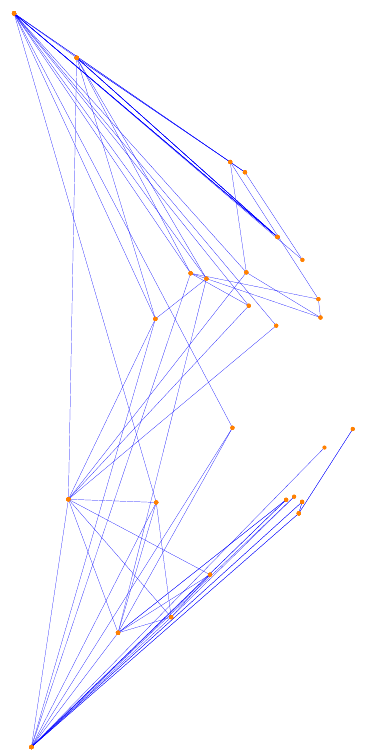}}
  \subfigure[Normalized adjacency matrix~$\mathbf N$]{
    \includegraphics[width=\wFour]{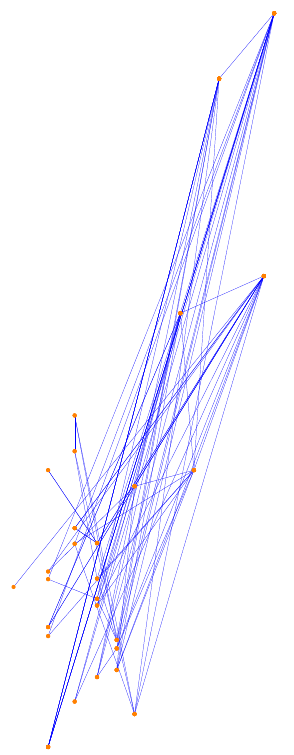}}
  \subfigure[Laplacian $\mathbf L$]{
    \includegraphics[width=\wFour]{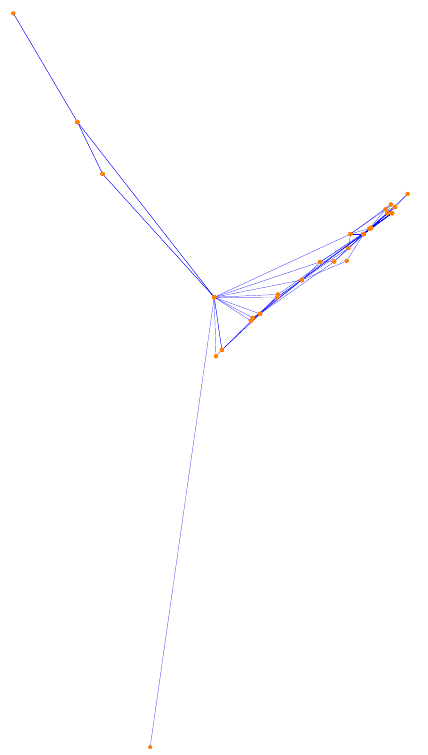}}
  \caption{ Drawings of the Zachary karate club social network
    (\href{http://konect.cc/networks/ucidata-zachary/}{\textsf{ZA}})
    using (a)~the adjacency matrix~$\mathbf A$, (b)~the normalized
    adjacency matrix~$\mathbf N$, (c)~the Laplacian matrix~$\mathbf L$.
    \label{fig:map.ax}
  }
\end{figure}

For the adjacency matrix $\mathbf A$ and the normalized adjacency matrix
$\mathbf N$, we use the two eigenvector with largest absolute eigevalue.
For the Laplacian matrix $\mathbf L$, we use the two eigenvectors with
smallest nonzero eigenvalue.  Examples for the Zachary karate club
social network
(\href{http://konect.cc/networks/ucidata-zachary/}{\textsf{ZA}})
are shown in Figure~\ref{fig:map.ax}.

\section{Other Definitions}
\label{sec:other}
This section contains other definitions used in KONECT. 

\subsection{Node Features}
\label{sec:features}
A feature is a numerical characteristic of a node, such as the degree
and the eccentricity.  Features have multiple uses, such as to measure
the centrality or the influence of a node in a network. 

The degree is defined as the number 
\marginpar{\texttt{degree}}
of neighbors of a node.  In directed networks, we can distinguish the
indegree, the outdegree and the degree difference (indegree minus
outdegree, notes \texttt{degreediff}). 

Certain features are spectral, i.e., they are defined as the
eigenvectors of certain matrices.  For instance, the PageRank vector
\marginpar{\texttt{pagerank}} is defined as the dominant eigenvector of
the matrix $\mathbf G = (1-\alpha) \mathbf P + \alpha\mathbf 1$, and the
eigenvector centrality as the component in the dominant eigenvector of
the adjacency matrix $\mathbf A$ \citep{todo-from-book}. 

The local clustering coefficients give the clustering coefficient
distribution \marginpar{\texttt{cluscod}} \citep{b865}. 

Other node features include:
\begin{itemize}
\item Average degree of neighbors
\item Betweenness centrality
\item $k$-Core number:  largest $k$ such that a node is still part of
  the graph's $k$-core.  The distribution of $k$-core values over all
  nodes is then the $k$-core distribution, which can also be seen as the
  size of the $k$-core for all $k \leq 0$. 
\end{itemize}

\subsection{Measures}
What is called a \emph{measure} in KONECT is a measure of link
prediction accuracy. 

\begin{itemize}
\item The \textbf{average precision}.  Measured over the whole test
  \marginpar{\texttt{ap}}
  set.  The best value is 1; the worst value depends on the ratio of
  true and false elements in the test set (for equally many, it is
  $1-\ln 2$.  When a proportion $p$ of pairs in
  the test set is true, then the average precision of a random guess is
  $p$. 
\item The \textbf{mean average precision}.  This is the average
  \marginpar{\texttt{map}}
  precision measured for each node separately, and then averaged over
  all nodes.  The value is between 0 and 1. This is very slow to
  compute. 
\item The \textbf{area under the curve} \citep{b366}.  This is the area under
  \marginpar{\texttt{auc}}
  the ROC curve.  The value is between 0 and 1. The area under the curve
  of a random is always $\frac 12$, regardless of the test
  set true/false ratio.  This is very similar to the average precision,
  but can be used when the relative size of the true/false test sets is
  different from run to run. 
\item The \textbf{mean area under the curve}.  This is the area
  \marginpar{\texttt{mauc}}
  under the curve computed for each node, and then averaged over all
  nodes. This is very slow to compute. 
\item The \textbf{Pearson correlation coefficient}.  This is probably the
  \marginpar{\texttt{corr}}
  fastest to compute, but is not justified by any model.  We used it
  successfully in an early paper
  \citep{kunegis:spectral-transformation}. 
\item The \textbf{Kendall rank correlation coefficient}.  Very slow to compute
  \marginpar{\texttt{kendall}}
  and not justified by any model. 
\item \textbf{Spearman's rank correlation coefficient}. 
  \marginpar{\texttt{spear}}
  Very slow to compute and not
  justified by any model.
\end{itemize}

Recommendations:  We performing machine learning, use \texttt{auc}.  When
evaluating a recommender system, use \texttt{mauc}.
If \texttt{mauc} is too slow, use \texttt{auc} instead. 

\section{The KONECT Toolbox}
\label{sec:toolbox}
The KONECT
Toolbox\footnote{\href{https://github.com/kunegis/konect-toolbox}{https://github.com/kunegis/konect-toolbox}}
for Matlab is a set of functions for the Matlab programming
language\footnote{\href{http://www.mathworks.com/products/matlab/}{www.mathworks.com/products/matlab}}
containing implementations of statistics, plots and other network
analysis methods.  The KONECT Toolbox is used to generate the numerical
statistics and plots in this handbook as well as on the KONECT website.

\paragraph{Installation}
The KONECT Toolbox is provided as a directory containing *.m files.  The
directory can be added to the Matlab path using addpath() to be used.

\paragraph{Usage}
All functions have names beginning with \texttt{konect\_}.

\subsection{Examples}
This section gives short example for using the toolbox.  The examples
can be executed in Matlab. 

\paragraph{Load a unipartite dataset}
This example loads the Slashdot signed social network. 

\begin{verbatim}
T = load('out.slashdot-zoo');
n = max(max(T(:,1:2)));
A = sparse(T(:,1), T(:,2), T(:,3), n, n); 
\end{verbatim}

This loads the weighted adjacency matrix of the Slashdot Zoo into the
matrix \texttt{A}. 

\subsection{Variables}
Naming variables can be quite complicated and hard to read in
Octave or MAtlab. Therefore KONECT code follows specific rules as laid
out in this section.  As a general rule, variables with the same name
always denote the same thing.  Thus a variable \texttt{network} will
always be a string denoting the internal name of a network (such as
\texttt{ucidata-zachary} for the famous Karate Club), and a
variable \texttt{statistic} will always be a string representing the
internal name of a statistic (such as \texttt{clusco} for the clustering
coefficient).   Likewise, \texttt{value} will always be a numerical
variable denoting the value of a statistic (such as 0.234 for a
clustering coefficient). 

Long variable names (containing full words) are in all-lowercase. Words
are separated by underscore.  When refering to a variable in comments,
the variable is written in all-uppercase.  Short variable names
(letters) are lowercase for numbers and vectors, and uppercase for
matrices.

\subsubsection{Strings}
Table~\ref{tab:string-variables} shows common variable names used for
string variables.
As a general rules, strings as used for what we call \emph{internal
  names}.  These are canonical names given to individual networks,
matrix decompositions, statistics, error measures, etc.  These internal
names are stable:  We never change them, because that will break
things.  As a result, many internal names may appear inaccurate,
inconsistent, or may simply look like having bad typography.  For
instance, some variants of the diameter statistic include the name
\texttt{diam}, while others include the name \texttt{diameter}.  This is
unfortunate, but backward compatibility is tantamount here.  In
``user-facing'' material such as the website and papers, we use pretty
names for all of these, which we can change as we want. 

\begin{table}
  \caption{ 
    Long variable names of string type used in KONECT.
    \label{tab:string-variables}
  } 
  \centering
  \begin{tabular}{>{\ttfamily}lp{0.7\textwidth}}
    \toprule 
    network & The internal network name, e.g., ``advogato''.
    The internal network name is used in the names of files related to
    the network.  These names do not contain periods, slashes, spaces,
    or colons, but may otherwise contain dashes, spaces, etc.  \\ 
    class &
    The internal name for a set of networks, e.g., ``test'', ``1'',
    ``2'', ``3''. The class ``N'' includes the $10\times N$ smallest
    networks. \\ 
    code & The 1/2/3-character code for a network, e.g.,
    ``EN'' for Enron.  \\ 
    curve & The internal name of a curve fitting method. \\ 
    decomposition & The internal of a matrix decomposition, as passed to the function
    \texttt{konect\_decomposition()}, e.g., ``sym'', ``asym'' and ``lap''. \\ 
    feature & The internal name of a
    feature, e.g., ``degree'' and ``decomp.sym''.  \\ 
    filename & A filename. \\ 
    format & The network format in lower case as defined in
    the function \texttt{konect\_consts()}, e.g., ``sym'' and ``bip''. \\ 
    label &
    The readable name of things used in plots, tables, etc.  \\ 
    measure & The internal name of a measure
    of link prediction accuracy, e.g., ``map'' and ``auc''. \\ 
    method & The internal name of a
    link prediction method.  \\ 
    statistic & The internal name of a network
    statistic, e.g., ``power'' and ``alcon''.  \\ 
    transform & The name of a transform, e.g. ``simple'' and ``lcc''. \\
    type & The internal name of
    the computation type. This can be ``split'' or ``full''. This
    decides which version of a network gets used, in particular for
    time-dependent analyses.  \\ 
    weights & The edge weight type as defined in the function
    \texttt{konect\_consts()}, e.g., ``unweighted'' and
    ``signed''. \\ 

    \bottomrule
\end{tabular}
\end{table}

\subsubsection{Scalars}
Table~\ref{tab:scalar-variables} shows variable names used for scalar
values.  
As a general rule, we try to use individual lowercase latin letters for
scalar variables, in keeping with Octave/Matlab usage.  This aligns
nicely with the mathematical aspects of KONECT, but can lead to
ambiguities, and there we also use longer variable names in certain
cases. 

\begin{table}
  \caption{ Variable names used for scalars in KONECT.
    \label{tab:scalar-variables}
  } \centering
  \begin{tabular}{>{\ttfamily}lp{0.7\textwidth}}
    \toprule n, n1, n2 & Row/column count in matrices, left/right vertex
    count \\ r & Rank of a decomposition \\ m & Edge count \\ i, j &
    Vertices as integer, i.e., indexes in rows and
    columns. \\ prediction & A link prediction score, i.e., a value
    returned by a link prediction algorithm for a given node
    pair. \\ precision & The prediction accuracy value, typically
    between 0 and 1.  \\ means & Values used for additive
    (de)normalization, as a structure. \\ \bottomrule
  \end{tabular}
\end{table}

\subsubsection{Matrices}
Table~\ref{tab:matrix-variables} shows variable names used for
matrix-valued variables. 
As a general rules, matrices have uppercase names consisting of
individual letters, as is the usage in Octave/Matlab, and in
mathematics. 

Note that when the adjacency matrix of an undirected graph is stored in
a variable, each edge is usually stored just once, instead of twice. In
other words, the variable \texttt{A} for undirected networks does not
equal the matrix $\mathbf A$, instead the expression \texttt{A + A'}
does.

\begin{table}[h]
  \caption{ 
    Variable names used for matrices and vectors in KONECT.
    As a general rule, matrices have upper-case names and vectors have
    lower-case names. 
    \label{tab:matrix-variables}
  } 
  \centering
  \begin{tabular}{>{\ttfamily}lp{0.7\textwidth}}
    \toprule 
    A & ($n \times n$) Adjacency matrix (in code where the
    adjacency and biadjacency matrix are distinguished) \\ 
    A & ($n \times n$ or $n_1 \times n_2$) Adjacency or biadjacency matrix (in
    code where the two are not distinguished) \\ 
    B & ($n_1 \times n_2$) Biadjacency matrix
    (in code where the adjacency and biadjacency matrix are
    distinguished) \\ 
    D & ($r \times r$) Central matrix; e.g.,
    eigenvalues; as matrix \\ 
    dd & ($r \times 1$) Diagonal of the
    central matrix \\ 
    E & ($e \times 2$) Test set for link prediction, stored in the same
    way as \texttt{T} \\
    L & ($n \times n$) Laplacian matrix \\ 
    M, N &
    Normalized (bi)adjacency matrix \\ 
    T & ($m \times 2$ or
    $m \times 3$ or $m \times 4$) Compact adjacency matrix, as stored in
    \texttt{out.*} files, and such that it can be converted to a sparse
    matrix using \texttt{konect\_spconvert()}. \\ 
    & First column: row
    IDs \\ 
    & Second column: column IDs \\ 
    & Third column (optional):
    edge weights (1 if not present) \\ 
    & Fourth column (optional):
    timestamps in Unix time \\ 
    U & ($n \times r$ or $n_1 \times
    r$) Left part of decomposition; e.g., left eigenvectors \\ 
    V & ($n
    \times r$ or $n_2 \times r$) Right part of decomposition; e.g.,
    right eigenvectors \\ 
    X & ($r \times r$) Central matrix, when
    explicitly nondiagonal \\ 
    Z & ($n \times n$) Normalized Laplacian
    matrix \\ 
    \bottomrule
\end{tabular}
\end{table}

\subsubsection{Compound Types}

A struct containing elements whose names are of a specific type are
named \texttt{[VALUETYPE]s\_[KEYTYPE]}.  For instance, a struct with
labels used for methods is named as follows:

\begin{verbatim}
labels_method.('auc') = 'Area under the curve';
\end{verbatim}

Note:
\begin{itemize}
\item The first element is the name of the content type.
\item The plural is used only for the content type.
\end{itemize}

\subsubsection{IDs}

Variables named \texttt{method}, \texttt{decomposition}, etc.\ are
always strings.  If a method, decomposition or any other type is
represented as an integer (e.g., as an index into an array), then
\texttt{\_id} is appended to the variable name. For instance:

\begin{verbatim}
decomposition = 'sym'; decomposition_id = 2;
\end{verbatim}

This means that an array of values by ID of keys is called for instance:

\begin{verbatim}
labels_decomposition_id{1} = 'Eigenvalue decomposition';
labels_decomposition_id{2} = 'Singular value decomposition';
\end{verbatim}

\section{File Formats}
\label{sec:format}
Due to the ubiquity of networks in many areas, there are a large number
of file formats for storing graphs and graph-like structures.  Some of
these are well-suited for accessibility from many different programming
languages (mostly line-oriented text formats), some are well-suited for
integration with other formats (semantic formats such as RDF and
XML-based ones), while other formats are optimized for efficient access
(binary formats).  In KONECT, we thus use three file formats covering
the three cases:
\begin{itemize}
\item TSV format: This format is text-based and uses tab- (or space-) separated 
  values.  This is the main KONECT data format from which the two others
  are derived.  The format has the advantage that it can be read easily
  from many different programming languages and environments.
\item RDF format: Datasets are also available as RDF, as text-based
  format used in the \emph{Semantic Web} community.  
  This format was deprecated as it did not have any users, and the files
  were generally an order of magnitude larger than other formats. 
\item Matlab format: To compute statistics and plots and perform
  experiments, we use Matlab's own binary format (version 7.3), which can be accessed
  efficiently from within Matlab.
\end{itemize}

\subsection{TSV Format}
In the following, we describe KONECT's TSV format.  Each network
\texttt{\$network} is represented by the file \texttt{out.\$network}. 

The edges are stored as tab-separated values (TSV).  The file is a text
file, and each line contains information about one edge.  Each line
contains two, three or four numbers represented textually, and separated
by any sequence of whitespace.  The preferred separator is a single tab (\texttt{\textbackslash{t}}).
The first two columns are mandatory and contain the source and
destination node ID of the edge.  The third column is optional and
contains the edge weight.  When the network is dynamic, the third column
contains $+1$ for added edges and $-1$ for removed edges.  For
unweighted, non-temporal networks, multiple edges may be aggregated into
a single line containing, in the third column, the number of aggregated
edges.  The fourth column is optional and contains the edge creation
time, and is stored as UNIX time, i.e., the number of seconds since 1
January 1970.  The fourth column is usually an integer, but may contain
floating point numbers.  If the fourth column is present, the third
column must also be given, and may be \texttt{1} if it is otherwise not
needed.  The beginning of the file contains  
additional comment lines with the following information:
\begin{verbatim}
   % FORMAT WEIGHTS 
   % RELATIONSHIP-COUNT SUBJECT-COUNT OBJECT-COUNT
\end{verbatim}
where \texttt{FORMAT} is the internal name for the format as given
in Table~\ref{tab:format}, \texttt{WEIGHTS} is the internal name for
the weight types as given in Table~\ref{tab:weights},
\texttt{RELATIONSHIP-COUNT} is the number of data lines in the file,
and \texttt{SUBJECT-COUNT} and \texttt{OBJECT-COUNT} both equal the
number of nodes $n$ in unipartite networks, and the number of left
and right nodes $n_1$ and $n_2$ in bipartite networks.  The first
line is mandatory; the second line is optional.

\subsection{Meta File}
In addition to the data files (TSV, etc.), each network in KONECT has a
file \texttt{meta.\$network} associated with it.  This is a short text
file that contains all information about the dataset that are not
structural.  In other words, all structural information is stored in the
TSV and other data formats, while meta information is stored in this
file. 
The file contains metadata about the
network that is independent of the mathematical structure of the
network.  
The file is a text file coded in UTF-8.  Each line
contains one key/value pair, written as the key, a colon and the
value.  Whitespace is allowed everywhere, and is ignored.  The following metadata fields are used:
\begin{itemize}
\item \texttt{name}: (obligatory) The name of the dataset.  This contains only the
  name of the source, without description the type or category,
  e.g., ``YouTube'', ``Wikipedia elections'').  The name uses
  sentence case.
  When the name refers to the name of entities represented by the
  dataset, we use the plural.  
  For instance, a dataset of relationships between monkeys should be named ``Monkeys'' rather than ``Monkey'',
  but note that many animal names are invariable in English.  
  Note also that words such as ``trust'' do no refer to individual
  edges, but to the \emph{type} of edge, and thus the plural is not used
  for them.  Thus, a dataset may be named ``MovieLens trust'' rather
  than ``MovieLens trusts''. 
  For networks with the same name, the source (e.g.,
  the conference or author name) is added in parentheses.  Within each category,
  all names must be distinct, but multiple networks can have the same
  \texttt{name} if they have different categories.  We use the
  conference name for disambiguation if the dataset is recent, and the
  author name if it is older; this is however just a rule of thumb.  The
  year number is used to disambiguate multiple versions of a dataset. 
\item \texttt{code}: (obligatory) The short code used in plots and narrow
  tables.  The code consists of two or three alphanumeric characters.  The first
  two characters are usually uppercase letters and denote the data
  source.  The last character, if present, usually distinguishes
  the different networks from one source.  The code is unique across all
  of KONECT, although it is not an error if they are not:  It will only
  lead to certain plots being confusing.  As KONECT grows, codes have become
  less and less unique; we may thus allow identical codes in the
  future. 
\item \texttt{category}: (obligatory) The internal name of the category, as given in the
  column ``Internal name'' in Table~\ref{tab:categories}.
  It is important that this field is present, and corresponds to a valid
  category. 
\item \texttt{long-description}: (optional, recommended) A long descriptive
  text consisting of full sentences, and describing the dataset in
  a verbose way.  HTML markup may be used sparingly (tags: \texttt{<I>}, \texttt{<CODE>},
  etc.), usually only for absolutely necessary typography, such as
  setting species names in italics.  Hyperlinks are not used, but URLs
  may be shown insode a \texttt{<CODE>} tags. 
\item \texttt{entity-names}: (optional, recommended) A comma-seperated list of
  entity names (e.g., ``user, movie'' or ``protein'').  Unipartite networks give
  a single name; bipartite networks give two.
  The names are in the singular and in lowercase. 
  We usually follow the nomenclature given by the data sources in naming
  these. 
\item \texttt{relationship-names}: (optional, recommended) The name of the
  relationship represented by edges, as a lowercase substantive in the singular (e.g.,
  ``friendship'', ``road'').
  Again, we usually follow the nomenclature given by the data sources in naming
  these.  The substantive may be a phrase, i.e., contain spaces (e.g., 
  ``tag assignment'', ``metabolic reaction'').  If there are multiple
  relationship types not otherwise distinguished by the dataset, they  
  are separated by a slash (e.g., ``answer/comment'').
\item \texttt{extr}: (optional, recommended) The name of the
  subdirectory under the directory \texttt{konect-extr/extr/} that
  contains the extraction code for this dataset in the KONECT
  Extraction package.\footnote{See
    \href{https://github.com/kunegis/konect-extr/tree/master/extr}{https://github.com/kunegis/konect-extr/tree/master/extr}} 
  If not given, it usually means that the dataset does not have
  extraction code, which is only the case with very old datasets.  These
  were added to KONECT by hand without making the extraction
  reproducible; Stu was a long way off then. 
\item \texttt{url}: (optional, recommended for almost all networks) The
  URL(s) of the data sources, as a comma 
  separated list.  Most datasets have a single URL.  
  The field is not present when datasets where given to us privately. 
\item \texttt{description}: (deprecated) A short description of the form
  ``User–movie ratings''.  Note that the file should contain an
  actual en dash, coded in UTF-8.
  The field is now deprecated as this information is now covered by the
  \texttt{entity-names} and \texttt{relationship-names} fields. 
\item \texttt{cite}: (optional, recommended for almost all networks) The
  bibtex key of the citation(s) for this dataset, as a  
  comma separated list.  The given bibtex keys must correspond to the
  publications listed in the file \texttt{konect.bib} in the KONECT-Extr
  package. 
  Most datasets have a single bibtex entry.  This
  is usually the actual paper about the dataset.  Many researchers have
  a preference about which of their papers should be cited for a dataset
  they have released, and we follow these. 
\item \texttt{timeiso}: (optional) A single ISO timestamp denoting
  the date of the dataset or two timestamps separated by a
  slash (/) for a time range. The format is:
  \texttt{YYYY[-MM[-DD]][/YYYY[-MM[-DD]]]}, e.g., \texttt{2005-10-08/2006-11-03}
  or \texttt{2007}.
  In general, this should denote when the actual edges where created,
  but it often denotes only the date of data aggregation by the source
  from which KONECT gets the data.  It does not denote the date at which
  the data is added to KONECT. 
\item \texttt{tags}: (optional) A space-separated list of hashtags
  describing the network.  The list of tags is given in
  Section~\ref{sec:tags}.  
  Properly chosen tags are important to define the unit tests that
  KONECT will perform, in order to detect erroneously detected
  datasets. 
\item \texttt{fullname}: (optional, semi-deprecated) A longer name to disambiguate
  different datasets from the same source, e.g., ``Youtube
  ratings'' and ``Youtube friendships''.  Uses sentence case.  These
  names are unique across KONECT, falling back to the \texttt{name} when
  the fullname is not given.  This field is only recommended when the
  \texttt{name} field is not unique. 
\item \texttt{n3-*}: (optional, deprecated) Metadata which is used for the
  generation of RDF files. The symbol \texttt{\{n\}} in the name
  of the meta key represents an order by unique, sequential
  numbers starting at one.  These are deprecated. 
  \begin{itemize}
  \item \texttt{n3-add-prefix\{n\}} (optional):
    Used to define additional N3 prefixes. The
    default prefixes are specified in this way.
  \item \texttt{n3-comment-\{n\}} (optional): Add
    commentary lines which are placed at the
    beginning of the N3 file.
  \item \texttt{n3-edgedata-\{n\}} (optional):
    Additional N3-data, to be displayed with each
    edge.
  \item \texttt{n3-nodedata-m-\{n\}} (optional):
    Additional N3-data, to be displayed with the
    first occurence of the source ID.
  \item \texttt{n3-nodedata-n-\{n\}} (optional):
    Additional N3-data, to be displayed with the
    first occurence of the target ID.
  \item \texttt{n3-prefix-m}: N3-prefix for the
    source IDs.
  \item \texttt{n3-prefix-n} (optional): N3-prefix
    for the target IDs. If this field is left out,
    the value of \{n3-prefix-m\} is used.
  \item \texttt{n3-prefix-j} (optional):
    Additional prefix which can be used with the
    source id, if there is an entity to be
    represented with the same id.
  \item \texttt{n3-prefix-k} (optional):
    Additional prefix which can be used with the
    target id, if there is an entity to be
    represented with the same id. This is used for
    example in meta.facebook-wosn-wall for the
    representation of users walls.
  \item \texttt{n3-prefix-l} (optional): N3-prefix
    for the edges, if they are to be represented
    by some N3-entity.
  \item \texttt{n3-type-l} (optional): RDF-type
    for the edges.
  \item \texttt{n3-type-m}: RDF-type for source
    IDs.
  \item \texttt{n3-type-n} (optional): RDF-type
    for target IDs.
  \end{itemize}
  The following fields are used in the n3-expressions for
  edgedata and nodedata:
  \begin{itemize}
  \item[\texttt{\$m}]: n3-prefix-m + source ID
  \item[\texttt{\$n}]: n3-prefix-n (or n3-prefix-m
    if the other is undefined) + target ID
  \item[\texttt{\$j}]: source ID
  \item[\texttt{\$k}]: target ID
  \item[\texttt{\$l}]: edge ID
  \item[\texttt{\$timestamp}]: edge timestamp
  \end{itemize}
\end{itemize}

\subsection{Semantic Data Format}
For datasets that contain multiple relationship types, we use a more
elaborative scheme.  This scheme is not used systematically in KONECT,
because the scope of KONECT is to handle each network separately.
However, this scheme is used by certain families of datasets, and in
particular by the extraction code. 

This data format is semantic in the sense that it is able to represent
almost all types of data in a structured fashion.  It should not be
confused with \emph{semantic web} formats.  This format is close in
spirit to relational databases, only that it systematically
distinguishes between entity types and relationship types. 

As all KONECT data formats, the goal is to enable batch processing, and
to allow reading and writing from many different programming languages,
and to allow multiple programming languages and libraries to be composed
easily.  This goes at the price of compactness (no binary
representation) and updatability (all IDs are continuous). 

In this format, a dataset consists of an arbitrary number of entities
types and relationship types.  For each entity types, all entities have
an integer ID ranging from one
to the number of entities of that type.  All information about entities
(i.e., what would be contained in the corresponding entity table in a
relational database) is contained in individual text files, separated by
entity type and attribute.  All information about relationships is
stored in text files with one line per relationship, with one file per
relationship type.  

In the following, \texttt{\$network} is the internal name of the semantic
network. 
For one semantic network, there are multiple \texttt{ent.*} files, and
multiple \texttt{rel.*} files, as described in the following. 

\paragraph{Entities}

For each attribute \texttt{\$att} of the entity type \texttt{\$ent}
related to the relationship type \texttt{\$rel}, there is one file named 
\texttt{ent.\$network\_\$rel[\_\$rel]\_\$ent\_\$att}.  This is not strictly
semantic, as the relationship type name should not appear. However, this
is done in order to facilitate conversion to individual datasets.

The \texttt{\$rel} string is present twice for files in \texttt{extr/} and only once for files in \texttt{\{uni/,dat/\}}.

The entity attributes file contain one line per entity, and do
\emph{not} contain the entity IDs -- these are implicit in the line
numbers.  In addition, there is a header:

\begin{verbatim}
   % <empty>
   [% COUNT]
\end{verbatim}

\texttt{COUNT} represents the biggest ID, and may be omitted.

An empty line indicates that there is no data available for this specific entity.
There is no need to quote or escape characters as there is always only one attribute per line.

\paragraph{Relationships}

For each relationship type \texttt{\$rel}, there is a file named
\texttt{rel.\$network\_\$rel}. 

The first line of the file is a comment line that defines the columns: 

\texttt{\% ent.NAME\_A ent.NAME\_B [WEIGHT\_TYPE.NAME] dat.NAME*}

\texttt{WEIGHT\_TYPE} can be:
\begin{itemize}
  \item \texttt{weight}: an unspecified type
  \item \texttt{double}, \texttt{float}, \texttt{int}, \texttt{short}, \texttt{byte}: the corresponding type as in
    C, except that \texttt{char} is written as \texttt{byte}.  (This
    last convention is influenced by the Java programming language.)
\end{itemize}

Data can be integers, floating point numbers or strings.  Only numbers
with well-defined numerical semantics are stored as numbers (e.g., zip codes are
stored as strings). 

\section*{Acknowledgments}
The KONECT project would not have been possible without the
effort of many people who have made datasets available and also helped
the project in many different ways.   
KONECT is maintained by Jérôme Kunegis at the University of Namur, Belgium. 
In the past, KONECT was also maintained by Daniel Dünker, Holger Heinz,
and Martina Sekulla at the University of Koblenz--Landau in Koblenz, Germany. 

We owe much of the datasets in KONECT to researchers of many different
fields of science from all around the world who contributed datasets to
the public, and to KONECT in particular.  The list of contributors would
be too long to list, so instead we refer to the online list of datasets
for links to dataset sources and citations relevant to each
network.\footnote{\url{http://konect.cc/networks/}}

KONECT was also supported by funding from multiple research projects. 
The research leading to
these results has received funding from the European Community's Seventh
Frame Programme under grant agreements n\textsuperscript{o}~257859,
\href{http://robust-project.eu/}{ROBUST}, 287975,
\href{http://www.socialsensor.eu/}{SocialSensor}, and 610928,
\href{http://revealproject.eu/}{REVEAL}, as well as the project
\href{http://nouvelles.unamur.be/upnews.2015-10-01.8995593781}{IDEES --
  L'Internet de Demain pour développer les Entreprises, l'Économie et la
  Société} (ERDF/FEDER -- Wallonia/Wallonie). 

The picture of a pair of cherries in Figure~\ref{fig:cherries} was
created by the authors of Wikimedia Commons and is released under the
Creative Commons CC-SA~3.0 license.\footnote{\url{https://commons.wikimedia.org/wiki/File:Cherry_Stella444.jpg}}

\let\oldbibliography\thebibliography
\renewcommand{\thebibliography}[1]{%
  \oldbibliography{#1}%
  \setlength{\itemsep}{0pt}%
}
\bibliographystyle{agsm}
\bibliography{kunegis,ref,konect}

\appendix

\section{Glossary of Terms}
Some terms related to graph theory are well established in mathematics,
network theory and computer science, while other terms do not have a
widely-used definition.  
The choices made in this work are those of the authors, and were chosen
to reflect best practices and to avoid confusion.

\begin{description}
  \item[Adjacency matrix]
    The matrix describing a network, usually denoted $\mathbf A$.  To be
    contrasted with the 
    half-adjacency matrix (for undirected unipartite networks, also
    denoted $\mathbf A$) and the
    biadjacency matrix (for bipartite networks, denoted $\mathbf B$). 
    The adjacency matrix is always square, and for undirected networks
    it is symmetric. 
  \item[All] The term \emph{all} in KONECT is the name of a group used when a certain method
    is applied to all KONECT networks individually.  When a method is
    applied to all networks at once, we use \emph{everything}, which is
    a class rather than a group. 
  \item[Arc] A directed edge.  In general, we consider arcs to be a
    special cases of edges, and thus we rarely use the term \emph{arc}
    in favor of \emph{directed edge}.  (In other texts, an edge is taken
    to be undirected by definition, and the term \emph{directed edge} is
    then a contradiction.)
  \item[Asymmetric] A directed graph is called asymmetric if it contains
    at least one unreciprocated edge. 
  \item[Biadjacency matrix]
    The characteristic matrix of a bipartite network, usually denoted
    $\mathbf B$.  The corresponding adjacency matrix is then $[\mathbf
      0, \mathbf B; \mathbf B^{\mathrm T}, \mathbf 0]$. 
  \item[Category] Networks have a category, which describes the domain
    they apply to:  social networks, transport networks, citation
    networks, etc. 
  \item[Central matrix] The matrix $\mathbf X$ in any decomposition
    of the form $\mathbf U \mathbf X \mathbf V^{\mathrm T}$, not necessarily
    diagonal or symmetric; a generalization of the diagonal eigenvalue
    matrix.
  \item[Class]
    A class is a set of networks determined by the number of edges they
    contain. 
    The networks of KONECT are divided into classes by their volume:
    Class~1 contains the ten smallest networks, Class~2 contains the
    next ten smallest networks, etc.  The class named `everything' contains all
    networks, and the class named `test' contains a set of small
    representative networks that cover most important network types. 
  \item[Claw]
    Three edges sharing a single vertex.  A claw can be understood as a 3-star. 
  \item[Code]
    The two- or three-character code representation of a network.  These
    are used in scatter plots that show many networks.  
  \item[Cross]
    A pattern of four edges sharing a single endpoint.  Also called a
    4-star.  
  \item[Curve]
    A curve fitting method used for link prediction, when using the link
    prediction method described
    in \citep{kunegis:spectral-transformation}.
  \item[Cycle] 
    A cyclic sequence of connected edges, not containing any edge twice.
    A cycle contrasts with a tour, in which a single vertex can appear
    multiple times.  
  \item[Decomposition] In KONECT the word \emph{decomposition} is used
    to denote the combination of a characteristic graph matrix (e.g. the
    adjacency matrix or Laplacian) with a matrix decomposition.  As an
    extension, some other constructions are also called
    \emph{decomposition}, such as LDA.
  \item[Density] This word is avoided in KONECT.  In the literature, it
    may refer to either the fill (probability that an edge exists), or to
    the average degree.  The former definition is typically used in mathematical
    contexts, while the latter is used in computer science contexts.
  \item[Directed] A directed graph contains directed edges.  A directed
    graph can be symmetric (if all edges are reciprocated), or
    asymmetric (otherwise). 
  \item[Edge] A connection between two nodes.  In mathematics, an edge
    is undirected and constrasts with an arc which is directed.  In the
    context of KONECT, all types of connections between nodes are called
    \emph{edges} and an arc is a special case of an edge. 
  \item[Everything] In KONECT, \emph{everything} is the class of all
    networks.  It is used when a method is applied to the set of all
    networks as a whole.  When a method is applied to all networks
    individually, we use \emph{all}, which is a group. 
  \item[Family] A family is a set of networks that were generated by the
    same extraction code.  Each family corresponds to a directory under \href{https://github.com/kunegis/konect-extr/tree/master/extr}{\texttt{konect-extr/extr/}}.
  \item[Feature] A node feature. I.e., a number assigned to each node.
    Examples are the degree, PageRank and the eccentricity. 
    Equivalently, a node vector.
  \item[Fill] The probability that two randomly chosen nodes are
    connected.  Also called the \emph{density} or \emph{edge density}, in particular in a
    mathematical context.  The fill is the sole parameter of the
    Erdős--Rényi random graph model.  The word \emph{fill} is specific
    to KONECT. 
  \item[Format] 
    The format of a network determines its general structure, and
    whether edges are directed.  There are three possible formats:
    unipartite and undirected; unipartite and directed; and bipartite.
    Directed bipartite networks are not possible in KONECT.  
  \item[Group]
    A group of networks is a set of networks that have similar
    attributes, as defined by their metadata such as format, weights,
    timestamps, etc.  Examples are unipartite networks, network with
    multiple edges, and signed temporal networks.  Each group has an
    internal name in KONECT, which is in all-uppercase, with words
    separated by underscore.  See Section~\ref{sec:groups}.  The purpose
    of groups is to be able to specify the set of networks to which a
    statistic, plot or other type of analysis applies to.
  \item[Half-adjacency matrix]
    The adjacency matrix $\mathbf A$ of an undirected graph contains two
    nonzero entries for each edge $\{i,j\}$:  $\mathbf A_{ij}$ and
    $\mathbf A_{ji}$.  To avoid this, KONECT code uses the
    half-adjacency matrix, which contains only one of the two nonzero
    entries.  The half-adjacency matrix is therefore not unique, i.e.,
    it is unspecified whether $\mathbf A_{ij}$ or $\mathbf A_{ji}$ is nonzero.  In
    code, the half-adjacency matrix is denoted \texttt{A}.  The term
    \emph{half-adjacency matrix} is specific to KONECT, but the use of
    such a representation is widespread. 
  \item[Loop] 
    A loop is an edge that connects a node with itself.  A loop may be
    directed or undirected.  They are often called \emph{self-loops} in
    the literature, but that expression is quite redundant. 
  \item[Measure]
    A measure of the accuracy of link prediction methods, for instance
    the area under the curve or the mean average precision.  
  \item[Method]
    A link prediction method. 
  \item[Normalization]
    In KONECT, \emph{normalization} is defined in its broad sense, in
    which some quantity is, at some level, divided by another to result
    in quantities that can be compared.  This may be performed on the
    level of individual numbers, in which case we call it
    \emph{relativization}, but also at higher levels such as on entire
    matrices.  For instance, the matrix $\mathbf N = \mathbf D^{-\frac 1 2}
    \mathbf A \mathbf D^{-\frac 1 2}$ is called the normalized adjacency
    matrix.  Normalization and relativization can often be contrasted.
    As an example, the largest absolute eigenvalue of the normalized
    matrix $\mathbf N$ is always one while the relative largest
    eigenvalue of the non-normalized adjacency matrix would be $\alpha /
    n$, a value in general not equal to one.  
  \item[PageRank] 
    A node-based feature of a directed network, defined as the dominant
    eigenvector of the matrix $\mathbf G = (1-\alpha) \mathbf P +
    \alpha\mathbf 1$, with eigenvalue one. 
  \item[Path]
    A sequence of connected nodes, in which each node can appear only
    once.  The extension that allows multiple nodes is called a walk.  A
    path with identical start and end nodes is called a \emph{cycle}. 
  \item[Relative]
    The adjective \emph{relative} is applied to statistics when they are
    derived by dividing a given statistic by a reference value, often
    the maximally possible attainable value.  For instance, the relative
    size of the largest connected component equals the size of the
    largest connected component divided by the number of nodes.  Such
    statistics have values in the range $[0,1]$ and are useful for
    comparing networks of different sizes.  Relative statistics are a
    form of normalization. 
  \item[Score]
    A numerical value given to a node pair.  Usually used for link
    prediction, but can also measure distance or similary between
    nodes. 
  \item[Size]
    The number of nodes in a network.  
  \item[Statistic]
    A statistic is a numerical measure of a network, i.e., a number that
    describes a network, such as the clustering coefficient, the
    diameter or the algebraic connectivity.  All statistics are real
    numbers.   
  \item[Symmetric] A directed graph is called symmetric if all its edges
    are reciprocated.  Symmetric directed graphs are equivalent to
    undirected graphs, but have double the node degrees.  Thus,
    equations between the two are in general \emph{not} exchangeable.
    For graphs, \emph{symmetric} only applies to directed graphs. 
  \item[Tour]
    A cyclic sequence of connected nodes which may contain a single
    vertex multiple times.  It can be considered a walk that returns to
    it starting point, or a generalization of a cycle that allows to
    visit nodes multiple times.  
  \item[Trail] A trail is a walk that does not visit any edge twice.
    Compare to a path, which must not visit any node twice.  Every path
    is a trail and a walk, and every trail is a walk, but not the other
    way around. 
  \item[Transform] 
    A transform is an operation that applies to a graph and that gives
    another graph.  Examples are taking the largest connected component,
    removing multiple edges, and making a bipartite graph unipartite.
    Certain graph properties can be expressed as other graph properties
    applied to graph transforms.  For instance, the size of the largest
    connected component is the size of the transform which keeps only
    the largest connected component. 
  \item[Triangle]
    Three nodes all connected with each other.  The number of triangles
    in a network is a commonly used statistic, used for instance as
    the basis to compute the clustering coefficient.  Counting the
    triangles in a network is a very common computational problem.  
  \item[Volume]
    The number of edges in a network.  
  \item[Walk]
    A sequence of connected nodes, which may contain a single node
    multiple times.  The restriction to include a single node only once
    is called a path.  If the endpoints of a walk are identical, then
    the walk is also a tour.  
  \item[Wedge] 
    Two edges sharing a common node, i.e., two incident edges.  The
    number of wedges in a network is an important network statistic,
    which characterizes that skewness of the degree distribution, and
    which can be easily calculated.  A wedge can be seen as a 2-star or
    a 2-path.  
  \item[Weights] (always in the plural)
    The weights of a network describe the range of edge weights it
    allows.  The list of possible edge weights is given in
    Table~\ref{tab:weights}.  
\end{description}

\section{Glossary of Mathematical Symbols}
The following symbols are used in mathematical expessions throughout
KONECT.  Due to the large number of different measures used in graph
theory and network analysis, many common symbols for measures overlap.
For many measures, there is more than one commonly-used notation; the
notation chosen in KONECT represents a reasonable balance between using established
notation when it exists, and having distinct symbols for different
measures. 

In KONECT and literature written by its authors, matrices are written as
bold uppercase letters, and vectors as bold lowercase letters.
Individual numbers and sets are written in non-bold symbols.  \\

\begin{longtable}{ll}
  \toprule
  \textbf{Symbol} & \textbf{Meaning} \\
  \midrule
  \endfirsthead
  \toprule
  \textbf{Symbol} & \textbf{Meaning} \\
  \midrule
  \endhead
  \bottomrule
  \endfoot
  \bottomrule
  \endlastfoot
  $a$ & algebraic connectivity \\
  $b$ & non-bipartivity \\
  $c$ & global clustering coefficient \\
  $c(u)$ & local clustering coefficient \\
  $d$ & average degree \\
  $d(u)$ & degree of a vertex \\
  $d(u,v)$ & shortest-path distance \\
  $e$ & edge \\
  $g$ & line count, data volume \\
  $l$ & loop count \\
  $m$ & volume, edge count \\
  $\tilde{m}$ & average multiplicity \\
  $\bar{\bar m}$ & number of unique edges \\
  $n$ & size, node count \\
  $p$ & fill \\
  $q$ & square count \\
  $r$ & rank of a decomposition \\
  $r$ & rating value \\
  $r$ & radius of a graph \\
  $s$ & wedge count \\
  $t$ & triangle count \\
  $u, v, w$ & vertices \\
  $w$ & edge weight \\ 
  $w$ & network weight \\
  $w(\ldots)$ & weight function \\
  $x$ & cross count \\
  $y$ & reciprocity \\
  $z$ & claw count \\
  \midrule
  $\alpha$ & spectral norm \\
  $\beta$  & preferential attachment exponent \\
  $\gamma$ & power law exponent \\
  $\delta$ & diameter \\
  $\epsilon$ & eccentricity \\
  $\zeta$ & negativity \\
  $\eta$ & dyadic conflict \\
  $\lambda$ & eigenvalue \\
  $\mu$  & average edge weight \\
  $\nu$  & operator 2-norm \\
  $\xi$  & algebraic conflict \\
  $\pi$  & cyclic eigenvalue \\
  $\rho$ & assortativity \\
  $\sigma$ & singular value \\
  $\tau$ & triadic conflict \\
  $\phi$ & spectral signed frustration \\
  $\chi$ & algebraic non-bipartivity \\
  \midrule
  $C$   & controllability \\
  $C_k$ & $k$-cycle count \\
  $E$   & edge set \\
  $F$ & frustration \\
  $G$ & graph \\
  $G$ & Gini coefficient \\
  $H$ & entropy \\
  $K_k$ & $k$-clique count \\
  $N$   & size of largest connected component\\
  $P_k$ & $k$-path count \\
  $S_k$ & $k$-star count \\
  $T_k$ & $k$-tour count \\
  $V$   & vertex set \\
  $W_k$ & $k$-walk count \\
  \midrule
  $\mathbf 0$ & zeroes matrix \\
  $\mathbf 1$ & ones matrix \\
  $\mathbf A$ & adjacency matrix \\
  $\mathbf B$ & biadjacency matrix \\
  $\mathbf D$ & degree matrix \\
  $\mathbf E$ & incidence matrix \\ 
  $\mathbf F$ & line matrix \\
  $\mathbf G$ & PageRank matrix (``Google matrix'') \\
  $\mathbf H$ & Hermitian adjacency matrix \\
  $\mathbf I$ & identity matrix \\
  $\mathbf K$ & signless Laplacian matrix \\
  $\mathbf L$ & Laplacian matrix \\
  $\mathbf M$ & normalized biadjacency matrix \\
  $\mathbf N$ & normalized adjacency matrix \\
  $\mathbf P$ & stochastic adjacency matrix \\
  $\mathbf S$ & stochastic Laplacian matrix \\
  $\mathbf T$ & magnetic Laplacian matrix \\
  $\mathbf U, \mathbf V$ & eigenvector matrices \\
  $\mathbf W$ & Seidel adjacency matrix \\
  $\mathbf X$ & central matrix \\
  $\mathbf Y$ & skew-symmetric adjacency matrix \\
  $\mathbf Z$ & normalized Laplacian matrix \\
  \midrule
  $\mathbf \Gamma$ & Moore--Penrose pseudoinverse of $\mathbf L$ \\
  $\mathbf \Lambda$ & eigenvalue matrix \\
  $\mathbf \Sigma$ & singular value matrix \\
  \midrule
  $\bar G$ & unweighted graph \\
  $\bar{\bar G}$ & graph with unique edges \\
  $|G|$ & unsigned graph \\ 
  $-G$  & negated graph \\
  \midrule
  $\bar{\mathbf X}$   & explicitly undirected or symmetrized version of a characteristic matrix \\
  $\hat{\mathbf X}$   & repulsive version of a characteristic matrix \\
  $\breve{\mathbf X}$ & Hermitian version of a characteristic matrix \\
  $\acute{\mathbf X}$ & skew-Hermitian version of a characteristic matrix \\
  $\vec{\mathbf X}$   & unidirectional version of a characteristic matrix \\
  $\grave{\mathbf X}$ & bipartite version of a characteristic matrix \\
\end{longtable}

\end{document}